\documentclass[twocolumn, prb, noshowpacs, notitlepage]{revtex4-1}

\usepackage{dcolumn}
\usepackage{bm}
\usepackage{amsmath}
\usepackage{amssymb}
\usepackage{revsymb}
\usepackage{graphics}
\usepackage{graphicx}
\usepackage{natbib}
\usepackage{color}
\usepackage{gensymb}
\usepackage{accents}
\usepackage{textcomp}
\usepackage[hidelinks]{hyperref}

\allowdisplaybreaks
\everymath{\displaystyle}

\begin{document}
\title{Near-field thermal radiative transfer between two coated spheres}
\author{Braden Czapla}
\affiliation{Department of Mechanical Engineering, Columbia University, New York, NY 10027, USA
}
\author{Arvind Narayanaswamy}
\affiliation{Department of Mechanical Engineering, Columbia University, New York, NY 10027, USA
}
\email{arvind.narayanaswamy@columbia.edu}

\begin{abstract}
In this work, we present an expression for the near-field thermal radiative transfer between two spheres with an arbitrary numbers of coatings. We numerically demonstrate that the spectrum of heat transfer between layered spheres exhibits novel features due to the newly introduced interfaces between coatings and cores. These features include broad super-Planckian peaks at non-resonant frequencies and near-field selective emission between metallic spheres with polar material coatings. Spheres with cores and coatings of two different polar materials are also shown to exceed the total conductance of homogeneous spheres in some cases.
\end{abstract}

\maketitle

\section{Introduction} \label{Introduction}

Optical metamaterials are a class of artificial materials which exhibit electromagnetic behavior not otherwise observed in nature, such as negative refractive index,\cite{Smith2004, Zhang2005, Soukoulis2007, Valentine2008} cloaking, \cite{Alu2005, Schurig2006, Cai2007, Valentine2009} and superlensing\cite{Pendry2000, Fang2005, Smolyaninov2007, Zhang2008} to name a few. Of particular interest are hyperbolic metamaterials (HMMs), those whose permissible wave-vector components form a hyperbolic isofrequency surface instead of the spherical surface found in typical isotropic materials. The simplest means of achieving HMM behavior is through layering different isotropic materials. When the layer thicknesses are much smaller than the free-space wavelength of light propagating through them, the entire nanocomposite can be viewed as a homogeneous material with hyperbolic effective optical properties.\cite{Halevi1999, Smith2003}

Hyperbolic metamaterials have found many uses in the field of near-field thermal radiative transfer. Appropriately designed HMMs have demonstrated the ability to tailor the spectrum of radiative transfer and to achieve heat transfer beyond that of Planck's blackbody limit.\cite{Francoeur2011, Liu2011, Mason2011, Biehs2012, Guo2012, Guo2013, Liu2013} To date, this has been achieved mostly by using layered planar surfaces.\cite{Biehs2007, Fu2009, Svetovoy2012} That configuration is attractive because the analytic solution to near-field thermal radiative transfer between two semi-infinite half spaces is well known, relatively straightforward to compute, and easily generalizes to include layered media.\cite{Polder1971, Francoeur2008, Francoeur2009, Song2016}

Determining analytic solutions to more complicated geometries opens up additional avenues of investigation for HMMs. Two geometries of interest are sphere-sphere and sphere-plane configurations. The formula for heat transfer between two homogeneous spheres has been determined\cite{Narayanaswamy2008, Mackowski2008, Kruger2012} and can be used to approximate the sphere-plane configuration\cite{Otey2011} in the limit that one sphere is much greater than the other.\cite{Sasihithlu2014} The solution method in Ref. \onlinecite{Narayanaswamy2008} cannot be easily extended to include coated spheres. Though the formalisms used in Refs. \onlinecite{Mackowski2008, Kruger2012}, which are based on the T-matrix method, \cite{Waterman1965, Peterson1974} can in principle be used to calculate near-field radiation between coated spheres, the authors did not apply them for that purpose.

In this work, we present an expression for the near-field thermal radiative transfer between two spheres, either of which may have any number of coatings. The expression is derived using the framework of fluctuational electrodynamics and a Green's function formalism. The key advance in this work is evaluating the radiative transfer between spheres using surface integrals instead of volume integrals, which permits the same formalism and numerical code to be used for coated as well as uncoated spheres, irrespective of the number of coatings. This approach is well suited for near-field radiative transfer analysis of coated spheres because, as in the case of two homogeneous spheres, the main geometric  parameters of interest are still the center-to-center distance between the spheres (or alternatively the minimum gap between spheres) and the overall dimensions of each sphere, not the details of the coatings themselves. The details of the coatings on a sphere are encoded in the effective Mie reflection coefficients of the various vector spherical waves at the interface between the outer-most coating of that sphere and the intervening medium (usually vacuum).

We will show that coated spheres have some advantages over homogeneous spheres. Coated spheres exhibit broad, super-Planckian peaks (peaks exceeding that of blackbodies) which are due neither to surface phonon polariton nor surface plasmon polariton resonances and are not commonly observed in homogeneous spheres (see Fig. \ref{fig:SpectralConductance}). We will also show that coated spheres with silver cores and silica coatings transition from silica-like behavior at gaps much smaller than the coating thickness to silver-like behavior for larger gaps (see Fig. \ref{fig:TotalConductance}). Last, we present coated spheres whose total conductance exceeds that of homogeneous spheres of either of its constitutive components (see Fig. \ref{fig:TwoDielectrics}). These advantages make coated spheres a promising platform for research and experimentation.

The structure of the paper is as follows: In Sec. \ref{Geometry}, the geometry of the two sphere problem is described. In Sec. \ref{MathematicalFormulation}, the expression for thermal conductance is derived using the fluctuation-dissipation theorem, resulting in an expression which includes a transmissivity function for energy transfer. In Sec. \ref{DeterminationOfTransmissivityFunction}, the transmissivity function is evaluated from the dyadic Green's functions for the two coated spheres. Finally, in Sec. \ref{NumericalResults}, numerical results are presented for the conductance between two coated spheres.

\section{Geometry} \label{Geometry}

\begin{figure}
\includegraphics[width=3 in]{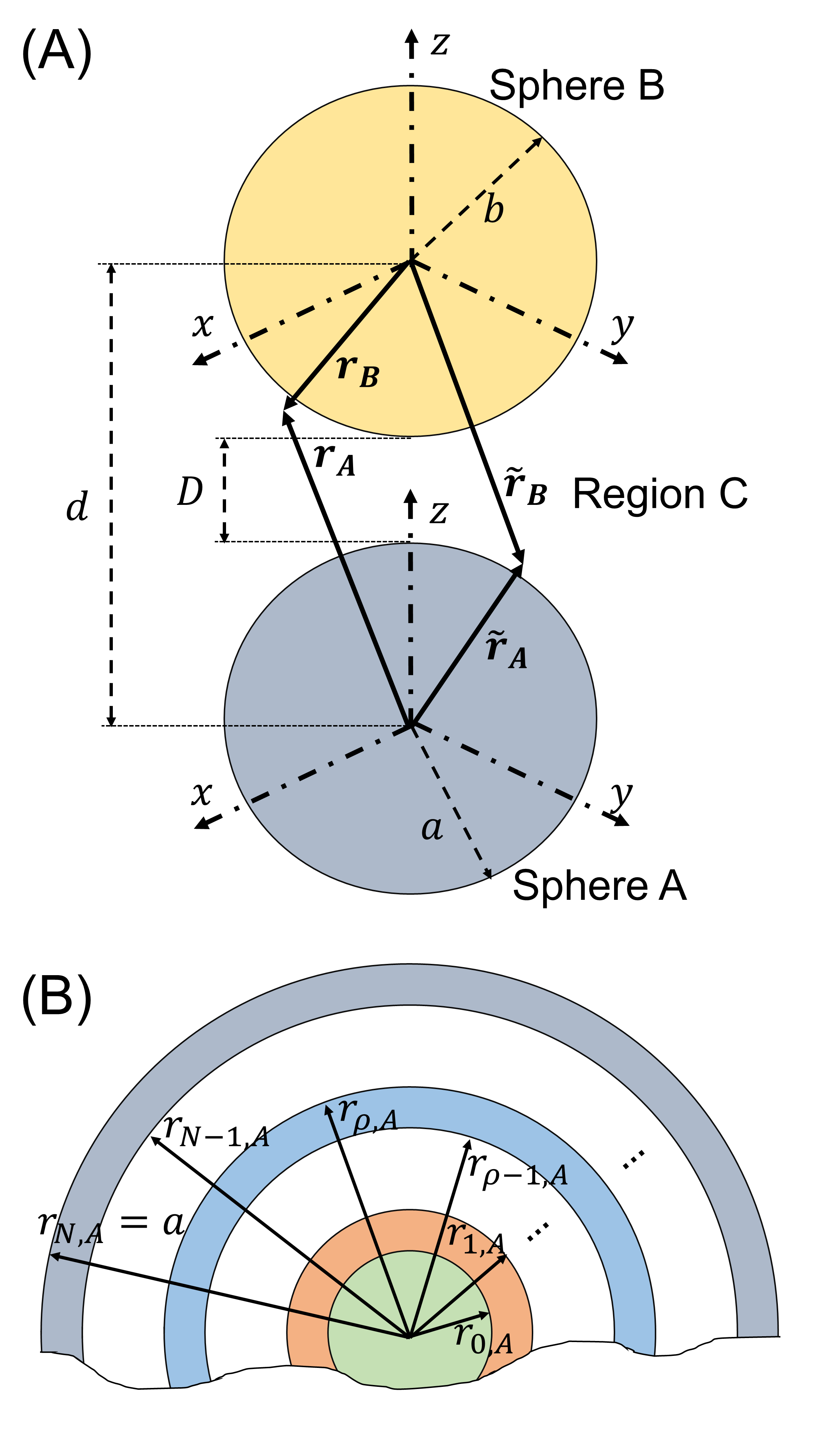}
\caption{\label{fig:CoatedSphereGeometry} Geometry of two-sphere problem: (A) exterior view of two sphere configuration and (B) interior view of sphere A. Sphere B is the similar, but with $M$ layers and outer radius $b$.}
\end{figure}

The geometry of the problem is shown in Fig. \ref{fig:CoatedSphereGeometry}. Two spheres, denoted $A$ and $B$, are composed of an arbitrary number of coatings, $N$ and $M$ coatings, respectively, atop a core. As shown in Fig \ref{fig:CoatedSphereGeometry}(B), the outer radius of any coating $\rho$ ($0 < \rho \le N$ for sphere $A$ and $0 < \rho \le M$ for sphere $B$) in sphere $s$ ($s=A$ or $B$) is given by $r_{\rho,s}$ and the outer radius of the core is given by $r_{0,s}$. For simplicity, the outermost radii of spheres $A$ and $B$ are denoted $r_{N,A}=a$ and $r_{M,B}=b$, respectively. Though only the internal structure of sphere $A$ is shown in Fig. \ref{fig:CoatedSphereGeometry}(B), sphere $B$ is similar, but with outer radius $b$ on layer $M$. The exterior region, denoted $C$, is vacuum.

As shown in Fig \ref{fig:CoatedSphereGeometry}(A), a coordinate system is placed with its origin at the center of sphere $A$, henceforth referred to as the $A$-coordinate system. Similarly, a second coordinate system is placed with its origin at the center of sphere $B$ and referred to as the $B$-coordinate system. The two coordinate systems are oriented such that their $z$ axes are aligned along the vector connecting the origin of the $A$-coordinate system to that of the $B$-coordinate system. The spheres are separated by a center-to-center gap $d$. The minimum gap separating spheres $A$ and $B$ is $D=d-a-b$.

The same location may be represented by the position vector $\boldsymbol{r_{A}}$ in the $A$-coordinate system or $\boldsymbol{r_{B}}$ in the $B$-coordinate system. In order to differentiate between two different locations, we will use $\boldsymbol{r}$ and $\boldsymbol{\widetilde{r}}$, the need for which arises from the use of Green's functions.

\section{Mathematical formulation} \label{MathematicalFormulation}

The radiative heat transfer from object A to object B, $Q_{A\rightarrow B}$, is given by
\begin{align}
Q_{A\rightarrow B} &= - \oint_{S_B} \left[ \widehat{\boldsymbol{n}}_{B}(\boldsymbol{\widetilde{r}}) \cdot \boldsymbol{P}(\boldsymbol{\widetilde{r}}) \right] d\boldsymbol{\widetilde{r}},
\label{eq:heattransfer}
\end{align}
where $\widehat{\boldsymbol{n}}(\boldsymbol{\widetilde{r}})$ and $\boldsymbol{P}(\boldsymbol{\widetilde{r}})$ are the unit outward normal of surface $B$ and the Poynting vector, respectively, at the location $\boldsymbol{\widetilde{r}}$. The Poynting vector is related to the cross-spectral density of the components of the electric and magnetic fields and is defined as
\begin{align}
\boldsymbol{P}(\boldsymbol{\widetilde{r}}) &= \int_{0}^{\infty} \frac{d\omega}{2\pi} \left< \boldsymbol{E}(\boldsymbol{\widetilde{r}}) \times \boldsymbol{H}^{*}(\boldsymbol{\widetilde{r}}) + \boldsymbol{E}^{*}(\boldsymbol{\widetilde{r}}) \times \boldsymbol{H}(\boldsymbol{\widetilde{r}}) \right>,
\end{align}
where $\omega$ is the angular frequency, $\left< \cdot \right>$ denotes an ensemble average, $\left(\cdot\right)^{*}$ is the complex conjugate, and $\boldsymbol{E}(\boldsymbol{\widetilde{r}})$ and $\boldsymbol{H}(\boldsymbol{\widetilde{r}})$ are the Fourier-transformed electric and magnetic fields, respectively. The frequency dependence of $\boldsymbol{E}(\boldsymbol{\widetilde{r}})$, $\boldsymbol{H}(\boldsymbol{\widetilde{r}})$, and other fields and Green's functions are suppressed for ease of notation. Inside each sphere, $\boldsymbol{E}(\boldsymbol{\widetilde{r}})$ and $\boldsymbol{H}(\boldsymbol{\widetilde{r}})$ are given by
\begin{subequations}
\begin{align}
\boldsymbol{E}(\boldsymbol{\widetilde{r}}) &= \! \! \int\displaylimits_{V} \! \left[ \boldsymbol{p}(\boldsymbol{r}) \! \cdot \overline{\overline{\boldsymbol{G}}}_{e}(\boldsymbol{r}; \boldsymbol{\widetilde{r}}) - \! \boldsymbol{J^{m}}(\boldsymbol{r}) \! \cdot \overline{\overline{\boldsymbol{G}}}_{E}(\boldsymbol{r}; \boldsymbol{\widetilde{r}}) \right] \! d\boldsymbol{r},
\\
\boldsymbol{H}(\boldsymbol{\widetilde{r}}) &= \! \! \int\displaylimits_{V} \! \left[ \boldsymbol{m}(\boldsymbol{r}) \! \cdot \overline{\overline{\boldsymbol{G}}}_{m}(\boldsymbol{r}; \boldsymbol{\widetilde{r}}) + \! \boldsymbol{J^{e}}(\boldsymbol{r}) \! \cdot \overline{\overline{\boldsymbol{G}}}_{M}(\boldsymbol{r}; \boldsymbol{\widetilde{r}}) \right] \! d\boldsymbol{r},
\end{align}
\end{subequations}
where $\boldsymbol{p}(\boldsymbol{r}) = i \omega \mu_{0} \mu \boldsymbol{J^{e}}(\boldsymbol{r})$, $\boldsymbol{m}(\boldsymbol{r}) = i \omega \varepsilon_{0} \varepsilon \boldsymbol{J^{m}}(\boldsymbol{r})$, $i$ is the imaginary unit, $\mu_{0}$ and $\varepsilon_{0}$ are the permeability and permittivity of free space, $\mu$ and $\varepsilon$ are the relative permeability and permittivity, and $\overline{\overline{\boldsymbol{G}}}(\boldsymbol{r}; \boldsymbol{\widetilde{r}})$ is the dyadic Green's function (DGF) relating sources and fields at $\boldsymbol{r}$ and $\boldsymbol{\widetilde{r}}$. The subscripts on the DGFs denote the electric ($e$) and magnetic ($m$) variants, and we define $\overline{\overline{\boldsymbol{G}}}_{E}(\boldsymbol{r}; \boldsymbol{\widetilde{r}}) = \nabla \times \overline{\overline{\boldsymbol{G}}}_{e}(\boldsymbol{r}; \boldsymbol{\widetilde{r}})$ and $\overline{\overline{\boldsymbol{G}}}_{M}(\boldsymbol{r}; \boldsymbol{\widetilde{r}}) = \nabla \times \overline{\overline{\boldsymbol{G}}}_{m}(\boldsymbol{r}; \boldsymbol{\widetilde{r}})$, where $\nabla$ operates on functions involving $\boldsymbol{r}$ alone. More details about the DGF are given in Appendix \ref{FormOfTheDyadicGreensFunctions}.

$\boldsymbol{J^{e}}(\boldsymbol{r})$ and $\boldsymbol{J^{m}}(\boldsymbol{r})$ are the Fourier transforms of the electric and magnetic current densities, respectively. The spectral densities of the components of $\boldsymbol{J^{e}}(\boldsymbol{r})$ and $\boldsymbol{J^{m}}(\boldsymbol{r})$ are related by the fluctuation-dissipation theorem of the second kind: \cite{Rytov1967, Eckhardt1982} 
\begin{subequations}
\begin{align}
\left< \! \boldsymbol{J_{p}^{e}}(\boldsymbol{r}) \boldsymbol{J_{q}^{e*}}(\boldsymbol{r}) \right> &= 2 \omega \varepsilon_{0} \Im{\left( \varepsilon \right)} \Theta(\omega,T) \delta \! \left( \boldsymbol{r} - \boldsymbol{\widetilde{r}} \right) \delta_{pq},
\label{eq:flucdisp1}
\\
\left< \! \boldsymbol{J_{p}^{m}}(\boldsymbol{r}) \boldsymbol{J_{q}^{m*}}(\boldsymbol{r}) \right> &= 2 \omega \mu_{0} \Im{\left( \mu \right)} \Theta(\omega,T) \delta \! \left( \boldsymbol{r} - \boldsymbol{\widetilde{r}} \right) \delta_{pq},
\label{eq:flucdisp2}
\\
\left< \! \boldsymbol{J_{p}^{e}}(\boldsymbol{r}) \boldsymbol{J_{q}^{m*}}(\boldsymbol{r}) \right> &= 0,
\label{eq:flucdisp3}
\end{align}
\end{subequations}
where $p,q=1,2,3$ are the Cartesian components of the current densities, $\Im{\left(\cdot\right)}$ denotes the imaginary component, $\hbar$ is the reduced Planck's constant, $k_{b}$ is Boltzmann's constant, $T$ is the thermodynamic temperature, and $\Theta= (\hbar\omega/2) \coth{\left( \hbar\omega/2 k_{b} T\right)}$ is the average energy of a harmonic oscillator of frequency $\omega$ at temperature $T$.

Using Eqs. (\ref{eq:heattransfer})-(\ref{eq:flucdisp3}), the linearized spectral conductance between two objects is given by \cite{Narayanaswamy2013a} 
\begin{align}
G(\omega,T) &= \lim_{T_{A},T_{B} \rightarrow T} \frac{Q_{A\rightarrow B}}{T_{A}-T_{B}} =  \frac{k_{b} X^2}{\sinh^{2}(X)} T_{A \rightarrow B}^{e} \! \left( \omega \right),
\label{eq:SpecCond}
\end{align}
where $X = \hbar \omega / 2 k_{b} T$, and $T_{A \rightarrow B}^{e}(\omega)$ is the transmissivity function for energy transfer. The total linearized conductance is obtained from
\begin{align}
\label{eq:totalconductancedef}
G_{t}(T) = \int_{0}^{\infty} \frac{d\omega}{2\pi} G(\omega,T) = \int_{0}^{\infty} d\lambda G(\lambda,T),
\end{align}
where $c$ is the speed of light in vacuum and $\lambda=2\pi c/\omega$ is the free-space wavelength. The definition of total conductance in the second integral in Eq. (\ref{eq:totalconductancedef}) is the result of a change of integration variables from frequency to wavelength. In the discussion of the numerical results (Sec. \ref{NumericalResults}), the wavelength-dependent definition of spectral conductance will be employed.

\section{Determination of transmissivity function} \label{DeterminationOfTransmissivityFunction}

Since emission and absorption of electromagnetic waves are volumetric phenomena, it might seem intuitive to expect the expression for the transmissivity function for energy transfer to contain two volume integrals. Our previous work\cite{Narayanaswamy2013a} has shown that if the two volumes are isothermal, the properties of the vector Helmholtz equation [shown later as  Eq. (\ref{eq:VectorHelmholtzEquation}) in Appendix \ref{DeterminationOfTheDyadicGreensFunctions}] allow us to reduce the volume integrals into two surface integrals, so that the transmissivity function is given by
\begin{align}
& T_{A \rightarrow B}^{e} (\omega) = 2 \Re \text{Tr} \oint\limits_{S_B} d\boldsymbol{r} \oint\limits_{S_A} d\boldsymbol{\widetilde{r}} \Bigg[
\nonumber \\
& \frac{\omega^2}{c^2}
\left[ \boldsymbol{\widehat{n}}(\boldsymbol{r}) \times \mu_{C} \overline{\overline{\boldsymbol{G}}}_{e}(\boldsymbol{\boldsymbol{r}, \widetilde{r}}) \right]
\cdot \left[ \boldsymbol{\widehat{n}}(\boldsymbol{\widetilde{r}}) \times \varepsilon_{C} \overline{\overline{\boldsymbol{G}}}_{m}^{T}(\boldsymbol{r}, \boldsymbol{\widetilde{r}}) \right]^{*}
\nonumber \\
& + \left[ \boldsymbol{\widehat{n}}(\boldsymbol{r}) \times \overline{\overline{\boldsymbol{G}}}_{M}(\boldsymbol{r}, \boldsymbol{\widetilde{r}}) \right]
\cdot \left[ \boldsymbol{\widehat{n}}(\boldsymbol{\widetilde{r}}) \times \overline{\overline{\boldsymbol{G}}}_{E}^{T}(\boldsymbol{r}, \boldsymbol{\widetilde{r}}) \right]^{*} \Bigg],
\label{eq:GeneralTransmissivity}
\end{align}
where $\Re{\left(\cdot\right)}$ denotes the real part, $\text{Tr}\left(\cdot\right)$ denotes the trace, $\left(\cdot\right)^{T}$ is the transpose, and $\boldsymbol{r}$ and $\boldsymbol{\widetilde{r}}$ are locations on the surfaces $S_B$ and $S_A$ of spheres $B$ and $A$, respectively.

Although $\boldsymbol{r}$ and $\boldsymbol{\widetilde{r}}$ will be integrated over the surfaces of spheres $B$ and $A$, respectively, both position vectors must be defined within a medium. This gives us the option of defining the position vectors as either approaching the surfaces of the spheres from the inside or the outside of the spheres. Because the fluctuating charges responsible for emission and absorption are contained within the spheres, it would be natural to assume $\boldsymbol{r}$ and $\boldsymbol{\widetilde{r}}$ approach the spheres' surfaces from the inside. This is the approach favored in Ref. \onlinecite{Narayanaswamy2008} and is referred to as the ``interior formula" by \citeauthor{Narayanaswamy2013a}.\cite{Narayanaswamy2013a}  In the approach we adopt here, $\boldsymbol{r}$ and $\boldsymbol{\widetilde{r}}$ instead approach the surfaces of their respective spheres from region $C$, which is referred to as the ``exterior method."\cite{Narayanaswamy2013a} The advantage of this approach is that the same formalism can be used for uncoated spheres or spheres with any number of coatings.

Since $\boldsymbol{r}$ is integrated over the surface of sphere $B$ and $\boldsymbol{\widetilde{r}}$ is integrated over the surface of sphere $A$, it is most convenient to evaluate Eq. \ref{eq:GeneralTransmissivity} using DGFs with $\boldsymbol{r}$ appearing as $\boldsymbol{r_B}$ and $\boldsymbol{\widetilde{r}}$ appearing as $\boldsymbol{\widetilde{r}_A}$. The appropriate DGFs when $ \boldsymbol{\widetilde{r}} , \boldsymbol{r} \in C$  are given by (see Appendix \ref{SimplificationOfDyadicGreensFunctions} for a full discussion on their determination):
\begin{widetext}
\begin{subequations}
\begin{align}
& \overline{\overline{\boldsymbol{G}}}_{e}(\boldsymbol{r}; \boldsymbol{\widetilde{r}}) = ik_{C} \sum\limits_{m=-\infty}^{\infty} \sum\limits_{l=\widetilde{m}}^{\infty} \sum_{\nu = \widetilde{m}}^{\infty} (-1)^{m}
\nonumber \\*
& \times \quad \left\{ \begin{array}{r} \left[ \begin{array}{r}
D_{\nu,m}^{lM} \left[ \boldsymbol{M}_{\nu m}^{(1)}(k_{C}\boldsymbol{r_{B}}) + R_{\nu}^{(M)}(b) \boldsymbol{M}_{\nu m}^{(3)}(k_{C}\boldsymbol{r_B}) \right]
\\[8 pt]
+ D_{\nu,m}^{lN} \left[ \boldsymbol{N}_{\nu m}^{(1)}(k_{C}\boldsymbol{r_{B}}) + R_{\nu}^{(N)}(b) \boldsymbol{N}_{\nu m}^{(3)}(k_{C}\boldsymbol{r_B}) \right]
\end{array} \right]
\left[ \begin{array}{l} \boldsymbol{M}_{l, -m}^{(1)}(k_{C}\boldsymbol{\widetilde{r}_{A}}) + R_{l}^{(M)}(a) \boldsymbol{M}_{l, -m}^{(3)}(k_{C}\boldsymbol{\widetilde{r}_A}) \end{array} \right]
\\[20 pt]
+ \left[ \begin{array}{r}
J_{\nu,m}^{lM} \left[ \boldsymbol{M}_{\nu m}^{(1)}(k_{C}\boldsymbol{r_{B}}) + R_{\nu}^{(M)}(b) \boldsymbol{M}_{\nu m}^{(3)}(k_{C}\boldsymbol{r_B}) \right]
\\[8 pt]
+ J_{\nu,m}^{lN} \left[ \boldsymbol{N}_{\nu m}^{(1)}(k_{C}\boldsymbol{r_{B}}) + R_{\nu}^{(N)}(b) \boldsymbol{N}_{\nu m}^{(3)}(k_{C}\boldsymbol{r_B}) \right]
\end{array} \right]
\left[ \begin{array}{l} \boldsymbol{N}_{l, -m}^{(1)}(k_{C}\boldsymbol{\widetilde{r}_{A}}) + R_{l}^{(N)}(a) \boldsymbol{N}_{l, -m}^{(3)}(k_{C}\boldsymbol{\widetilde{r}_A}) \end{array} \right]
\end{array} \right\},
\label{eq:FinalGe}
\\[10 pt]
& \overline{\overline{\boldsymbol{G}}}_{E}(\boldsymbol{r}; \boldsymbol{\widetilde{r}}) = ik_{C}^{2} \sum\limits_{m=-\infty}^{\infty} \sum\limits_{l=\widetilde{m}}^{\infty} \sum_{\nu = \widetilde{m}}^{\infty} (-1)^{m}
\nonumber \\*
& \times \quad \left\{ \begin{array}{r} \left[ \begin{array}{r}
D_{\nu,m}^{lN} \left[ \boldsymbol{M}_{\nu m}^{(1)}(k_{C}\boldsymbol{r_{B}}) + R_{\nu}^{(N)}(b) \boldsymbol{M}_{\nu m}^{(3)}(k_{C}\boldsymbol{r_B}) \right]
\\[8 pt]
+ D_{\nu,m}^{lM} \left[ \boldsymbol{N}_{\nu m}^{(1)}(k_{C}\boldsymbol{r_{B}}) + R_{\nu}^{(M)}(b) \boldsymbol{N}_{\nu m}^{(3)}(k_{C}\boldsymbol{r_B}) \right]
\end{array} \right]
\left[ \begin{array}{l} \boldsymbol{M}_{l, -m}^{(1)}(k_{C}\boldsymbol{\widetilde{r}_{A}}) + R_{l}^{(M)}(a) \boldsymbol{M}_{l, -m}^{(3)}(k_{C}\boldsymbol{\widetilde{r}_A}) \end{array} \right]
\\[20 pt]
+ \left[ \begin{array}{r}
J_{\nu,m}^{lN} \left[ \boldsymbol{M}_{\nu m}^{(1)}(k_{C}\boldsymbol{r_{B}}) + R_{\nu}^{(N)}(b) \boldsymbol{M}_{\nu m}^{(3)}(k_{C}\boldsymbol{r_B}) \right]
\\[8 pt]
+ J_{\nu,m}^{lM} \left[ \boldsymbol{N}_{\nu m}^{(1)}(k_{C}\boldsymbol{r_{B}}) + R_{\nu}^{(M)}(b) \boldsymbol{N}_{\nu m}^{(3)}(k_{C}\boldsymbol{r_B}) \right]
\end{array} \right]
\left[ \begin{array}{l} \boldsymbol{N}_{l, -m}^{(1)}(k_{C}\boldsymbol{\widetilde{r}_{A}}) + R_{l}^{(N)}(a) \boldsymbol{N}_{l, -m}^{(3)}(k_{C}\boldsymbol{\widetilde{r}_A}) \end{array} \right]
\end{array} \right\},
\label{eq:FinalGE}
\\[10 pt]
& \overline{\overline{\boldsymbol{G}}}_{m}(\boldsymbol{r}; \boldsymbol{\widetilde{r}}) = ik_{C} \sum\limits_{m=-\infty}^{\infty} \sum\limits_{l=\widetilde{m}}^{\infty} \sum_{\nu = \widetilde{m}}^{\infty} (-1)^{m}
\nonumber \\*
& \times \quad \left\{ \begin{array}{r} \left[ \begin{array}{r}
J_{\nu,m}^{lN} \left[ \boldsymbol{M}_{\nu m}^{(1)}(k_{C}\boldsymbol{r_{B}}) + R_{\nu}^{(N)}(b) \boldsymbol{M}_{\nu m}^{(3)}(k_{C}\boldsymbol{r_B}) \right]
\\[8 pt]
+ J_{\nu,m}^{lM} \left[ \boldsymbol{N}_{\nu m}^{(1)}(k_{C}\boldsymbol{r_{B}}) + R_{\nu}^{(M)}(b) \boldsymbol{N}_{\nu m}^{(3)}(k_{C}\boldsymbol{r_B}) \right]
\end{array} \right]
\left[ \begin{array}{l} \boldsymbol{M}_{l, -m}^{(1)}(k_{C}\boldsymbol{\widetilde{r}_{A}}) + R_{l}^{(N)}(a) \boldsymbol{M}_{l, -m}^{(3)}(k_{C}\boldsymbol{\widetilde{r}_A}) \end{array} \right]
\\[20 pt]
+ \left[ \begin{array}{r}
D_{\nu,m}^{lN} \left[ \boldsymbol{M}_{\nu m}^{(1)}(k_{C}\boldsymbol{r_{B}}) + R_{\nu}^{(N)}(b) \boldsymbol{M}_{\nu m}^{(3)}(k_{C}\boldsymbol{r_B}) \right]
\\[8 pt]
+ D_{\nu,m}^{lM} \left[ \boldsymbol{N}_{\nu m}^{(1)}(k_{C}\boldsymbol{r_{B}}) + R_{\nu}^{(M)}(b) \boldsymbol{N}_{\nu m}^{(3)}(k_{C}\boldsymbol{r_B}) \right]
\end{array} \right]
\left[ \begin{array}{l} \boldsymbol{N}_{l, -m}^{(1)}(k_{C}\boldsymbol{\widetilde{r}_{A}}) + R_{l}^{(M)}(a) \boldsymbol{N}_{l, -m}^{(3)}(k_{C}\boldsymbol{\widetilde{r}_A}) \end{array} \right]
\end{array} \right\},
\label{eq:FinalGm}
\\[10 pt]
& \overline{\overline{\boldsymbol{G}}}_{M}(\boldsymbol{r}; \boldsymbol{\widetilde{r}}) = ik_{C}^{2} \sum\limits_{m=-\infty}^{\infty} \sum\limits_{l=\widetilde{m}}^{\infty} \sum_{\nu = \widetilde{m}}^{\infty} (-1)^{m}
\nonumber \\*
& \times \left\{ \begin{array}{r} \left[ \begin{array}{r}
J_{\nu,m}^{lM} \left[ \boldsymbol{M}_{\nu m}^{(1)}(k_{C}\boldsymbol{r_{B}}) + R_{\nu}^{(M)}(b) \boldsymbol{M}_{\nu m}^{(3)}(k_{C}\boldsymbol{r_B}) \right]
\\[8 pt]
+ J_{\nu,m}^{lN} \left[ \boldsymbol{N}_{\nu m}^{(1)}(k_{C}\boldsymbol{r_{B}}) + R_{\nu}^{(N)}(b) \boldsymbol{N}_{\nu m}^{(3)}(k_{C}\boldsymbol{r_B}) \right]
\end{array} \right]
\left[ \begin{array}{l} \boldsymbol{M}_{l, -m}^{(1)}(k_{C}\boldsymbol{\widetilde{r}_{A}}) + R_{l}^{(N)}(a) \boldsymbol{M}_{l, -m}^{(3)}(k_{C}\boldsymbol{\widetilde{r}_A}) \end{array} \right]
\\[20 pt]
+ \left[ \begin{array}{r}
D_{\nu,m}^{lM} \left[ \boldsymbol{M}_{\nu m}^{(1)}(k_{C}\boldsymbol{r_{B}}) + R_{\nu}^{(M)}(b) \boldsymbol{M}_{\nu m}^{(3)}(k_{C}\boldsymbol{r_B}) \right]
\\[8 pt]
+ D_{\nu,m}^{lN} \left[ \boldsymbol{N}_{\nu m}^{(1)}(k_{C}\boldsymbol{r_{B}}) + R_{\nu}^{(N)}(b) \boldsymbol{N}_{\nu m}^{(3)}(k_{C}\boldsymbol{r_B}) \right]
\end{array} \right]
\left[ \begin{array}{l} \boldsymbol{N}_{l, -m}^{(1)}(k_{C}\boldsymbol{\widetilde{r}_{A}}) + R_{l}^{(M)}(a) \boldsymbol{N}_{l, -m}^{(3)}(k_{C}\boldsymbol{\widetilde{r}_A}) \end{array} \right]
\end{array} \right\},
\label{eq:FinalGM}
\end{align}
\end{subequations}
\end{widetext}
where we define $\widetilde{m} = \max{\left\{ \left| m \right|, 1 \right\}}$ for compactness, $k = (\omega/c) \sqrt{\varepsilon \mu}$ is the magnitude of the wavevector, $\boldsymbol{M}_{l m}^{(p)}(k \boldsymbol{r})$  and $\boldsymbol{N}_{l m}^{(p)}(k \boldsymbol{r})$ ($p=1$ or $3$) are vector spherical waves (VSWs) of order $(l,m)$, and $R_{\nu}^{(M)}(r)$ and $R_{\nu}^{(N)}(r)$ are the Mie reflection coefficients at $r=\left| \boldsymbol{r} \right|$ for $\boldsymbol{M}_{\nu m}^{(1)}(k \boldsymbol{r})$ and $\boldsymbol{N}_{\nu m}^{(1)}(k \boldsymbol{r})$ waves, respectively (see Appendix \ref{ComputationOfEffectiveMieReflectionCoefficients} for definitions).

The value of $p$ determines the behavior of the VSWs in the radial direction. For $p=1$, $\boldsymbol{M}_{l m}^{(p)}(k \boldsymbol{r})$ is an incoming wave scaled radially by $z_{l}^{(1)}(kr)$, the spherical Bessel function of the first kind, where $r = |\boldsymbol{r}|$. For $p=3$, $\boldsymbol{M}_{l m}^{(p)}(k \boldsymbol{r})$ is an outgoing wave scaled radially by $z_{l}^{(3)}(kr)$, the spherical Hankel function of the first kind. The definition of $\boldsymbol{N}_{l m}^{(p)}(k \boldsymbol{r})$ and the relations between VSWs and vector spherical harmonics are given in Appendix \ref{MathematicalDefinitionsAndUsefulRelations}. $\boldsymbol{M}_{l m}^{(p)}(k \boldsymbol{r})$  and $\boldsymbol{N}_{l m}^{(p)}(k \boldsymbol{r})$ are related to one another by $k \boldsymbol{M}_{l m}^{(p)}(k \boldsymbol{r}) = \boldsymbol{\nabla} \times \boldsymbol{N}_{l m}^{(p)}(k \boldsymbol{r})$ and $k \boldsymbol{N}_{l m}^{(p)}(k \boldsymbol{r}) = \boldsymbol{\nabla} \times \boldsymbol{M}_{l m}^{(p)}(k \boldsymbol{r})$.   For any two VSWs $\boldsymbol{P}$ and $\boldsymbol{Q}$, the term $\boldsymbol{P} \boldsymbol{Q}$, such as those appearing in Eqs. (\ref{eq:FinalGe})-(\ref{eq:FinalGM}), denotes the dyadic product of the two vectors.  \cite{Lai2009}

$C^{M}$, $C^{N}$,  $D^{M}$, $D^{N}$, $F^{M}$, $F^{N}$, $J^{M}$, and $J^{N}$ are unknown coefficients multiplying the VSWs. They are related to one another through a set of coupled linear equations, given by
\begin{subequations}
\begin{align}
C_{n,m}^{lM} 
& -
\sum\limits_{\nu=\widetilde{m}}^{\infty} \left[ \begin{array}{r}D_{\nu,m}^{lM} R_{\nu}^{(M)}(b) A_{n m}^{\nu m}(k_{C} d_{BA}) \\ + D_{\nu,m}^{lN} R_{\nu}^{(N)}(b) B_{n m}^{\nu m}(k_{C} d_{BA}) \end{array} \right]
= 0,
\label{eq:BC1}
\\
C_{n,m}^{lN}
& - 
\sum\limits_{\nu=\widetilde{m}}^{\infty} \left[ \begin{array}{r} D_{\nu,m}^{lN} R_{\nu}^{(N)}(b) A_{n m}^{\nu m}(k_{C} d_{BA}) \\ + D_{\nu,m}^{lM} R_{\nu}^{(M)}(b) B_{n m}^{\nu m}(k_{C} d_{BA}) \end{array} \right]
= 0,
\\
\begin{split}
D_{n,m}^{lM}
& -
\sum\limits_{\nu=\widetilde{m}}^{\infty} \left[ \begin{array}{r}C_{\nu,m}^{lM} R_{\nu}^{(M)}(a) A_{n m}^{\nu m}(k_{C} d_{AB}) \\ + C_{\nu,m}^{lN} R_{\nu}^{(N)}(a) B_{n m}^{\nu m}(k_{C} d_{AB}) \end{array} \right] \\
&= A_{n m}^{l m}(k_{C}d_{AB}),
\end{split}
\label{eq:BC3}
\\
\begin{split}
D_{n,m}^{lN}
& - 
\sum\limits_{\nu=\widetilde{m}}^{\infty} \left[ \begin{array}{r} C_{\nu,m}^{lN} R_{\nu}^{(N)}(a) A_{n m}^{\nu m}(k_{C} d_{AB}) \\ + C_{\nu,m}^{lM} R_{\nu}^{(M)}(a) B_{n m}^{\nu m}(k_{C} d_{AB}) \end{array}\right] \\
& = B_{n m}^{l m}(k_{C}d_{AB}),
\label{eq:BC4}
\end{split}
\\
F_{n,m}^{lM}
& -
\sum\limits_{\nu=\widetilde{m}}^{\infty} \left[\begin{array}{r} J_{\nu,m}^{lM} R_{\nu}^{(M)}(b) A_{n m}^{\nu m}(k_{C} d_{BA}) \\ + J_{\nu,m}^{lN} R_{\nu}^{(N)}(b) B_{n m}^{\nu m}(k_{C} d_{BA}) \end{array} \right]
= 0,
\\
F_{n,m}^{lN}
& - 
\sum\limits_{\nu=\widetilde{m}}^{\infty} \left[\begin{array}{r} J_{\nu,m}^{lN} R_{\nu}^{(N)}(b) A_{n m}^{\nu m}(k_{C} d_{BA}) \\ + J_{\nu,m}^{lM} R_{\nu}^{(M)}(b) B_{n m}^{\nu m}(k_{C} d_{BA}) \end{array}\right]
= 0,
\\ 
\begin{split}
J_{n,m}^{lM}
& -
\sum\limits_{\nu=\widetilde{m}}^{\infty} \left[\begin{array}{r} F_{\nu,m}^{lM} R_{\nu}^{(M)}(a) A_{n m}^{\nu m}(k_{C} d_{AB}) \\ + F_{\nu,m}^{lN} R_{\nu}^{(N)}(a) B_{n m}^{\nu m}(k_{C} d_{AB}) \end{array} \right] \\
& = B_{n m}^{l m}(k_{C}d_{AB}),
\end{split}
\label{eq:BC7}
\\
\begin{split}
J_{n,m}^{lN}
& -
\sum\limits_{\nu=\widetilde{m}}^{\infty} \left[\begin{array}{r} F_{\nu,m}^{lN} R_{\nu}^{(N)}(a) A_{n m}^{\nu m}(k_{C} d_{AB}) \\ + F_{\nu,m}^{lM} R_{\nu}^{(M)}(a) B_{n m}^{\nu m}(k_{C} d_{AB})\end{array} \right] \\
&= A_{n m}^{l m}(k_{C}d_{AB}),
\end{split}
\label{eq:BC8}
\end{align}
\end{subequations}
where $A_{nm}^{\nu m}$ and $B_{nm}^{\nu m}$ are well-known translation coefficients \cite{Chew1992, Chew1995, Kim2004a, Dufva2008} and $d_{AB} = - d_{BA} = d$. The linear system is identical to that given by \citet{Mackowski1991} (see Eqs. (11) and (12) of Ref. [\onlinecite{Mackowski1991}]) but for one small difference: the scattered field coefficients used in this work, such as $C_{n,m}^{lM}$ and $C_{n,m}^{lN}$, are analogous to $-a_{mn}^{i}/\alpha_{n}^{i}$ and $-b_{nm}^{i}/\beta_{n}^{i}$ in \citeauthor{Mackowski1991}'s work. 

The linear system is obtained by writing the DGFs as expansions of their VSW eigenfunctions in the coordinate systems of both spheres and then reexpanding some of the VSWs such that they may all be written into a consistent coordinate system. This allows for boundary conditions on the DGFs to be enforced. The details of this procedure may be found in Appendix \ref{DeterminationOfTheDyadicGreensFunctions}. It is important to note that the principles of this procedure are well established for dealing with pairs or even clusters of spheres. Previous works have used this technique in not only a wide range of electromagnetic scattering problems,\cite{Mackowski1994, Xu1996} but also in other wave scattering problems, such as those found in acoustics.\cite{Gumerov2002, Letourneau2017} Though the linear system of equations in Eqs. (\ref{eq:BC1})-(\ref{eq:BC8}) is analogous to Waterman's T-matrix representation, \cite{Mishchenko1996, Mishchenko2013} additional difficulties are introduced  due to our interest in near-field phenomena. The criterion for convergence when modeling near-field phenomena is more stringent than for far-field phenomena alone.\cite{Sasihithlu2011a}

For the DGFs given by Eqs. \ref{eq:FinalGe}-\ref{eq:FinalGM}, the transmissivity function can be evaluated as
\begin{align}
& T_{A \rightarrow B}^{e}\left( \omega \right)
= \left( k_{C} a \right)^{2} \left( k_{C} b \right)^{2} \sum\limits_{m=-\infty}^{\infty} \sum\limits_{l=\widetilde{m}}^{\infty}\sum_{\nu = \widetilde{m}}^{\infty}
\nonumber \\*
& \times \left[ \begin{array}{r}
\left[
\epsilon_{\nu}^{(M)}(b)
\left| D_{\nu,m}^{lM} \right|^{2}
+ \epsilon_{\nu}^{(N)}(b)
\left| D_{\nu,m}^{lN} \right|^{2}
\right]
\epsilon_{l}^{(M)}(a)
\\ \\
+ \left[
\epsilon_{\nu}^{(M)}(b)
\left| J_{\nu,m}^{lM} \right|^{2}
+ \epsilon_{\nu}^{(N)}(b)
\left| J_{\nu,m}^{lN} \right|^{2}
\right]
\epsilon_{l}^{(N)}(a)
\end{array} \right],
\label{eq:Transmissivity}
\end{align}
where
\begin{align}
\epsilon_{\nu}^{(P)}(w) &= \frac{-2}{\left( k_{C} w \right)^{2}} \left[ \Re ( R_{\nu}^{(P)}(w) ) + \left| R_{\nu}^{(P)}(w) \right|^{2} \right],
\end{align}
and $P$ may be $M$ or $N$ and $w=a$ or $b$. Equation (\ref{eq:Transmissivity}) obeys the reciprocity principle. See Appendix \ref{ReciprocityofTransmissivityFunction} for proof.

It is worth noting that our choices of symbol and definition of $\epsilon_{\nu}^{(P)}(w)$ were not arbitrary. The spectral emissivity of an isolated sphere of radius $w$ is given by\cite{Kattawar1970}
\begin{align}
\epsilon\left( \omega \right)
= \sum\limits_{m=-\infty}^{\infty} \sum\limits_{l=\widetilde{m}}^{\infty}
\left[ \epsilon_{l}^{(M)}(w) + \epsilon_{l}^{(N)}(w) \right].
\end{align}

\section{Numerical results} \label{NumericalResults}

In this section, all numerical results will be shown for $T=300$ K. In order to compute the transmissivity function efficiently, it must be rewritten into a more computationally efficient form. See Appendix \ref{ComputationalImplementation} for details. Additionally, simulated values of spectral and total conductance will be normalized by the conductance between the same two spheres, assuming them to be blackbodies. The spectral conductance between two blackbodies is given by
\begin{align}
G_{BB}(\lambda, T) &= \left[ \frac{ \left( \frac{8 \pi^{3} c^{3} \hbar^{2}}{k_{b} T^{2} \lambda^{6}} \right) \exp{\left( \frac{2 \pi \hbar c}{k_{b} T \lambda} \right)} }{ \left[ \exp{\left( \frac{2 \pi \hbar c}{k_{b} T \lambda} \right)} - 1 \right]^{2} } \right] A_{A} F_{A \rightarrow B},
\end{align}
and the total conductance is given by
\begin{align}
G_{t,BB}(T) &= \int_{0}^{\infty} G_{BB}(\lambda, T) d\lambda = 4 \sigma T^{3} A_{A} F_{A \rightarrow B},
\end{align}
where $A_{A}$ is the surface area of object A, $F_{A \rightarrow B}$ is the radiative view factor from object A to object B, and $\sigma=\pi^{2} k_{b}^{4}/60c^{2}\hbar^{3}$ is the Stefan-Boltzmann constant.

\subsection{Dielectric coating atop metal core}

Planar stratified HMMs have previously been investigated for heat transfer applications due to their broadband super-Planckian thermal emission properties.\cite{Guo2012, Guo2013} Though use of non-planar layered media is relatively rare in the study of near-field heat transfer, a thermal microelectromechanical systems (MEMS) device with a layer of polar material atop a curved chromium sensor has been used in extreme near-field experiments by \citeauthor{Kim2015}.\cite{Kim2015} It is important to note that a single layer of material does not make the device an HMM. Regardless, our work may still give insight into the behavior of the device. Despite their device having coatings, \citeauthor{Kim2015} modeled their curved probe as homogeneous, composed of the polar material only. The authors provided only a \textit{post hoc} justification of this assumption: the seeming agreement between modeled and measured results.

Numerical investigation of heat transfer can shed light on the validity of such an assumption. The simplest test case is to simulate the heat transfer between two identical single-coated spheres. For the simulations here we use a metallic core and a dielectric coating, composed of silver\cite{Yang2015} and silica,\cite{Palik1985} respectively. Varying the spheres' dimensions, the position of the metal/dielectric interface and the separation gap allows for characterization of the impact of dielectric coatings atop metallic cores.

\begin{figure}
\includegraphics[width=3.5 in]{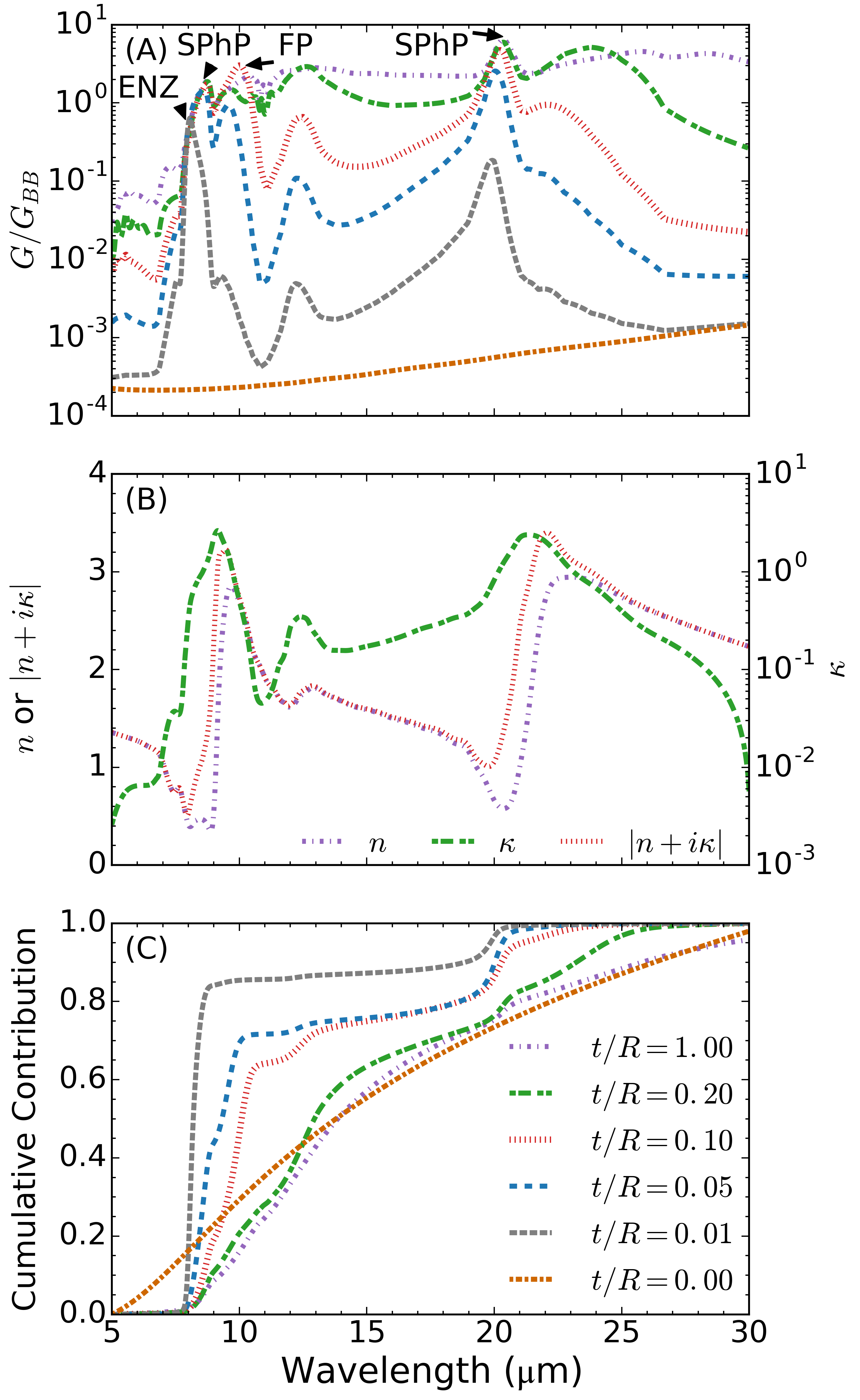}
\caption{\label{fig:SpectralConductance} (A) Spectral conductance between sets of identical coated spheres with varying core/coating interface positions (normalized by that of two blackbody spheres). Spheres have a silver core and silica layer with outer radii of 10 $\mu$m and minimum separation gap of 1 $\mu$m. Surface phonon polariton (SPhP), Fabry-Perot (FP), and epsilon near zero (ENZ) peaks are labeled. Legend appearing in (C) also applies to curves appearing in (A). (B) Real ($n$) and imaginary ($\kappa$) components and magnitude ($|n+i\kappa|$) of the complex refractive index of silica. (C) Cumulative spectral contribution to conductance for the curves depicted in (A). Select data points may be found in tabular form in the Supplemental Materials.\cite{SupplementalMaterials} }
\end{figure}

Figure \ref{fig:SpectralConductance} shows the effect of altering the position of the metal/dielectric interface on the spectrum of radiative heat transfer. The coated spheres have an outer radius, coating thickness, and core radius of $R$, $t$, and $R-t$, respectively. Their geometries are fixed such that $R=10$ $\mu$m and the minimum separation gap is $D=1$ $\mu$m.

The spectral conductance is shown in Fig. \ref{fig:SpectralConductance}(A). For the case of a homogeneous silica sphere ($t/R=1$), the result is a relatively wideband distribution of frequencies contributing to the radiative transfer which exactly reproduces the results of previous work. \cite{Narayanaswamy2008} The two well-known surface phonon polariton (SPhP) peaks are present at 8.75 $\mu$m and 20.3 $\mu$m [labeled in Fig. \ref{fig:SpectralConductance}(A)]. As the silver core is allowed to grow, the spectrum incrementally changes into the case of two bare silver spheres. Spheres with $t/R=0.1$, $0.05$, and $0.01$ exhibit spectral conductances which appear to be roughly scaled versions of each other, the scaling proportional to the thickness of the coating. 

The sequence in which the spectrum of radiative transfer for silica spheres transitions to that of silver spheres is not uniform across the spectrum. Although the magnitude of the spectral conductance of silver is always lower than that of silica, increasing the proportion of silver to silica may actually increase the spectral conductance for some wavelengths at some intermediate coating thicknesses. This is evident in the spectrum of spheres with $t/R=0.2$ at 12.5 $\mu$m and 23.5 $\mu$m and $t/R=0.1$ at the wavelength of 10 $\mu$m, where a broad super-Planckian peak, not associated with a SPhP, manifests. At these wavelengths, the conductance of the coated spheres exceeds that of pure silica spheres. 

When a thin layer of polar material (actually, the class of materials is broader and any material which has narrow absorption bands may serve as such a thin-film material) is coated on a metallic substrate, the wavelength at which the magnitude of the dielectric function (or equivalently the complex refractive index) of the polar material reaches a minimum, $\lambda_{ENZ}$ (ENZ denoting epsilon near zero), takes special significance.\cite{Narayanaswamy2014} At this wavelength alone, the interface between the coating and vacuum behaves as a highly reflective mirror. The interface between the metallic substrate and the thin film is highly reflective at all wavelengths considered here because of the high dielectric function of metals for mid-infrared wavelengths.

Near $\lambda_{ENZ}$, electromagnetic waves experience reflective conditions at both interfaces, leading to a larger number of reflections than at other wavelengths, if the thin film is not too absorptive. The result of a greater number of reflections is the appearance of an optically thicker film. Because amorphous silica has a relatively high damping, these interesting effects manifest themselves in the near field only when the thickness becomes very small. Amorphous silica has a $\lambda_{ENZ}$ point at 7.95 $\mu$m [see Fig. \ref{fig:SpectralConductance}(B)]. Hence, the stand-alone peak in Fig. \ref{fig:SpectralConductance}(A) for $t/R = 0.01$ at 8.06 $\mu$m is an epsilon near zero mode. As the thickness is increased, this peak can no longer be resolved because of its proximity to a SPhP peak.

Another class of peaks which appears in the spectrum of thermal radiative transfer of coated structures is the Fabry-Perot--like resonance. This type of resonance results from the interference of the multiple reflections of waves within a thin film. Because Fabry-Perot-like resonances require the constructive interference of waves, the location of the peak will drift as the thickness of the coating changes. This type of peak is evident in Fig. \ref{fig:SpectralConductance}(A) for $t/R = 0.1$ at 10.0 $\mu$m and $t/R = 0.05$ at 9.55 $\mu$m.

The cumulative spectral contribution to conductance is shown in Fig. \ref{fig:SpectralConductance}(C). The cumulative contribution at wavelength $\lambda$ is given by 
\begin{align}
CSC(\lambda) &= \frac{\int_{0}^{\lambda} d\lambda' G(\lambda',T)}{\int_{0}^{\infty} d\lambda' G(\lambda',T)},
\end{align}
where $\lambda'$ is a dummy integration variable. The slopes of the cumulative contribution curves indicate how relatively dominant a wavelength is in contributing to the total conductance. A greater slope indicates a greater relative contribution to the total conductance and vice versa.

\begin{figure}
\includegraphics[width=3.5 in]{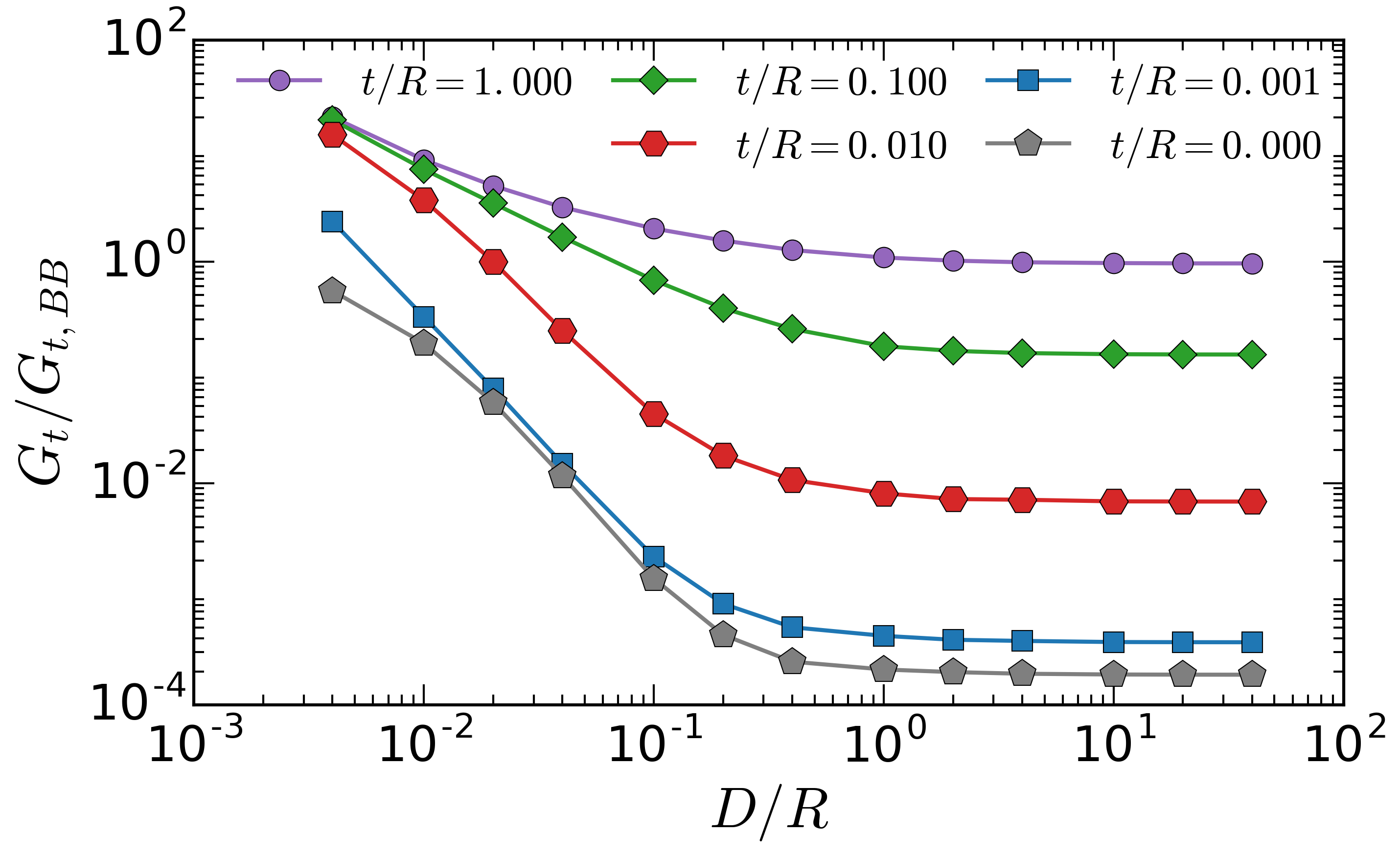}
\caption{\label{fig:TotalConductance} Distance dependence of total conductance between two identical coated spheres. Spheres have a silver core and silica layer with outer radii of 5 $\mu$m. Total conductance is normalized by that of two blackbodies, and the minimum separation gap is normalized by the outer radii of the spheres. Data points may be found in tabular form in the Supplemental Materials.\cite{SupplementalMaterials}}
\end{figure}

The curves for spheres with $t/R \ge 0.2$ have relatively small slopes across most wavelengths. This is consistent with the fairly wideband behavior demonstrated in Fig. \ref{fig:SpectralConductance}(A). As $t/R$ decreases, however, wavelengths differentiate into two categories: those with nearly zero slope and those with a very steep slope. For $t/R =0.10$ and $t/R =0.05$, the curves become nearly vertical at the SPhP wavelengths. In the most extreme case, for $t/R=0.01$, the curve is nearly vertical at 7.95 $\mu$m and 19.79 $\mu$m and nearly horizontal elsewhere. These wavelengths correspond to ENZ points of the silica layer.  If the minimum separation gap between the spheres were to be decreased, SPhP peaks would grow and eventually dominate over ENZ peaks. The dominance of SPhP modes in the extreme near field is made clear in the discussion of Fig. \ref{fig:TotalConductance}.

As we have shown, adding a very thin layer of a material supporting surface polaritons to a metallic substrate creates a selective near-field emitter. (This was already known to be true in the far field\cite{Granqvist1980, Narayanaswamy2014}.) Experimental measurement of spheres with very thin coatings would allow for probing of resonant heat transfer, some of which may be due to SPhPs, while suppressing heat transfer at other wavelengths. Because SPhPs are known to dominate heat transfer between polar materials in the extreme near field, isolating the contributions from SPhPs by using a coated sphere would serve as a superior experimental method compared to measuring the effects of SPhPs with homogeneous spheres as in past experiments.\cite{Narayanaswamy2008a, Shen2009, Guha2012}

Figure \ref{fig:TotalConductance} shows the effect of varying the separation gap between spheres with outer radii of 5 $\mu$m on total conductance. According to classical radiative transfer, the distance dependence in the far field is due to changes in view factor. Indeed, for gaps such that $D/R \ge 2$, all cases are well approximated as graybodies, as indicated by the curves' near-zero slopes. In that regime, the total conductance of spheres of constant radius increases as the fraction of silica increases.

As the separation gap decreases, the conductance between spheres with a silica coating begins to be dominated by the surface phonon polaritonic contributions. At a separation gap such that $D/R=0.004$, a coating of just 50 nm of silica can achieve 70\% of the conductance of a fully silica sphere. This allows for the creation of spheres with silicalike behavior in the near field but tunable radiative transfer behavior in the far-field. As a simple rule of thumb, the conductance between two silica coated silver spheres exceeds 70\% of that between two homogeneous silica spheres for $D/t \lesssim 1/4$. For larger gaps, the conductance is more like that of silver.

This observation partially validates the assumption made by \citeauthor{Kim2015}.\cite{Kim2015} Their device had a 100-nm silica coating atop an optically opaque chromium thermocouple. Their measurements were performed in the extreme near-field, at gaps ranging from 1 nm to 50 nm. For the smallest gaps, we have shown that the SPhP contributions will be dominant and modeling the whole body as homogeneous silica is a reasonable approximation. However, should the gaps of interest be larger or the materials not be dominated by SPhPs, more care must be taken to properly approximate near-field thermal radiative transfer of coated bodies.

\subsection{Dielectric coating atop dielectric core}

\begin{figure}
\includegraphics[width=3.5 in]{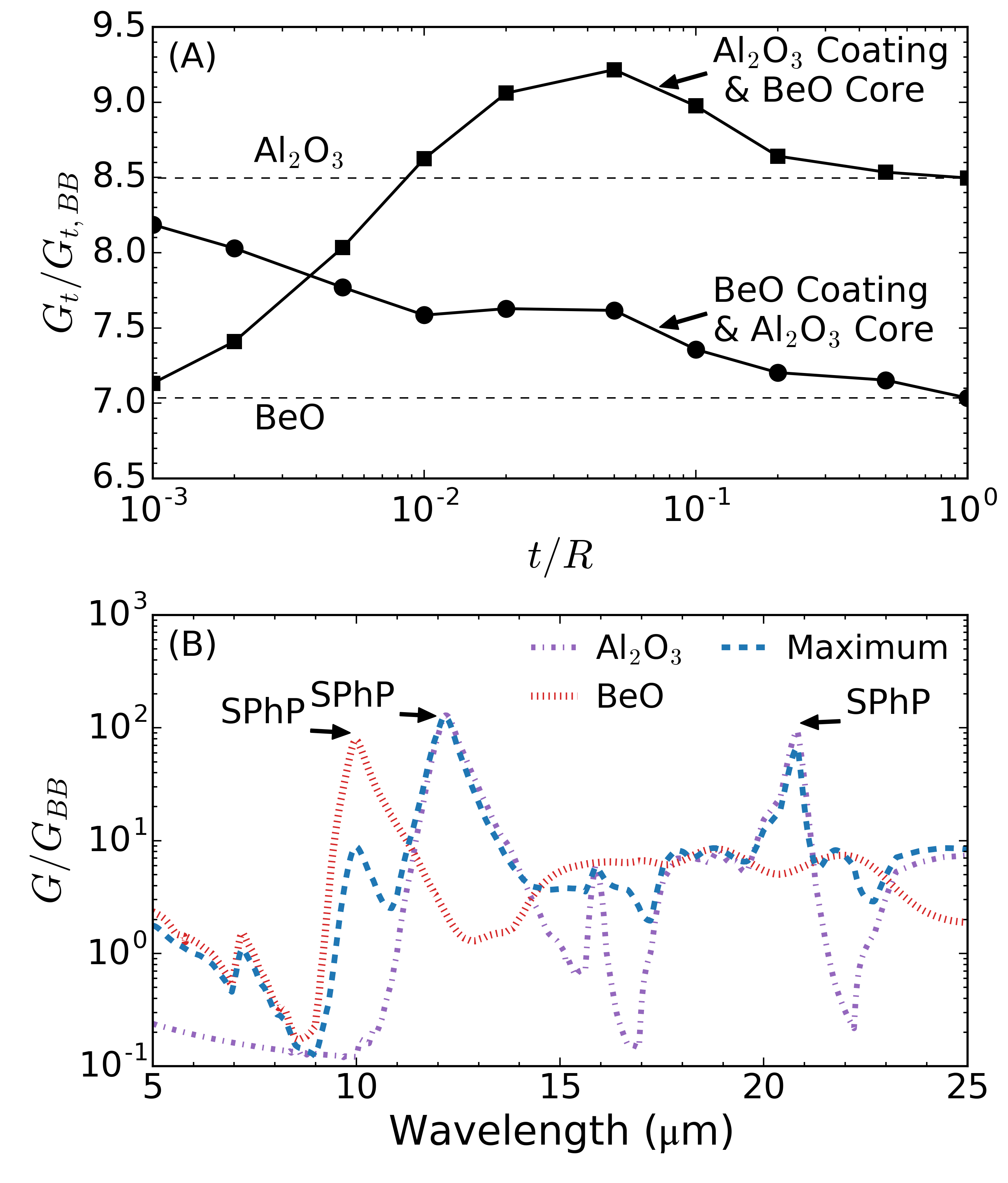}
\caption{\label{fig:TwoDielectrics} (A) Total conductance between two identical coated spheres (normalized by that of two blackbody spheres) as a function of the core/coating interface position. Spheres have a beryllia core and an alumina coating, or vice versa. The spheres have outer radii of 5 $\mu$m and a minimum separation gap of 100 nm. Dashed lines represent the total conductance of homogeneous beryllia and alumina spheres of the same geometry. Surface phonon polaritonic peaks (SPhP) for the homogeneous spheres are labeled. (B) Spectral conductance of spheres from (A) for homogeneous beryllia, homogeneous alumina, and the coated sphere with the maximum total conductance (alumina coating and beryllia core with $t/R=0.05$). All spectral conductances are normalized by that of two blackbody spheres. Select data points may be found in tabular form in the Supplemental Materials.\cite{SupplementalMaterials} }
\end{figure}

We have observed that two spheres with dielectric coatings and metallic cores have their total conductance effectively capped at that of two homogeneous spheres with the same dielectric material. A natural question is whether or not two coated spheres can ever exceed the total conductance of two homogeneous spheres which are composed of any of the coated spheres' constitutive materials. Because of the importance of SPhPs in near-field radiative heat transfer, we simulate the conductance between two coated spheres whose cores and coatings both support SPhPs. As shown in Fig. \ref{fig:TwoDielectrics}(A), we simulate identical spheres with beryllia\cite{Palik1985} cores and alumina\cite{Palik1985} coatings, or vice versa, with outer radii of 5 $\mu$m and a minimum separation gap of 100 nm. Those spheres simulated with an alumina coating and a beryllia core such that $0.01 \le t/R \le 0.5$ all exceed the total conductance between two homogeneous alumina spheres (which themselves exceed that of two homogeneous beryllia spheres). The maximum occurs at $t/R \approx 0.05$. Although spheres with beryllia coatings and alumina cores never exceed the total conductance of homogeneous alumina spheres, they too exhibit a slight local maximum at the same value of $t/R$. The maximum total conductance of the coated spheres outperforms homogeneous alumina spheres by 8.5\%.

When looking at the spectral conductance of the homogeneous spheres and the coated sphere with the maximum conductance, it becomes apparent how the coated spheres are able to outperform the homogeneous spheres. As shown in Fig. \ref{fig:TwoDielectrics}(B), the coated spheres exhibit spectral features similar to features found in the spectra of their components. Most importantly, the coated spheres strongly reproduce the SPhP peaks of homogeneous alumina at 12.2 $\mu$m and 20.8 $\mu$m while capturing a portion of the enhancement due to the SPhP peak of homogeneous beryllia at 10.0 $\mu$m (SPhP peaks labeled in Fig. \ref{fig:TwoDielectrics}). This suggests that it may be possible to ``stack" the effect of SPhPs at multiple wavelengths by choosing coatings of materials with spectrally spread SPhP peaks.

\section{Conclusions} \label{Conclusions}
In this paper, we have presented a formula for calculating the near-field thermal radiative transfer between two spheres, which allows for the inclusion of any number of coatings. Since we lack the formalism to analyze the radiative transfer between two homogeneous spheres with anisotropic properties, we are unable, at this stage, to replace the two coated spheres with equivalent hyperbolic metamaterial spheres. Instead, the effective properties of the coated spheres are characterized by the Mie reflection coefficients.

We have demonstrated that spheres with metallic cores and coatings of polar materials also exhibit super-Planckian peaks at wavelengths not corresponding to surface phonon polaritons. Such spheres behave like polar materials in the extreme near field but like metals in the far field. This creates the possibility of measuring near-field radiative transfer due to surface phonon polaritons while suppressing contributions from other modes of heat transfer. We have also shown that a coating of alumina atop a beryllia core can outperform homogeneous alumina or beryllia for an optimal coating thickness, demonstrating that the surface phonon polariton contribution from material inside the sphere may also contribute to the conductance between two spheres.

This work will also be useful in the development of methods to estimate the heat transfer between curved surfaces such as proximity approximations\cite{Rousseau2009, Sasihithlu2011, Song2015a} or the thermal discrete dipole approximation.\cite{Edalatpour2014, Edalatpour2016} Future work should be devoted to developing a formalism of near-field radiative transfer between spheres which permits anisotropic dielectric properties. 

\section{Acknowledgments}

This work was funded partially by ONR Grant N00014-12-1-0996 and NSF IGERT DGE-1069240.

\appendix

\section{Mathematical definitions and useful relations} \label{MathematicalDefinitionsAndUsefulRelations}

The vector spherical waves of interest are given by
\begin{align}
\boldsymbol{M}_{l m}^{(p)}(k \boldsymbol{r}) = & z_{l}^{(p)}(kr) \boldsymbol{V}_{lm}^{(2)}(\theta, \phi),
\\
\boldsymbol{N}_{l m}^{(p)}(k \boldsymbol{r}) = & \zeta_{l}^{(p)}(kr) \boldsymbol{V}_{lm}^{(3)}(\theta, \phi) \nonumber \\* &+ \frac{\sqrt{l(l+1)}}{kr} z_{l}^{(p)}(kr) \boldsymbol{V}_{lm}^{(1)}(\theta, \phi),
\end{align}
where $z_{l}^{(p)}(kr)$ is the spherical Bessel ($p=1$) or spherical Hankel ($p=3$) function of the first kind and $ \zeta_{l}^{(p)}(kr) = \frac{1}{kr}\frac{d}{d(kr)} ( kr z_{l}^{(p)}(kr) )$.  $\boldsymbol{V}_{lm}^{(1)}(\theta, \phi)$, $\boldsymbol{V}_{lm}^{(2)}(\theta, \phi)$, and $\boldsymbol{V}_{lm}^{(3)}(\theta, \phi)$ are the vector spherical harmonics of order $(l,m)$ for polar and azimuthal angles $\theta$ and $\phi$, respectively. They are defined as
\begin{align}
\boldsymbol{V}_{lm}^{(1)}(\theta, \phi) &= Y_{lm}(\theta, \phi) \boldsymbol{\widehat{r}},
\\
\boldsymbol{V}_{lm}^{(2)}(\theta, \phi) &= \frac{r}{\sqrt{l(l+1)}} \boldsymbol{\nabla} Y_{lm}(\theta, \phi) \times \boldsymbol{\widehat{r}},
\\
\boldsymbol{V}_{lm}^{(3)}(\theta, \phi) &= \frac{r}{\sqrt{l(l+1)}} \boldsymbol{\nabla} Y_{lm}(\theta, \phi),
\end{align}
where $Y_{lm}(\theta, \phi)$ is the scalar spherical harmonic of order $(l,m)$ and $\boldsymbol{\widehat{r}}$, $\boldsymbol{\widehat{\theta}}$, and $\boldsymbol{\widehat{\phi}}$ are the unit vectors of the spherical coordinate system. Scalar spherical harmonics are given by
\begin{align}
Y_{lm}(\theta, \phi) &= \sqrt{\frac{2l+1}{4\pi} \frac{(l-m)!}{(l+m)!}} P_{l}^{m}(\cos{\theta}) e^{im\phi},
\end{align}
where $P_{l}^{m}(x)$ is the associated Legendre polynomial. \cite{Olver2017}

The Wronskian of $z_{n}^{(1)}(x)$ and $z_{n}^{(3)}(x)$ is given by
\begin{align}
W[z_{n}^{(1)}(x), z_{n}^{(3)}(x)] &= z_{n}^{(1)}(x) \zeta_{n}^{(3)}(x) - z_{n}^{(3)}(x) \zeta_{n}^{(1)}(x) \nonumber
\\*
&= i / x^{2}.
\label{eq:Wronskian13}
\end{align}

The following integral identities are useful in evaluating surface integrals on the sphere:
\begin{align}
& \oint_{S} \boldsymbol{\widehat{r}} \cdot \left[ \boldsymbol{M}_{lm}^{(u)}(k_{1} \boldsymbol{r}) \times \boldsymbol{M}_{pq}^{(v)*}(k_{2} \boldsymbol{r}) \right] d\boldsymbol{r},
= 0
\\
& \oint_{S} \boldsymbol{\widehat{r}} \cdot \left[ \boldsymbol{N}_{lm}^{(u)}(k_{1} \boldsymbol{r}) \times \boldsymbol{N}_{pq}^{(v)*}(k_{2} \boldsymbol{r}) \right] d\boldsymbol{r},
= 0
\\
& \oint_{S} \boldsymbol{\widehat{r}} \cdot \left[ \boldsymbol{M}_{lm}^{(u)}(k_{1}\boldsymbol{r}) \times \boldsymbol{N}_{pq}^{(v)*}(k_{2} \boldsymbol{r}) \right] d\boldsymbol{r} \nonumber
\\*
&= r^2 z_{l}^{(u)}(k_{1} r) \zeta_{p}^{(v)*}(k_{2} r) \delta_{lp} \delta_{mq},
\\
& \oint_{S} \boldsymbol{\widehat{r}} \cdot \left[ \boldsymbol{N}_{lm}^{(v)}(k_{1}\boldsymbol{r}) \times \boldsymbol{M}_{pq}^{(v)*}(k_{2} \boldsymbol{r}) \right] d\boldsymbol{r} \nonumber
\\*
&= - r^2 \zeta_{l}^{(u)}(k_{1} r) z_{p}^{(v)*}(k_{2} r) \delta_{lp} \delta_{mq},
\end{align}

\section{Determination of the dyadic Green's functions} \label{DeterminationOfTheDyadicGreensFunctions}

\subsection{Form of the dyadic Green's functions} \label{FormOfTheDyadicGreensFunctions}

A dyadic Green's function gives the vectorial response at a location due to a vector source at another location, the two positions being the arguments of the DGF. A convenient method to compute the DGFs in Eq. (\ref{eq:GeneralTransmissivity}) is to expand them in terms of the eigenfunction solutions to the vector Helmholtz equation, given by
\begin{align}
\boldsymbol{\nabla} \times \boldsymbol{\nabla} \times \boldsymbol{F}(\boldsymbol{r}) - k^{2} \boldsymbol{F}(\boldsymbol{r}) = 0,
\label{eq:VectorHelmholtzEquation}
\end{align}
where $\boldsymbol{F}(\boldsymbol{r})$ is the electric or magnetic field at location $\boldsymbol{r}$ and $k = (\omega/c) \sqrt{\varepsilon \mu}$. In spherical coordinates, the eigenfunctions of the vector Helmholtz equation are the vector spherical waves $\boldsymbol{M}_{l m}^{(p)}(k \boldsymbol{r})$  and $\boldsymbol{N}_{l m}^{(p)}(k \boldsymbol{r})$.

Each region has a DGF composed of two parts: a homogeneous DGF, $\overline{\overline{\boldsymbol{G}}}_{0}(\boldsymbol{r}; \boldsymbol{\widetilde{r}})$, corresponding to waves which travel directly from $\boldsymbol{\widetilde{r}}$ to $\boldsymbol{r}$, and a scattered DGF corresponding to waves which have experienced scattering at inhomogeneities. When expanding a DGF into its VSW eigenfunctions, the choice of coordinate system for $\boldsymbol{r}$ and $\boldsymbol{\widetilde{r}}$ becomes important because they appear in the arguments of vector spherical waves. Assuming that $\boldsymbol{r}, \boldsymbol{\widetilde{r}} \in C$, $\overline{\overline{\boldsymbol{G}}}_{0}(\boldsymbol{r}; \boldsymbol{\widetilde{r}})$ can be written in the $A$-coordinate system as
\begin{widetext}
\begin{align}
\frac{\overline{\overline{\boldsymbol{G}}}_{0}(\boldsymbol{r}; \boldsymbol{\widetilde{r}})}{i k_{C}} &= \left\{ \begin{array}{lll}
\sum\limits_{m=-\infty}^{\infty} \sum\limits_{l=\widetilde{m}}^{\infty} (-1)^{m} \left\{ \boldsymbol{M}_{l m}^{(3)}(k_{C} \boldsymbol{r_A}) \boldsymbol{M}_{l, -m}^{(1)}(k_{C} \boldsymbol{\widetilde{r}_A}) + \boldsymbol{N}_{l m}^{(3)}(k_{C} \boldsymbol{r_A}) \boldsymbol{N}_{l, -m}^{(1)}(k_{C} \boldsymbol{\widetilde{r}_A}) \right\}
& \text{for } r_{A} > \widetilde{r}_{A}
\\
\sum\limits_{m=-\infty}^{\infty} \sum\limits_{l=\widetilde{m}}^{\infty} (-1)^{m} \left\{ \boldsymbol{M}_{l m}^{(1)}(k_{C} \boldsymbol{r_A}) \boldsymbol{M}_{l, -m}^{(3)}(k_{C} \boldsymbol{\widetilde{r}_A}) + \boldsymbol{N}_{l m}^{(1)}(k_{C} \boldsymbol{r_A}) \boldsymbol{N}_{l, -m}^{(3)}(k_{C} \boldsymbol{\widetilde{r}_A}) \right\}
& \text{for } r_{A} < \widetilde{r}_{A}
\end{array} \right. ,
\label{eq:DGF_homogeneous}
\end{align} 
\end{widetext}
where we define $\widetilde{m} = \max{\left\{ \left| m \right|, 1 \right\}}$. With respect to the center of the $A$-coordinate system, the homogeneous DGF for $r_{A} > \widetilde{r}_{A}$ ($r_{A} < \widetilde{r}_{A}$) corresponds to outgoing (incoming) VSWs at $\boldsymbol{r_A}$. The double surface integral in Eq. (\ref{eq:GeneralTransmissivity}) for computing $T_{A \rightarrow B}^{e}$ requires $\boldsymbol{r_A} \in S_B$ and $\boldsymbol{\widetilde{r}}_A \in S_A$.  Hence, we must choose the branch of $\overline{\overline{\boldsymbol{G}}}_{0}(\boldsymbol{r}; \boldsymbol{\widetilde{r}})$ for which $r_{A} > \widetilde{r}_{A}$.

The scattered DGF captures the collective effect of all scattering events at interfaces. The scattered DGF splits naturally into two parts: a part representing waves scattered off of a single sphere only, $\overline{\overline{\boldsymbol{G}}}_{e}^{(sc)\prime}(\boldsymbol{r}; \boldsymbol{\widetilde{r}})$, and a part representing waves scattered off of both spheres, $\overline{\overline{\boldsymbol{G}}}_{e}^{(sc)\prime\prime}(\boldsymbol{r}; \boldsymbol{\widetilde{r}})$. $\overline{\overline{\boldsymbol{G}}}_{e}^{(sc)\prime\prime}(\boldsymbol{r}; \boldsymbol{\widetilde{r}})$  obviously includes multiple reflections between the two spheres. Because we chose to write Eq. (\ref{eq:DGF_homogeneous}) in the $A$-coordinate system, we must also express $\overline{\overline{\boldsymbol{G}}}_{e}^{(sc)\prime}(\boldsymbol{r}; \boldsymbol{\widetilde{r}})$, representing waves scattered by sphere A only, in the $A$-coordinate system. That part of the scattered DGF is related to the branch of $\overline{\overline{\boldsymbol{G}}}_{0}(\boldsymbol{r}; \boldsymbol{\widetilde{r}})$ for $r_{A} < \widetilde{r}_{A}$. In that case, $\overline{\overline{\boldsymbol{G}}}_{e}^{(sc)\prime}(\boldsymbol{r}; \boldsymbol{\widetilde{r}})$ can be thought of as arising from VSWs emitted at $\boldsymbol{\widetilde{r}}$ which travel inward before reflecting off of sphere A and proceeding to $\boldsymbol{r}$. It is given by
\begin{align}
& \overline{\overline{\boldsymbol{G}}}_{e}^{(sc)\prime}(\boldsymbol{r}; \boldsymbol{\widetilde{r}}) = i k_{C} \sum\limits_{m=-\infty}^{\infty} \sum\limits_{l=\widetilde{m}}^{\infty} (-1)^{m} \nonumber \\*
& \times \left\{\begin{array}{l} R_{l}^{(M)}(a) \boldsymbol{M}_{l m}^{(3)}(k_{C} \boldsymbol{r_A}) \boldsymbol{M}_{l, -m}^{(3)}(k_{C} \boldsymbol{\widetilde{r}_A}) \\[8 pt] + R_{l}^{(N)}(a) \boldsymbol{N}_{l m}^{(3)}(k_{C} \boldsymbol{r_A}) \boldsymbol{N}_{l, -m}^{(3)}(k_{C} \boldsymbol{\widetilde{r}_A}) \end{array} \right\},
\label{eq:DGF_scattered1}
\end{align} 
where $R_{l}^{(M)}(a)$ and $R_{l}^{(N)}(a)$ are the Mie reflection coefficients at the surface of sphere A for $\boldsymbol{M}_{l m}^{(1)}(k_{C} \boldsymbol{\widetilde{r}_A})$ and $\boldsymbol{N}_{l m}^{(1)}(k_{C} \boldsymbol{\widetilde{r}_A})$ waves, respectively.

Some waves may reflect off of both spheres multiple times on their journey from $\boldsymbol{\widetilde{r}}$ to $\boldsymbol{r}$. The DGF which takes into account those multiple scatterings is given by
\begin{widetext}
\begin{align}
& \overline{\overline{\boldsymbol{G}}}_{e}^{(sc)\prime\prime}(\boldsymbol{r}; \boldsymbol{\widetilde{r}}) = i k_{C} \sum\limits_{m=-\infty}^{\infty} \sum\limits_{l=\widetilde{m}}^{\infty} \sum\limits_{\nu=\widetilde{m}}^{\infty} (-1)^{m} \times \nonumber
\\*
& \left\{ \! \begin{array}{r}
\! \! \left[ \begin{array}{r}
\! C_{\nu,m}^{lM^{(1)}} \boldsymbol{M}_{\nu m}^{(3)}(k_{C} \boldsymbol{r_A}) 
+ C_{\nu,m}^{lN^{(1)}} \boldsymbol{N}_{\nu m}^{(3)}(k_{C} \boldsymbol{r_A})
+ D_{\nu,m}^{lM^{(1)}} \boldsymbol{M}_{\nu m}^{(3)}(k_{C} \boldsymbol{r_B})
+ D_{\nu,m}^{lN^{(1)}} \boldsymbol{N}_{\nu m}^{(3)}(k_{C} \boldsymbol{r_B}) \!
\end{array} \right]
\! \boldsymbol{M}_{l, -m}^{(1)}(k_{C} \boldsymbol{\widetilde{r}_A}) \!
\\[8 pt]
\! \! + \left[ \begin{array}{r}
\! F_{\nu,m}^{lM^{(1)}} \boldsymbol{M}_{\nu m}^{(3)}(k_{C} \boldsymbol{r_A})
+ F_{\nu,m}^{lN^{(1)}} \boldsymbol{N}_{\nu m}^{(3)}(k_{C} \boldsymbol{r_A})
+ J_{\nu,m}^{lM^{(1)}} \boldsymbol{M}_{\nu m}^{(3)}(k_{C} \boldsymbol{r_B})
+ J_{\nu,m}^{lN^{(1)}} \boldsymbol{N}_{\nu m}^{(3)}(k_{C} \boldsymbol{r_B}) \!
\end{array} \right]
\! \boldsymbol{N}_{l, -m}^{(1)}(k_{C} \boldsymbol{\widetilde{r}_A}) \!
\\[8 pt]
\! \! + \left[ \begin{array}{r}
\! C_{\nu,m}^{lM^{(3)}} \boldsymbol{M}_{\nu m}^{(3)}(k_{C} \boldsymbol{r_A})
+ C_{\nu,m}^{lN^{(3)}} \boldsymbol{N}_{\nu m}^{(3)}(k_{C} \boldsymbol{r_A})
+ D_{\nu,m}^{lM^{(3)}} \boldsymbol{M}_{\nu m}^{(3)}(k_{C} \boldsymbol{r_B})
+ D_{\nu,m}^{lN^{(3)}} \boldsymbol{N}_{\nu m}^{(3)}(k_{C} \boldsymbol{r_B}) \!
\end{array} \right]
\! \boldsymbol{M}_{l, -m}^{(3)}(k_{C} \boldsymbol{\widetilde{r}_A}) \!
\\[8 pt]
\! \! + \left[ \begin{array}{r}
\! F_{\nu,m}^{lM^{(3)}} \boldsymbol{M}_{\nu m}^{(3)}(k_{C} \boldsymbol{r_A})
+ F_{\nu,m}^{lN^{(3)}} \boldsymbol{N}_{\nu m}^{(3)}(k_{C} \boldsymbol{r_A})
+ J_{\nu,m}^{lM^{(3)}} \boldsymbol{M}_{\nu m}^{(3)}(k_{C} \boldsymbol{r_B})
+ J_{\nu,m}^{lN^{(3)}} \boldsymbol{N}_{\nu m}^{(3)}(k_{C} \boldsymbol{r_B}) \!
\end{array} \right]
\! \boldsymbol{N}_{l, -m}^{(3)}(k_{C} \boldsymbol{\widetilde{r}_A}) \!
\end{array} \! \right\},
\label{eq:DGF_scattered2}
\end{align}
\end{widetext}
where the coefficients on the VSWs are unknowns to be determined from the boundary conditions, which is discussed shortly.

All DGFs of the form discussed in this work are composed of dyadic products\cite{Lai2009} of VSWs. For any dyadic product, the VSW to the right can be any of the VSWs to the right in  $\overline{\overline{\boldsymbol{G}}}_{0}(\boldsymbol{r}; \boldsymbol{\widetilde{r}})$, i.e., $\boldsymbol{M}_{l, -m}^{(1)}(k_{C} \boldsymbol{\widetilde{r}_A})$, $\boldsymbol{N}_{l, -m}^{(1)}(k_{C} \boldsymbol{\widetilde{r}_A})$, $\boldsymbol{M}_{l, -m}^{(3)}(k_{C} \boldsymbol{\widetilde{r}_A})$, or $\boldsymbol{N}_{l, -m}^{(3)}(k_{C} \boldsymbol{\widetilde{r}_A})$. The vector to the left has to be an outgoing VSW in either of the coordinate systems in order to satisfy the far-field boundary conditions. Hence, the vector to the left can be a linear combination of the following VSWs: $\boldsymbol{M}_{\nu m}^{(3)}(k_{C} \boldsymbol{r_{A}})$, $\boldsymbol{N}_{\nu m}^{(3)}(k_{C} \boldsymbol{r_A})$, $\boldsymbol{M}_{\nu m}^{(3)}(k_{C} \boldsymbol{r_B})$, and $\boldsymbol{N}_{\nu m}^{(3)}(k_{C} \boldsymbol{r_B})$. The expression in Eq. (\ref{eq:DGF_scattered2}) takes into account all these possibilities.

The magnetic DGF takes the same form as the electric DGF but with a corresponding set of unknown magnetic coefficients, denoted with a tilde. That is to say that each $C$, $D$, $F$, and $J$   has a corresponding magnetic counterpart: $\widetilde{C}$, $\widetilde{D}$, $\widetilde{F}$, and $\widetilde{J}$. The same holds true for the Mie reflection coefficients --- every $R$ has a counterpart $\widetilde{R}$. As will become apparent, only the waves in the DGFs representing scattering off of both sphere $A$ and sphere $B$ will contribute to the heat transfer between the two spheres (see Appendix \ref{SimplificationOfDyadicGreensFunctions} for more details). This is completely analogous to the result for heat transfer between two semi-infinite half spaces.\cite{Narayanaswamy2013a}

In order to simplify the DGFs and determine the unknown coefficients, two steps must be taken. First, the scattered fields must be converted into a single coordinate system. Second, boundary conditions on the DGFs must be enforced at the interfaces between different media. This will allow us to express the boundary conditions in terms of the coefficients multiplying the VSWs.

The vector addition translation theorem is used to convert the coordinate system of VSWs.\cite{Chew1992, Chew1995, Kim2004a, Dufva2008} To convert an outgoing VSW in the $B$-coordinate system into VSWs in the $A$-coordinate system,  the following expression can be used:
\begin{align}
\boldsymbol{M}_{\nu m}^{(3)}(k \boldsymbol{r_B}) &= \! \! \! \sum\limits_{n=\widetilde{m}}^{\infty} \! \left[ \! \begin{array}{l} A_{nm}^{\nu m}(k d_{BA}) \boldsymbol{M}_{nm}^{(1)}(k \boldsymbol{r_A}) \\ + B_{nm}^{\nu m}(k d_{BA}) \boldsymbol{N}_{nm}^{(1)}(k \boldsymbol{r_A}) \end{array} \! \right], \label{eq:Translation1} \\
\boldsymbol{N}_{\nu m}^{(3)}(k \boldsymbol{r_B}) &= \! \! \! \sum\limits_{n=\widetilde{m}}^{\infty} \! \left[ \! \begin{array}{l}  B_{nm}^{\nu m}(k d_{BA}) \boldsymbol{M}_{nm}^{(1)}(k \boldsymbol{r_A}) \\ + A_{nm}^{\nu m}(k d_{BA}) \boldsymbol{N}_{nm}^{(1)}(k \boldsymbol{r_A}) \end{array} \! \right], \label{eq:Translation2}
\end{align}
when the translation $\boldsymbol{r_B} - \boldsymbol{r_A}  = \hat{z} d_{BA} $ is restricted to the $z$ axis and $\left| \boldsymbol{r_A} \right| < |d_{BA}| = d$. 

Similarly, for conversion of outgoing VSWs from the $A$-coordinate system to VSWs in the $B$-coordinate system, we use
\begin{align}
\boldsymbol{M}_{\nu m}^{(3)}(k \boldsymbol{r_A}) &= \! \! \! \sum\limits_{n=\widetilde{m}}^{\infty} \! \left[ \! \begin{array}{l} A_{nm}^{\nu m}(k d_{AB}) \boldsymbol{M}_{nm}^{(1)}(k \boldsymbol{r_B}) \\ + B_{nm}^{\nu m}(k d_{AB}) \boldsymbol{N}_{nm}^{(1)}(k \boldsymbol{r_B}) \end{array} \! \right], \label{eq:Translation3} \\
\boldsymbol{N}_{\nu m}^{(3)}(k \boldsymbol{r_A}) &= \! \! \! \sum\limits_{n=\widetilde{m}}^{\infty} \! \left[ \! \begin{array}{l} B_{nm}^{\nu m}(k d_{AB}) \boldsymbol{M}_{nm}^{(1)}(k \boldsymbol{r_B}) \\ + A_{nm}^{\nu m}(k d_{AB}) \boldsymbol{N}_{nm}^{(1)}(k \boldsymbol{r_B}) \end{array} \! \right], \label{eq:Translation4}
\end{align}
where $\boldsymbol{r_A} - \boldsymbol{r_B}  = \hat{z}  d_{AB} $ and $\left| \boldsymbol{r_B} \right| < |d_{AB}| = d$.

At any location $\boldsymbol{r}$ on a boundary and for any $\boldsymbol{\widetilde{r}}$, the DGFs must satisfy
\begin{align}
& \text{Continuity of } \boldsymbol{\widehat{n}}(\boldsymbol{r}) \times \mu(\boldsymbol{r}) \overline{\overline{\boldsymbol{G}}}_{e}(\boldsymbol{r}; \boldsymbol{\widetilde{r}}), \label{eq:GDF_BC1}
\\
& \text{Continuity of } \boldsymbol{\widehat{n}}(\boldsymbol{r}) \times \overline{\overline{\boldsymbol{G}}}_{E}(\boldsymbol{r}; \boldsymbol{\widetilde{r}}),
\\
& \text{Continuity of } \boldsymbol{\widehat{n}}(\boldsymbol{r}) \times \varepsilon(\boldsymbol{r}) \overline{\overline{\boldsymbol{G}}}_{m}(\boldsymbol{r}; \boldsymbol{\widetilde{r}}),
\\
& \text{Continuity of } \boldsymbol{\widehat{n}}(\boldsymbol{r}) \times \overline{\overline{\boldsymbol{G}}}_{M}(\boldsymbol{r}; \boldsymbol{\widetilde{r}}), \label{eq:GDF_BC4}
\end{align}

Utilizing Eqs. (\ref{eq:DGF_homogeneous})-(\ref{eq:GDF_BC4}) leads to a set of linear equations between the coefficients multiplying the various VSWs that appear in Eq. (\ref{eq:DGF_scattered2}). Some details have been omitted here. For example, the DGFs used to derive the linear equations not only include those when $\boldsymbol{r}$ is in region $C$ [Eqs. (\ref{eq:DGF_homogeneous})-(\ref{eq:DGF_scattered2})], but also those when $\boldsymbol{r}$ is inside spheres $A$ and $B$ (not given in this paper). The resultant set of linear equations is given in Eqs. (\ref{eq:BC1})-(\ref{eq:BC8}) where
\begin{align}
D_{\nu,m}^{lM}
&= \frac{D_{\nu,m}^{lM^{(1)}}}{R_{\nu}^{(M)}(b)}
= \frac{D_{\nu,m}^{lM^{(3)}}}{R_{l}^{(M)}(a) R_{\nu}^{(M)}(b)},
\label{eq:VSWCoeff1}
\\
D_{\nu,m}^{lN}
&= \frac{D_{\nu,m}^{lN^{(1)}}}{R_{\nu}^{(N)}(b)}
= \frac{D_{\nu,m}^{lN^{(3)}}}{R_{l}^{(M)}(a) R_{\nu}^{(N)}(b)},
\\
J_{\nu,m}^{lM}
&= \frac{J_{\nu,m}^{lM^{(1)}}}{R_{\nu}^{(M)}(b)}
= \frac{J_{\nu,m}^{lM^{(3)}}}{R_{l}^{(N)}(a) R_{\nu}^{(M)}(b)},
\\
J_{\nu,m}^{lN}
&= \frac{J_{\nu,m}^{lN^{(1)}}}{R_{\nu}^{(N)}(b)}
= \frac{J_{\nu,m}^{lN^{(3)}}}{R_{l}^{(N)}(a) R_{\nu}^{(N)}(b)},
\\
C_{\nu,m}^{lM}
&= \frac{C_{\nu,m}^{lM^{(1)}}}{R_{\nu}^{(M)}(a)}
= \frac{C_{\nu,m}^{lM^{(3)}}}{R_{l}^{(M)}(a)R_{\nu}^{(M)}(a)},
\\
C_{\nu,m}^{lN}
&= \frac{C_{\nu,m}^{lN^{(1)}}}{R_{\nu}^{(N)}(a)}
= \frac{C_{\nu,m}^{lN^{(3)}}}{R_{l}^{(M)}(a)R_{\nu}^{(N)}(a)},
\\
F_{\nu,m}^{lM}
&= \frac{F_{\nu,m}^{lM^{(1)}}}{R_{\nu}^{(M)}(a)}
= \frac{F_{\nu,m}^{lM^{(3)}}}{R_{l}^{(N)}(a)R_{\nu}^{(M)}(a)},
\\
F_{\nu,m}^{lN}
&= \frac{F_{\nu,m}^{lN^{(1)}}}{R_{\nu}^{(N)}(a)}
= \frac{F_{\nu,m}^{lN^{(3)}}}{R_{l}^{(N)}(a)R_{\nu}^{(N)}(a)}.
\label{eq:VSWCoeff8}
\end{align}

The coefficients for the magnetic DGFs, $\widetilde{C}$, $\widetilde{D}$, $\widetilde{F}$, and  $\widetilde{J}$ [see discussion following Eq. (\ref{eq:DGF_scattered2})], are related to the coefficients of the electric DGFs by interchanging $\varepsilon$ and $\mu$ and noticing $\widetilde{R}_{\nu}^{(M)}(r) = R_{\nu}^{(N)}(r)$ and $\widetilde{R}_{\nu}^{(N)}(r) = R_{\nu}^{(M)}(r)$. Accordingly, we get
\begin{align}
\widetilde{D}_{\nu,m}^{lM}
&= J_{\nu,m}^{lN},
\label{eq:MagVSWCoeff1}
\\
\widetilde{D}_{\nu,m}^{lN}
&= J_{\nu,m}^{lM},
\\
\widetilde{J}_{\nu,m}^{lM}
&= D_{\nu,m}^{lN},
\\
\widetilde{J}_{\nu,m}^{lN}
&= D_{\nu,m}^{lM},
\\
\widetilde{C}_{\nu,m}^{lM}
&= F_{\nu,m}^{lN},
\\
\widetilde{C}_{\nu,m}^{lN}
&= F_{\nu,m}^{lM},
\\
\widetilde{F}_{\nu,m}^{lM}
&= C_{\nu,m}^{lN},
\\
\widetilde{F}_{\nu,m}^{lN}
&= C_{\nu,m}^{lM}.
\label{eq:MagVSWCoeff8}
\end{align}

\subsection{Simplification of dyadic Green's function} \label{SimplificationOfDyadicGreensFunctions}

The derivation of the simplified electric DGF, previously presented as Eq. (\ref{eq:FinalGe}) in its final form, is given below. The electric DGF is given by
\begin{align}
\overline{\overline{\boldsymbol{G}}}_{e}(\boldsymbol{r}; \boldsymbol{\widetilde{r}}) &= \overline{\overline{\boldsymbol{G}}}_{0}(\boldsymbol{r}; \boldsymbol{\widetilde{r}}) + \overline{\overline{\boldsymbol{G}}}_{e}^{(sc)\prime}(\boldsymbol{r}; \boldsymbol{\widetilde{r}}) + \overline{\overline{\boldsymbol{G}}}_{e}^{(sc)\prime\prime}(\boldsymbol{r}; \boldsymbol{\widetilde{r}}),
\end{align}
 where $\overline{\overline{\boldsymbol{G}}}_{0}(\boldsymbol{r}; \boldsymbol{\widetilde{r}})$, $\overline{\overline{\boldsymbol{G}}}_{e}^{(sc)\prime}(\boldsymbol{r}; \boldsymbol{\widetilde{r}})$, and $\overline{\overline{\boldsymbol{G}}}_{e}^{(sc)\prime\prime}(\boldsymbol{r}; \boldsymbol{\widetilde{r}})$ are given by Eqs. (\ref{eq:DGF_homogeneous})-(\ref{eq:DGF_scattered2}), and represent scattering off of neither sphere, only sphere $A$, and spheres $A$ and $B$, respectively. For reasons which will become apparent shortly, we will group $\overline{\overline{\boldsymbol{G}}}_{0}(\boldsymbol{r}; \boldsymbol{\widetilde{r}})$ and $\overline{\overline{\boldsymbol{G}}}_{e}^{(sc)\prime}(\boldsymbol{r}; \boldsymbol{\widetilde{r}})$ together. $\overline{\overline{\boldsymbol{G}}}_{0}(\boldsymbol{r}; \boldsymbol{\widetilde{r}}) + \overline{\overline{\boldsymbol{G}}}_{e}^{(sc)\prime}(\boldsymbol{r}; \boldsymbol{\widetilde{r}})$ is  the DGF for an isolated sphere $A$ in the absence of sphere $B$.

Using Eqs. (\ref{eq:DGF_homogeneous}) and (\ref{eq:DGF_scattered1}), we get
\begin{align}
& \overline{\overline{\boldsymbol{G}}}_{0}(\boldsymbol{r}; \boldsymbol{\widetilde{r}}) + \overline{\overline{\boldsymbol{G}}}_{e}^{(sc)\prime}(\boldsymbol{r}; \boldsymbol{\widetilde{r}}) = i k_{C} \!\! \sum\limits_{m=-\infty}^{\infty} \sum\limits_{l=\widetilde{m}}^{\infty} \! (-1)^{m} \nonumber
\\*
& \times \left\{ \! \begin{array}{l}
\boldsymbol{M}_{l m}^{(3)}(k_{C} \boldsymbol{r_A})
\left[ \begin{array}{l} \boldsymbol{M}_{l, -m}^{(1)}(k_{C} \boldsymbol{\widetilde{r}_A}) \\[8 pt] + R_{l}^{(M)}(a) \boldsymbol{M}_{l, -m}^{(3)}(k_{C} \boldsymbol{\widetilde{r}_A}) \end{array} \right]
\\[16 pt]
+ \boldsymbol{N}_{l m}^{(3)}(k_{C} \boldsymbol{r_A})
\left[ \begin{array}{l} \boldsymbol{N}_{l, -m}^{(1)}(k_{C} \boldsymbol{\widetilde{r}_A}) \\[8 pt] + R_{l}^{(N)}(a) \boldsymbol{N}_{l, -m}^{(3)}(k_{C} \boldsymbol{\widetilde{r}_A}) \end{array} \right]
\end{array} \! \right\}.
\label{eq:G0andGeSc1}
\end{align}  
Using  Eqs. (\ref{eq:VSWCoeff1})-(\ref{eq:VSWCoeff8}), Eq. (\ref{eq:DGF_scattered2}) simplifies to
\begin{widetext}
\begin{align}
& \overline{\overline{\boldsymbol{G}}}_{e}^{(sc)\prime\prime}(\boldsymbol{r}; \boldsymbol{\widetilde{r}}) = i k_{C} \sum\limits_{m=-\infty}^{\infty} \sum\limits_{l=\widetilde{m}}^{\infty} \sum\limits_{\nu=\widetilde{m}}^{\infty} (-1)^{m} \times \nonumber
\\*
& \quad \left\{ \begin{array}{r}
\left[ \begin{array}{r}
C_{\nu,m}^{lM} R_{\nu}^{(M)}(a) \boldsymbol{M}_{\nu m}^{(3)}(k_{C} \boldsymbol{r_A}) 
+ C_{\nu,m}^{lN} R_{\nu}^{(N)}(a) \boldsymbol{N}_{\nu m}^{(3)}(k_{C} \boldsymbol{r_A})
\\[8 pt]
+ D_{\nu,m}^{lM} R_{\nu}^{(M)}(b) \boldsymbol{M}_{\nu m}^{(3)}(k_{C} \boldsymbol{r_B})
+ D_{\nu,m}^{lN} R_{\nu}^{(N)}(b) \boldsymbol{N}_{\nu m}^{(3)}(k_{C} \boldsymbol{r_B})
\end{array} \right]
\left[ \begin{array}{l} \boldsymbol{M}_{l, -m}^{(1)}(k_{C} \boldsymbol{\widetilde{r}_A}) \\[8 pt]+ R_{l}^{(M)}(a) \boldsymbol{M}_{l, -m}^{(3)}(k_{C} \boldsymbol{\widetilde{r}_A}) \end{array}\right]
\\[16 pt]
+ \left[ \begin{array}{r}
F_{\nu,m}^{lM} R_{\nu}^{(M)}(a) \boldsymbol{M}_{\nu m}^{(3)}(k_{C} \boldsymbol{r_A})
+ F_{\nu,m}^{lN} R_{\nu}^{(N)}(a) \boldsymbol{N}_{\nu m}^{(3)}(k_{C} \boldsymbol{r_A})
\\[8 pt]
+ J_{\nu,m}^{lM} R_{\nu}^{(M)}(b) \boldsymbol{M}_{\nu m}^{(3)}(k_{C} \boldsymbol{r_B})
+ J_{\nu,m}^{lN} R_{\nu}^{(N)}(b) \boldsymbol{N}_{\nu m}^{(3)}(k_{C} \boldsymbol{r_B})
\end{array} \right]
\left[ \begin{array}{l} \boldsymbol{N}_{l, -m}^{(1)}(k_{C} \boldsymbol{\widetilde{r}_A}) \\[8 pt]+ R_{l}^{(N)}(a) \boldsymbol{N}_{l, -m}^{(3)}(k_{C} \boldsymbol{\widetilde{r}_A}) \end{array} \right]
\end{array} \right\}.
\end{align}
\end{widetext}

 In order to conveniently evaluate $T_{A \rightarrow B}^e$ [Eq. (\ref{eq:GeneralTransmissivity})], it will prove useful to convert the VSWs on the left in the dyadic products of the DGF to the $B$-coordinate system. To do so, we employ the vector addition translation theorem, given in Eqs. (\ref{eq:Translation3}) and (\ref{eq:Translation4}). Accordingly, we get
\begin{widetext}
\begin{align}
& \overline{\overline{\boldsymbol{G}}}_{e}^{(sc)\prime\prime}(\boldsymbol{r}; \boldsymbol{\widetilde{r}}) = i k_{C} \sum\limits_{m=-\infty}^{\infty} \sum\limits_{l=\widetilde{m}}^{\infty} \sum\limits_{\nu=\widetilde{m}}^{\infty} (-1)^{m}
\nonumber \\*
& \times \quad \left\{ \begin{array}{l}
\left[ \begin{array}{r}
D_{\nu,m}^{lM} R_{\nu}^{(M)}(b) \boldsymbol{M}_{\nu m}^{(3)}(k_{C} \boldsymbol{r_B})
+ D_{\nu,m}^{lN} R_{\nu}^{(N)}(b) \boldsymbol{N}_{\nu m}^{(3)}(k_{C} \boldsymbol{r_B})
\end{array} \right]
\left[ \begin{array}{l} \boldsymbol{M}_{l, -m}^{(1)}(k_{C} \boldsymbol{\widetilde{r}_A}) \\[8 pt]+ R_{l}^{(M)}(a) \boldsymbol{M}_{l, -m}^{(3)}(k_{C} \boldsymbol{\widetilde{r}_A}) \end{array}\right]
\\[20 pt]
+ \left[ \begin{array}{r}
J_{\nu,m}^{lM} R_{\nu}^{(M)}(b) \boldsymbol{M}_{\nu m}^{(3)}(k_{C} \boldsymbol{r_B})
+ J_{\nu,m}^{lN} R_{\nu}^{(N)}(b) \boldsymbol{N}_{\nu m}^{(3)}(k_{C} \boldsymbol{r_B})
\end{array} \right]
\left[ \begin{array}{l} \boldsymbol{N}_{l, -m}^{(1)}(k_{C} \boldsymbol{\widetilde{r}_A}) \\[8 pt]+ R_{l}^{(N)}(a) \boldsymbol{N}_{l, -m}^{(3)}(k_{C} \boldsymbol{\widetilde{r}_A}) \end{array} \right]
\\[20 pt]
+ \sum\limits_{n=\widetilde{m}}^{\infty} \left[ \begin{array}{l} \left( \begin{array}{l} C_{\nu,m}^{lM} R_{\nu}^{(M)}(a) A_{nm}^{\nu m}(k_{C} d_{AB}) \\[8 pt]+ C_{\nu,m}^{lN} R_{\nu}^{(N)}(a) B_{nm}^{\nu m}(k_{C} d_{AB}) \end{array} \right) \boldsymbol{M}_{nm}^{(1)}(k_{C} \boldsymbol{r_B})
\\[8 pt]
+ \left( \begin{array}{l} C_{\nu,m}^{lN} R_{\nu}^{(N)}(a) A_{nm}^{\nu m}(k_{C} d_{AB}) \\[8 pt]+ C_{\nu,m}^{lM} R_{\nu}^{(M)}(a) B_{nm}^{\nu m}(k_{C} d_{AB}) \end{array} \right) \boldsymbol{N}_{nm}^{(1)}(k_{C} \boldsymbol{r_B}) \end{array} \right] 
\left[ \begin{array}{l} \boldsymbol{M}_{l, -m}^{(1)}(k_{C} \boldsymbol{\widetilde{r}_A}) \\[8 pt]+ R_{l}^{(M)}(a) \boldsymbol{M}_{l, -m}^{(3)}(k_{C} \boldsymbol{\widetilde{r}_A}) \end{array}\right]
\\[40 pt]
+ \sum\limits_{n=\widetilde{m}}^{\infty} \left[ \begin{array}{l} \left( \begin{array}{l} F_{\nu,m}^{lM} R_{\nu}^{(M)}(a) A_{nm}^{\nu m}(k_{C} d_{AB}) \\[8 pt]+ F_{\nu,m}^{lN} R_{\nu}^{(N)}(a) B_{nm}^{\nu m}(k_{C} d_{AB}) \end{array} \right) \boldsymbol{M}_{nm}^{(1)}(k_{C} \boldsymbol{r_B})
\\[8 pt]
+ \left( \begin{array}{l} F_{\nu,m}^{lN} R_{\nu}^{(N)}(a) A_{nm}^{\nu m}(k_{C} d_{AB}) \\[8 pt]+ F_{\nu,m}^{lM} R_{\nu}^{(M)}(a) B_{nm}^{\nu m}(k_{C} d_{AB}) \end{array} \right) \boldsymbol{N}_{nm}^{(1)}(k_{C} \boldsymbol{r_B}) \end{array} \right] 
\left[ \begin{array}{l} \boldsymbol{N}_{l, -m}^{(1)}(k_{C} \boldsymbol{\widetilde{r}_A}) \\[8 pt]+ R_{l}^{(N)}(a) \boldsymbol{N}_{l, -m}^{(3)}(k_{C} \boldsymbol{\widetilde{r}_A}) \end{array}\right]
\end{array} \right\}.
\label{eq:DGF_scattered3}
\end{align}
\end{widetext}

Equations (\ref{eq:BC3})-(\ref{eq:BC4}) and (\ref{eq:BC7})-(\ref{eq:BC8}) may then be used to eliminate summation over the index $\nu$ of the $C$ and $F$ coefficients in Eq. \ref{eq:DGF_scattered3}. After further simplification, we get
\begin{widetext}
\begin{align}
& \overline{\overline{\boldsymbol{G}}}_{e}^{(sc)\prime\prime}(\boldsymbol{r}; \boldsymbol{\widetilde{r}}) = i k_{C} \sum\limits_{m=-\infty}^{\infty} \sum\limits_{l=\widetilde{m}}^{\infty} \sum\limits_{\nu=\widetilde{m}}^{\infty} (-1)^{m} \nonumber
\\*
& \times \quad \left\{ \begin{array}{r}
\left[ \begin{array}{r}
D_{\nu,m}^{lM} \left[ \boldsymbol{M}_{\nu m}^{(1)}(k_{C} \boldsymbol{r_B}) + R_{\nu}^{(M)}(b) \boldsymbol{M}_{\nu m}^{(3)}(k_{C} \boldsymbol{r_B}) \right]
\\[8 pt]
+ D_{\nu,m}^{lN} \left[ \boldsymbol{N}_{\nu m}^{(1)}(k_{C} \boldsymbol{r_B}) + R_{\nu}^{(N)}(b) \boldsymbol{N}_{\nu m}^{(3)}(k_{C} \boldsymbol{r_B}) \right]
\end{array} \right]
\left[ \boldsymbol{M}_{l, -m}^{(1)}(k_{C} \boldsymbol{\widetilde{r}_A}) + R_{l}^{(M)}(a) \boldsymbol{M}_{l, -m}^{(3)}(k_{C} \boldsymbol{\widetilde{r}_A}) \right]
\\[20 pt]
+ \left[ \begin{array}{r}
J_{\nu,m}^{lM} \left[ \boldsymbol{M}_{\nu m}^{(1)}(k_{C} \boldsymbol{r_B}) + R_{\nu}^{(M)}(b) \boldsymbol{M}_{\nu m}^{(3)}(k_{C} \boldsymbol{r_B}) \right]
\\[8 pt]
+ J_{\nu,m}^{lN} \left[ \boldsymbol{N}_{\nu m}^{(1)}(k_{C} \boldsymbol{r_B}) + R_{\nu}^{(N)}(b) \boldsymbol{N}_{\nu m}^{(3)}(k_{C} \boldsymbol{r_B}) \right]
\end{array} \right]
\left[ \boldsymbol{N}_{l, -m}^{(1)}(k_{C} \boldsymbol{\widetilde{r}_A}) + R_{l}^{(N)}(a) \boldsymbol{N}_{l, -m}^{(3)}(k_{C} \boldsymbol{\widetilde{r}_A}) \right]
\end{array} \right\}
\nonumber \\*
& \quad - i k_{C} \sum\limits_{m=-\infty}^{\infty} \sum\limits_{l=\widetilde{m}}^{\infty} (-1)^{m}
\left\{ \begin{array}{r}
\boldsymbol{M}_{l m}^{(3)}(k_{C} \boldsymbol{r_A})
\left[ \boldsymbol{M}_{l, -m}^{(1)}(k_{C} \boldsymbol{\widetilde{r}_A}) + R_{l}^{(M)}(a) \boldsymbol{M}_{l, -m}^{(3)}(k_{C} \boldsymbol{\widetilde{r}_A}) \right]
\\[8 pt]
+ \boldsymbol{N}_{l m}^{(3)}(k_{C} \boldsymbol{r_A})
\left[ \boldsymbol{N}_{l, -m}^{(1)}(k_{C} \boldsymbol{\widetilde{r}_A}) + R_{l}^{(N)}(a) \boldsymbol{N}_{l, -m}^{(3)}(k_{C} \boldsymbol{\widetilde{r}_A}) \right]
\end{array} \right\}
\label{eq:GeSc2}
\end{align}
\end{widetext}

Adding Eq. (\ref{eq:G0andGeSc1}) and Eq. (\ref{eq:GeSc2}), we obtain the Eq. (\ref{eq:FinalGe}) for $\overline{\overline{\boldsymbol{G}}}_{e}(\boldsymbol{r}; \boldsymbol{\widetilde{r}})$. The rationale for splitting $\overline{\overline{\boldsymbol{G}}}_{e}(\boldsymbol{r}; \boldsymbol{\widetilde{r}}) $ into $ \overline{\overline{\boldsymbol{G}}}_{0}(\boldsymbol{r}; \boldsymbol{\widetilde{r}}) + \overline{\overline{\boldsymbol{G}}}_{e}^{(sc)\prime}(\boldsymbol{r}; \boldsymbol{\widetilde{r}})$ and $\overline{\overline{\boldsymbol{G}}}_{e}^{(sc)\prime\prime}(\boldsymbol{r}; \boldsymbol{\widetilde{r}})$ should be clearer now: only the coefficients of VSWs appearing in $\overline{\overline{\boldsymbol{G}}}_{e}^{(sc)\prime\prime}(\boldsymbol{r}; \boldsymbol{\widetilde{r}})$ manifest in the final  expression for $T_{A \rightarrow B}^{e}$ [Eq. (\ref{eq:GeneralTransmissivity})].

The form of the magnetic DGF, $\overline{\overline{\boldsymbol{G}}}_{m}(\boldsymbol{r}; \boldsymbol{\widetilde{r}})$, is identical, but for $\widetilde{D},\widetilde{J},\widetilde{R}$ in place of $D,J,R$ in Eq. (\ref{eq:FinalGe}). See Eqs. (\ref{eq:MagVSWCoeff1})-(\ref{eq:MagVSWCoeff8}) for relations between coefficients of VSWs in $\overline{\overline{\boldsymbol{G}}}_{m}$ and $\overline{\overline{\boldsymbol{G}}}_{e}$. $\overline{\overline{\boldsymbol{G}}}_{E}(\boldsymbol{r}; \boldsymbol{\widetilde{r}})$ and $\overline{\overline{\boldsymbol{G}}}_{M}(\boldsymbol{r}; \boldsymbol{\widetilde{r}})$ are obtained from their definitions: $\overline{\overline{\boldsymbol{G}}}_{E}(\boldsymbol{r}; \boldsymbol{\widetilde{r}}) = \nabla \times \overline{\overline{\boldsymbol{G}}}_{e}(\boldsymbol{r}; \boldsymbol{\widetilde{r}})$ and $\overline{\overline{\boldsymbol{G}}}_{M}(\boldsymbol{r}; \boldsymbol{\widetilde{r}}) = \nabla \times \overline{\overline{\boldsymbol{G}}}_{m}(\boldsymbol{r}; \boldsymbol{\widetilde{r}})$

\section{Computation of effective Mie reflection coefficients} \label{ComputationOfEffectiveMieReflectionCoefficients}

The expressions for Mie reflection coefficients, $R_{\nu}^{(M)}(r_{\rho,s})$ and $R_{\nu}^{(N)}(r_{\rho,s})$ (see Sec. \ref{Geometry} for the geometry as well as definition of terms), are well known from the literature for both uncoated and single-coated spheres. \cite{Kaiser1993, Bohren2004, Zhou2006, Zheng2015} The full, multilayered Mie reflection coefficients can be determined recursively in a manner similar to that of Fresnel reflection coefficients for planar stratified media.\cite{Narayanaswamy2013b} The recurrence relation is given by
\begin{align}
R_{\nu}^{(M)}(r_{\rho,s}) \! &= - \frac{ u_{\nu}(r_{\rho,s}) + R_{\nu}^{(M)}(r_{\rho-1,s}) \alpha_{\nu}^{(M)}(r_{\rho,s}) }{ 1 + R_{\nu}^{(M)}(r_{\rho-1,s}) \beta_{\nu}^{(M)}(r_{\rho,s}) },
\\
R_{\nu}^{(N)}(r_{\rho,s}) \! &= - \frac{ v_{\nu}(r_{\rho,s}) + R_{\nu}^{(N)}(r_{\rho-1,s}) \alpha_{\nu}^{(N)}(r_{\rho,s}) }{ 1 + R_{\nu}^{(N)}(r_{\rho-1,s}) \beta_{\nu}^{(N)}(r_{\rho,s}) },
\end{align}
where
\begin{align}
u_{\nu}(r_{\rho,s})
&= \frac{z_{\nu}^{(1)}(k_{\rho+1,s}r_{\rho,s})}{z_{\nu}^{(3)}(k_{\rho+1,s}r_{\rho,s})}
\nonumber \\*
& \times \left[ \frac{ \frac{1}{Z_{\rho+1}} \frac{\zeta_{\nu}^{(1)}(k_{\rho+1}r_{\rho,s})}{z_{\nu}^{(1)}(k_{\rho+1}r_{\rho,s})} - \frac{1}{Z_{\rho}} \frac{\zeta_{\nu}^{(1)}(k_{\rho}r_{\rho,s})}{z_{\nu}^{(1)}(k_{\rho}r_{\rho,s})} }{ \frac{1}{Z_{\rho+1}} \frac{\zeta_{\nu}^{(3)}(k_{\rho+1}r_{\rho,s})}{z_{\nu}^{(3)}(k_{\rho+1}r_{\rho,s})} - \frac{1}{Z_{\rho}} \frac{\zeta_{\nu}^{(1)}(k_{\rho}r_{\rho,s})}{z_{\nu}^{(1)}(k_{\rho}r_{\rho,s})} } \right],
\label{eq:ReflCoeff3}
\end{align}
\begin{align}
v_{\nu}(r_{\rho,s})
&= \frac{\zeta_{\nu}^{(1)}(k_{\rho+1}r_{\rho,s})}{\zeta_{\nu}^{(3)}(k_{\rho+1}r_{\rho,s})}
\nonumber \\*
& \times \left[ \frac{ \frac{1}{Z_{\rho+1}} \frac{z_{\nu}^{(1)}(k_{\rho+1}r_{\rho,s})}{\zeta_{\nu}^{(1)}(k_{\rho+1}r_{\rho,s})} - \frac{1}{Z_{\rho}} \frac{z_{\nu}^{(1)}(k_{\rho}r_{\rho,s})}{\zeta_{\nu}^{(1)}(k_{\rho}r_{\rho,s})} }{ \frac{1}{Z_{\rho+1}} \frac{z_{\nu}^{(3)}(k_{\rho+1}r_{\rho,s})}{\zeta_{\nu}^{(3)}(k_{\rho+1}r_{\rho,s})} - \frac{1}{Z_{\rho}} \frac{z_{\nu}^{(1)}(k_{\rho}r_{\rho,s})}{\zeta_{\nu}^{(1)}(k_{\rho}r_{\rho,s})} } \right],
\end{align}
\begin{align}
\alpha_{\nu}^{(M)}(r_{\rho,s})
&= \frac{z_{\nu}^{(1)}(k_{\rho+1}r_{\rho,s})}{z_{\nu}^{(3)}(k_{\rho+1}r_{\rho,s})} \frac{z_{\nu}^{(3)}(k_{\rho}r_{\rho,s})}{z_{\nu}^{(1)}(k_{\rho}r_{\rho,s})}
\nonumber \\*
& \times \left[ \frac{ \frac{1}{Z_{\rho+1}} \frac{\zeta_{\nu}^{(1)}(k_{\rho+1}r_{\rho,s})}{z_{\nu}^{(1)}(k_{\rho+1}r_{\rho,s})} - \frac{1}{Z_{\rho}} \frac{\zeta_{\nu}^{(3)}(k_{\rho}r_{\rho,s})}{z_{\nu}^{(3)}(k_{\rho}r_{\rho,s})} }{ \frac{1}{Z_{\rho+1}} \frac{\zeta_{\nu}^{(3)}(k_{\rho+1}r_{\rho,s})}{z_{\nu}^{(3)}(k_{\rho+1}r_{\rho,s})} - \frac{1}{Z_{\rho}} \frac{\zeta_{\nu}^{(1)}(k_{\rho}r_{\rho,s})}{z_{\nu}^{(1)}(k_{\rho}r_{\rho,s})} } \right],
\end{align}
\begin{align}
\alpha_{\nu}^{(N)}(r_{\rho,s})
&= \frac{\zeta_{\nu}^{(1)}(k_{\rho+1}r_{\rho,s})}{\zeta_{\nu}^{(3)}(k_{\rho+1}r_{\rho,s})} \frac{\zeta_{\nu}^{(3)}(k_{\rho}r_{\rho,s})}{\zeta_{\nu}^{(1)}(k_{\rho}r_{\rho,s})}
\nonumber \\*
& \times \left[ \frac{ \frac{1}{Z_{\rho+1}} \frac{z_{\nu}^{(1)}(k_{\rho+1}r_{\rho,s})}{\zeta_{\nu}^{(1)}(k_{\rho+1}r_{\rho,s})} - \frac{1}{Z_{\rho}} \frac{z_{\nu}^{(3)}(k_{\rho}r_{\rho,s})}{\zeta_{\nu}^{(3)}(k_{\rho}r_{\rho,s})} }{ \frac{1}{Z_{\rho+1}} \frac{z_{\nu}^{(3)}(k_{\rho+1}r_{\rho,s})}{\zeta_{\nu}^{(3)}(k_{\rho+1}r_{\rho,s})} - \frac{1}{Z_{\rho}} \frac{z_{\nu}^{(1)}(k_{\rho}r_{\rho,s})}{\zeta_{\nu}^{(1)}(k_{\rho}r_{\rho,s})} } \right],
\end{align}
\begin{align}
\beta_{\nu}^{(M)}(r_{\rho,s})
&= \frac{z_{\nu}^{(3)}(k_{\rho}r_{\rho,s})}{z_{\nu}^{(1)}(k_{\rho}r_{\rho,s})}
\nonumber \\*
& \times \left[ \frac{ \frac{1}{Z_{\rho+1}} \frac{\zeta_{\nu}^{(3)}(k_{\rho+1}r_{\rho,s})}{z_{\nu}^{(3)}(k_{\rho+1}r_{\rho,s})} - \frac{1}{Z_{\rho}} \frac{\zeta_{\nu}^{(3)}(k_{\rho}r_{\rho,s})}{z_{\nu}^{(3)}(k_{\rho}r_{\rho,s})} }{ \frac{1}{Z_{\rho+1}} \frac{\zeta_{\nu}^{(3)}(k_{\rho+1}r_{\rho,s})}{z_{\nu}^{(3)}(k_{\rho+1}r_{\rho,s})} - \frac{1}{Z_{\rho}} \frac{\zeta_{\nu}^{(1)}(k_{\rho}r_{\rho,s})}{z_{\nu}^{(1)}(k_{\rho}r_{\rho,s})} } \right],
\end{align}
\begin{align}
\beta_{\nu}^{(N)}(r_{\rho,s})
&= \frac{\zeta_{\nu}^{(3)}(k_{\rho}r_{\rho,s})}{\zeta_{\nu}^{(1)}(k_{\rho}r_{\rho,s})}
\nonumber \\*
& \times \left[ \frac{ \frac{1}{Z_{\rho+1}} \frac{z_{\nu}^{(3)}(k_{\rho+1}r_{\rho,s})}{\zeta_{\nu}^{(3)}(k_{\rho+1}r_{\rho,s})} - \frac{1}{Z_{\rho}} \frac{z_{\nu}^{(3)}(k_{\rho}r_{\rho,s})}{\zeta_{\nu}^{(3)}(k_{\rho}r_{\rho,s})} }{ \frac{1}{Z_{\rho+1}} \frac{z_{\nu}^{(3)}(k_{\rho+1}r_{\rho,s})}{\zeta_{\nu}^{(3)}(k_{\rho+1}r_{\rho,s})} - \frac{1}{Z_{\rho}} \frac{z_{\nu}^{(1)}(k_{\rho}r_{\rho,s})}{\zeta_{\nu}^{(1)}(k_{\rho}r_{\rho,s})} } \right].
\label{eq:ReflCoeff8}
\end{align}

In Eqs. (\ref{eq:ReflCoeff3})-(\ref{eq:ReflCoeff8}), $Z=Z_0 \sqrt{\mu/\varepsilon}$ is the electromagnetic impedance for a dielectric material and $Z_0=\sqrt{\mu_0/\varepsilon_0}$ is the impedance of free space. Additionally, we define $k_{N+1,A} = k_{M+1,B} = k_{C}$. The recursion relation is terminated by $R_{\nu}^{(M)}(r_{0,s}) = -u_{\nu}(r_{0,s})$ and $R_{\nu}^{(N)}(r_{0,s}) = -v_{\nu}(r_{0,s})$.

\section{Reciprocity of transmissivity function} \label{ReciprocityofTransmissivityFunction}

The transmissivity function obeys the reciprocity relation $T_{A \rightarrow B}^{e}\left( \omega \right) = T_{B \rightarrow A}^{e}\left( \omega \right)$. Proof of this property for Eq. (\ref{eq:Transmissivity}) is a useful check of the validity of our derived transmissivity function. Because of the properties of electric and magnetic fields, DGFs must obey the following reciprocity relations: \cite{Tai1994}
\begin{align}
\mu(\boldsymbol{r}) \overline{\overline{\boldsymbol{G}}}_{e}^{T}(\boldsymbol{r}; \boldsymbol{\widetilde{r}}) &= \mu(\boldsymbol{\widetilde{r}}) \overline{\overline{\boldsymbol{G}}}_{e}(\boldsymbol{\widetilde{r}}; \boldsymbol{r})
\label{eq:DGFReciprocity1}
\\
\varepsilon(\boldsymbol{r}) \overline{\overline{\boldsymbol{G}}}_{m}^{T}(\boldsymbol{r}; \boldsymbol{\widetilde{r}}) &= \varepsilon(\boldsymbol{\widetilde{r}}) \overline{\overline{\boldsymbol{G}}}_{m}(\boldsymbol{\widetilde{r}}; \boldsymbol{r})
\label{eq:DGFReciprocity2}
\\
\overline{\overline{\boldsymbol{G}}}_{E}^{T}(\boldsymbol{r}; \boldsymbol{\widetilde{r}}) &= \overline{\overline{\boldsymbol{G}}}_{M}(\boldsymbol{\widetilde{r}}; \boldsymbol{r})
\label{eq:DGFReciprocity3}
\end{align}

To use these relations, the locations of the $\boldsymbol{r}$ and $\boldsymbol{\widetilde{r}}$ must be interchanged while holding the locations of the spheres fixed. The DGFs from the right-hand sides of Eqs. (\ref{eq:DGFReciprocity1})-(\ref{eq:DGFReciprocity3}) can be written with their own set of unknown coefficients, which we accent with a ``$\text{ }$ \textasciibreve $\text{ }$"  symbol. They arise when considering heat transfer from sphere $B$ to $A$. From Eqs. (\ref{eq:DGFReciprocity1})-(\ref{eq:DGFReciprocity3}), 
\begin{align}
\accentset{\smile}{D}_{l,-m}^{\nu M} &= D_{\nu,m}^{lM},
\label{eq:ReciprocityCoeff1}
\\
\accentset{\smile}{D}_{l,-m}^{\nu N} &= J_{\nu,m}^{lM},
\label{eq:ReciprocityCoeff2}
\\
\accentset{\smile}{J}_{l,-m}^{\nu M} &= D_{\nu,m}^{lN},
\label{eq:ReciprocityCoeff3}
\\
\accentset{\smile}{J}_{l,-m}^{\nu N} &= J_{\nu,m}^{lN}.
\label{eq:ReciprocityCoeff4}
\end{align}

Using Eqs. (\ref{eq:ReciprocityCoeff1})-(\ref{eq:ReciprocityCoeff4}), showing $T_{A \rightarrow B}^{e}\left( \omega \right) = T_{B \rightarrow A}^{e}\left( \omega \right)$ follows naturally from Eq. (\ref{eq:Transmissivity}). We get
\begin{align}
& T_{B \rightarrow A}^{e}\left( \omega \right)
= \left( k_{C} b \right)^{2} \left( k_{C} a \right)^{2} \sum\limits_{m=-\infty}^{\infty} \sum\limits_{l=\widetilde{m}}^{\infty}\sum_{\nu = \widetilde{m}}^{\infty}
\nonumber \\*
& \times \left[ \begin{array}{r}
\left[
\epsilon_{l}^{(M)}(a)
\left| \accentset{\smile}{D}_{l,-m}^{\nu M} \right|^{2}
+ \epsilon_{l}^{(N)}(a)
\left| \accentset{\smile}{D}_{l,-m}^{\nu N} \right|^{2}
\right]
\epsilon_{\nu}^{(M)}(b)
\\ \\
+ \left[
\epsilon_{l}^{(M)}(a)
\left| \accentset{\smile}{J}_{l,-m}^{\nu M} \right|^{2}
+ \epsilon_{l}^{(N)}(a)
\left| \accentset{\smile}{J}_{l,-m}^{\nu N} \right|^{2}
\right]
\epsilon_{\nu}^{(N)}(b)
\end{array} \right] \nonumber
\\
&= T_{A \rightarrow B}^{e}\left( \omega \right).
\end{align}
Here, the $l$ and $\nu$ indices and $m$ and $-m$ swap roles, but the overall sum is equal.

\section{Computational implementation} \label{ComputationalImplementation}

To be of any practical use, the transmissivity function must be numerically calculable. A number of numerical problems are introduced, however, due to the presence of spherical Bessel and Hankel functions of high order, which may experience issues with underflow and overflow on computers.\cite{Sasihithlu2014} To avoid such problems, we make two changes in the calculation of the transmissivity function.

First, instead of using the set of coefficients multiplying the VSWs in Eqs. (\ref{eq:BC1})-(\ref{eq:BC8}), we use the related coefficients defined in Eqs. \ref{eq:VSWCoeff1}-\ref{eq:VSWCoeff8} denoted with a $(1)$ superscript. These coefficients are analogous to the unknown coefficients used in Ref. [\onlinecite{Narayanaswamy2008}] and the scattering coefficients used in Ref. \onlinecite{Mackowski2008}. Second, we introduce prefactors to the translation and VSW coefficients which help stabilize the linear system used to solve for the VSW coefficients. The details of this procedure are discussed in greater detail by \citeauthor{Sasihithlu2014}.\cite{Sasihithlu2014} The prefactors take the form of a ratio of spherical Bessel or Hankel functions. For example, the prefactor for $D_{\nu,m}^{lM^{(1)}}$ is $z_{l}^{(1)}(k_{C} a) / z_{\nu}^{(1)}(k_{C} b)$. The linear system is modified in such a way that we solve directly for the value of $\left[ z_{l}^{(1)}(k_{C} a) / z_{\nu}^{(1)}(k_{C} b) \right] D_{\nu,m}^{lM^{(1)}}$, and other such coefficients. The expression for the transmissivity function may then be modified such that the VSW coefficients, with the appropriate prefactors, appear explicitly.

After some manipulation, we get
\begin{align}
& T_{A \rightarrow B}^{e}\left( \omega \right)
= 4 \left( \frac{a}{b} \right)^{2} \sum\limits_{m=-\infty}^{\infty} \sum\limits_{l=\widetilde{m}}^{\infty}\sum_{\nu = \widetilde{m}}^{\infty}
\nonumber \\* \quad
& \times \left[ \begin{array}{l}
\left[ \begin{array}{r}
\frac{ \Im{\left( \mathfrak{F}_{\nu}^{(M)}(b) \right)} }{ \left| \mathfrak{E}_{\nu}^{(M)}(b) \right|^{2} }
\left| \frac{z_{l}^{(1)}(k_{C} a)}{z_{\nu}^{(1)}(k_{C} b)} D_{\nu,m}^{lM^{(1)}} \right|^{2}
\\
+ \frac{ \Im{\left( \mathfrak{F}_{\nu}^{(N)}(b) \right)} }{ \left| \mathfrak{E}_{\nu}^{(N)}(b) \right|^{2} }
\left| \frac{z_{l}^{(1)}(k_{C} a)}{\zeta_{\nu}^{(1)}(k_{C} b)} D_{\nu,m}^{lN^{(1)}} \right|^{2}
\end{array} \right]
\\
\times \Im{\left( \mathfrak{F}_{l}^{(M)}(a) \right)}
\\ \\
+ \left[ \begin{array}{r}
\frac{ \Im{\left( \mathfrak{F}_{\nu}^{(M)}(b) \right)} }{ \left| \mathfrak{E}_{\nu}^{(M)}(b) \right|^{2} }
\left| \frac{\zeta_{l}^{(1)}(k_{C} a)}{z_{\nu}^{(1)}(k_{C} b)} J_{\nu,m}^{lM^{(1)}} \right|^{2}
\\
+ \frac{ \Im{\left( \mathfrak{F}_{\nu}^{(N)}(b) \right)} }{ \left| \mathfrak{E}_{\nu}^{(N)}(b) \right|^{2} }
\left| \frac{\zeta_{l}^{(1)}(k_{C} a)}{\zeta_{\nu}^{(1)}(k_{C} b)} J_{\nu,m}^{lN^{(1)}} \right|^{2}
\end{array} \right]
\\
\times \Im{\left( \mathfrak{F}_{l}^{(N)}(a) \right)}
\end{array} \right],
\label{eq:Trans_calc}
\end{align}
where
\begin{align}
\overline{R}_{\nu}^{(M)}(w) &= \frac{z_{\nu}^{(3)}(k_{C}w)}{z_{\nu}^{(1)}(k_{C}w)} R_{\nu}^{(M)}(w),
\\
\overline{R}_{\nu}^{(N)}(w) &= \frac{\zeta_{\nu}^{(3)}(k_{C}w)}{\zeta_{\nu}^{(1)}(k_{C}w)} R_{\nu}^{(N)}(w),
\\
\mathfrak{F}_{\nu}^{(M)}(w) &= \left( 1 + \overline{R}_{\nu}^{(M)}(w) \right) \nonumber \\* & \times \left( \frac{\zeta_{\nu}^{(1)}(k_{C}w)}{z_{\nu}^{(1)}(k_{C}w)} + \frac{\zeta_{\nu}^{(3)}(k_{C}w)}{z_{\nu}^{(3)}(k_{C}w)} \overline{R}_{\nu}^{(M)}(w) \right)^{*},
\\
\mathfrak{F}_{\nu}^{(N)}(w) &= \left( 1 + \overline{R}_{\nu}^{(N)}(w) \right)^{*} \nonumber \\* & \times \left( \frac{z_{\nu}^{(1)}(k_{C}w)}{\zeta_{\nu}^{(1)}(k_{C}w)} + \frac{z_{\nu}^{(3)}(k_{C}w)}{\zeta_{\nu}^{(3)}(k_{C}w)} \overline{R}_{\nu}^{(N)}(w) \right),
\\
\mathfrak{E}_{\nu}^{(M)}(b) &= \overline{R}_{\nu}^{(M)}(b) \left( \frac{\zeta_{\nu}^{(3)}(k_{C} b)}{z_{\nu}^{(3)}(k_{C} b)} - \frac{\zeta_{\nu}^{(1)}(k_{C} b)}{z_{\nu}^{(1)}(k_{C} b)} \right),
\\
\mathfrak{E}_{\nu}^{(N)}(b) &= \overline{R}_{\nu}^{(N)}(b) \left( \frac{z_{\nu}^{(3)}(k_{C} b)}{\zeta_{\nu}^{(3)}(k_{C} b)} - \frac{z_{\nu}^{(1)}(k_{C} b)}{\zeta_{\nu}^{(1)}(k_{C} b)} \right),
\end{align}
and $w=a$ or $b$.

The above results are obtained using the relation
\begin{align}
\epsilon_{\nu}^{(P)}(w) \! &= \!
2 \Im \! \left( \! \! \begin{array}{l} \left[ z_{\nu}^{(1)}(k_{C} w) \! + \! R_{\nu}^{(P)}(w) z_{\nu}^{(3)}(k_{C} w) \right] \! \\[10pt] \times \left[ \zeta_{\nu}^{(1)}(k_{C} w) \! + \! R_{\nu}^{(P)}(w) \zeta_{\nu}^{(3)}(k_{C} w) \right]^{*} \end{array} \! \! \right),
\end{align}
and simplified using Wronskian relations [see Eq. (\ref{eq:Wronskian13})].



\begin{thebibliography}{72}%
\makeatletter
\providecommand \@ifxundefined [1]{%
 \@ifx{#1\undefined}
}%
\providecommand \@ifnum [1]{%
 \ifnum #1\expandafter \@firstoftwo
 \else \expandafter \@secondoftwo
 \fi
}%
\providecommand \@ifx [1]{%
 \ifx #1\expandafter \@firstoftwo
 \else \expandafter \@secondoftwo
 \fi
}%
\providecommand \natexlab [1]{#1}%
\providecommand \enquote  [1]{``#1''}%
\providecommand \bibnamefont  [1]{#1}%
\providecommand \bibfnamefont [1]{#1}%
\providecommand \citenamefont [1]{#1}%
\providecommand \href@noop [0]{\@secondoftwo}%
\providecommand \href [0]{\begingroup \@sanitize@url \@href}%
\providecommand \@href[1]{\@@startlink{#1}\@@href}%
\providecommand \@@href[1]{\endgroup#1\@@endlink}%
\providecommand \@sanitize@url [0]{\catcode `\\12\catcode `\$12\catcode
  `\&12\catcode `\#12\catcode `\^12\catcode `\_12\catcode `\%12\relax}%
\providecommand \@@startlink[1]{}%
\providecommand \@@endlink[0]{}%
\providecommand \url  [0]{\begingroup\@sanitize@url \@url }%
\providecommand \@url [1]{\endgroup\@href {#1}{\urlprefix }}%
\providecommand \urlprefix  [0]{URL }%
\providecommand \Eprint [0]{\href }%
\providecommand \doibase [0]{http://dx.doi.org/}%
\providecommand \selectlanguage [0]{\@gobble}%
\providecommand \bibinfo  [0]{\@secondoftwo}%
\providecommand \bibfield  [0]{\@secondoftwo}%
\providecommand \translation [1]{[#1]}%
\providecommand \BibitemOpen [0]{}%
\providecommand \bibitemStop [0]{}%
\providecommand \bibitemNoStop [0]{.\EOS\space}%
\providecommand \EOS [0]{\spacefactor3000\relax}%
\providecommand \BibitemShut  [1]{\csname bibitem#1\endcsname}%
\let\auto@bib@innerbib\@empty
\bibitem [{\citenamefont {Smith}\ \emph {et~al.}(2004)\citenamefont {Smith},
  \citenamefont {Pendry},\ and\ \citenamefont {Wiltshire}}]{Smith2004}%
  \BibitemOpen
  \bibfield  {author} {\bibinfo {author} {\bibfnamefont {D.~R.}\ \bibnamefont
  {Smith}}, \bibinfo {author} {\bibfnamefont {J.~B.}\ \bibnamefont {Pendry}}, \
  and\ \bibinfo {author} {\bibfnamefont {M.~C.~K.}\ \bibnamefont {Wiltshire}},\
  }\href {\doibase 10.1126/science.1096796} {\bibfield  {journal} {\bibinfo
  {journal} {Science (New York, N.Y.)}\ }\textbf {\bibinfo {volume} {305}},\
  \bibinfo {pages} {788} (\bibinfo {year} {2004})}\BibitemShut {NoStop}%
\bibitem [{\citenamefont {Zhang}\ \emph {et~al.}(2005)\citenamefont {Zhang},
  \citenamefont {Fan}, \citenamefont {Panoiu}, \citenamefont {Malloy},
  \citenamefont {Osgood},\ and\ \citenamefont {Brueck}}]{Zhang2005}%
  \BibitemOpen
  \bibfield  {author} {\bibinfo {author} {\bibfnamefont {S.}~\bibnamefont
  {Zhang}}, \bibinfo {author} {\bibfnamefont {W.}~\bibnamefont {Fan}}, \bibinfo
  {author} {\bibfnamefont {N.~C.}\ \bibnamefont {Panoiu}}, \bibinfo {author}
  {\bibfnamefont {K.~J.}\ \bibnamefont {Malloy}}, \bibinfo {author}
  {\bibfnamefont {R.~M.}\ \bibnamefont {Osgood}}, \ and\ \bibinfo {author}
  {\bibfnamefont {S.~R.~J.}\ \bibnamefont {Brueck}},\ }\href {\doibase
  10.1103/PhysRevLett.95.137404} {\bibfield  {journal} {\bibinfo  {journal}
  {Phys Rev Lett}\ }\textbf {\bibinfo {volume} {95}},\ \bibinfo
  {pages} {137404} (\bibinfo {year} {2005})}\BibitemShut {NoStop}%
\bibitem [{\citenamefont {Soukoulis}\ \emph {et~al.}(2007)\citenamefont
  {Soukoulis}, \citenamefont {Linden},\ and\ \citenamefont
  {Wegener}}]{Soukoulis2007}%
  \BibitemOpen
  \bibfield  {author} {\bibinfo {author} {\bibfnamefont {C.~M.}\ \bibnamefont
  {Soukoulis}}, \bibinfo {author} {\bibfnamefont {S.}~\bibnamefont {Linden}}, \
  and\ \bibinfo {author} {\bibfnamefont {M.}~\bibnamefont {Wegener}},\ }\href
  {\doibase 10.1126/science.1136481} {\bibfield  {journal} {\bibinfo  {journal}
  {Science (New York, N.Y.)}\ }\textbf {\bibinfo {volume} {315}},\ \bibinfo
  {pages} {47} (\bibinfo {year} {2007})}\BibitemShut {NoStop}%
\bibitem [{\citenamefont {Valentine}\ \emph {et~al.}(2008)\citenamefont
  {Valentine}, \citenamefont {Zhang}, \citenamefont {Zentgraf}, \citenamefont
  {Ulin-Avila}, \citenamefont {Genov}, \citenamefont {Bartal},\ and\
  \citenamefont {Zhang}}]{Valentine2008}%
  \BibitemOpen
  \bibfield  {author} {\bibinfo {author} {\bibfnamefont {J.}~\bibnamefont
  {Valentine}}, \bibinfo {author} {\bibfnamefont {S.}~\bibnamefont {Zhang}},
  \bibinfo {author} {\bibfnamefont {T.}~\bibnamefont {Zentgraf}}, \bibinfo
  {author} {\bibfnamefont {E.}~\bibnamefont {Ulin-Avila}}, \bibinfo {author}
  {\bibfnamefont {D.~A.}\ \bibnamefont {Genov}}, \bibinfo {author}
  {\bibfnamefont {G.}~\bibnamefont {Bartal}}, \ and\ \bibinfo {author}
  {\bibfnamefont {X.}~\bibnamefont {Zhang}},\ }\href {\doibase
  10.1038/nature07247} {\bibfield  {journal} {\bibinfo  {journal} {Nature}\
  }\textbf {\bibinfo {volume} {455}},\ \bibinfo {pages} {376} (\bibinfo {year}
  {2008})}\BibitemShut {NoStop}%
\bibitem [{\citenamefont {Al{\`{u}}}\ and\ \citenamefont
  {Engheta}(2005)}]{Alu2005}%
  \BibitemOpen
  \bibfield  {author} {\bibinfo {author} {\bibfnamefont {A.}~\bibnamefont
  {Al{\`{u}}}}\ and\ \bibinfo {author} {\bibfnamefont {N.}~\bibnamefont
  {Engheta}},\ }\href {\doibase 10.1103/PhysRevE.72.016623} {\bibfield
  {journal} {\bibinfo  {journal} {Phys Rev E}\ }\textbf {\bibinfo
  {volume} {72}},\ \bibinfo {pages} {016623} (\bibinfo {year}
  {2005})}\BibitemShut {NoStop}%
\bibitem [{\citenamefont {Schurig}\ \emph {et~al.}(2006)\citenamefont
  {Schurig}, \citenamefont {Mock}, \citenamefont {Justice}, \citenamefont
  {Cummer}, \citenamefont {Pendry}, \citenamefont {Starr},\ and\ \citenamefont
  {Smith}}]{Schurig2006}%
  \BibitemOpen
  \bibfield  {author} {\bibinfo {author} {\bibfnamefont {D.}~\bibnamefont
  {Schurig}}, \bibinfo {author} {\bibfnamefont {J.~J.}\ \bibnamefont {Mock}},
  \bibinfo {author} {\bibfnamefont {B.~J.}\ \bibnamefont {Justice}}, \bibinfo
  {author} {\bibfnamefont {S.~A.}\ \bibnamefont {Cummer}}, \bibinfo {author}
  {\bibfnamefont {J.~B.}\ \bibnamefont {Pendry}}, \bibinfo {author}
  {\bibfnamefont {A.~F.}\ \bibnamefont {Starr}}, \ and\ \bibinfo {author}
  {\bibfnamefont {D.~R.}\ \bibnamefont {Smith}},\ }\href {\doibase
  10.1126/science.1133628} {\bibfield  {journal} {\bibinfo  {journal} {Science
  (New York, N.Y.)}\ }\textbf {\bibinfo {volume} {314}},\ \bibinfo {pages}
  {977} (\bibinfo {year} {2006})}\BibitemShut {NoStop}%
\bibitem [{\citenamefont {Cai}\ \emph {et~al.}(2007)\citenamefont {Cai},
  \citenamefont {Chettiar}, \citenamefont {Kildishev},\ and\ \citenamefont
  {Shalaev}}]{Cai2007}%
  \BibitemOpen
  \bibfield  {author} {\bibinfo {author} {\bibfnamefont {W.}~\bibnamefont
  {Cai}}, \bibinfo {author} {\bibfnamefont {U.~K.}\ \bibnamefont {Chettiar}},
  \bibinfo {author} {\bibfnamefont {A.~V.}\ \bibnamefont {Kildishev}}, \ and\
  \bibinfo {author} {\bibfnamefont {V.~M.}\ \bibnamefont {Shalaev}},\ }\href
  {\doibase 10.1038/nphoton.2007.28} {\bibfield  {journal} {\bibinfo  {journal}
  {Nat Photonics}\ }\textbf {\bibinfo {volume} {1}},\ \bibinfo {pages} {224}
  (\bibinfo {year} {2007})}\BibitemShut {NoStop}%
\bibitem [{\citenamefont {Valentine}\ \emph {et~al.}(2009)\citenamefont
  {Valentine}, \citenamefont {Li}, \citenamefont {Zentgraf}, \citenamefont
  {Bartal},\ and\ \citenamefont {Zhang}}]{Valentine2009}%
  \BibitemOpen
  \bibfield  {author} {\bibinfo {author} {\bibfnamefont {J.}~\bibnamefont
  {Valentine}}, \bibinfo {author} {\bibfnamefont {J.}~\bibnamefont {Li}},
  \bibinfo {author} {\bibfnamefont {T.}~\bibnamefont {Zentgraf}}, \bibinfo
  {author} {\bibfnamefont {G.}~\bibnamefont {Bartal}}, \ and\ \bibinfo {author}
  {\bibfnamefont {X.}~\bibnamefont {Zhang}},\ }\href {\doibase
  10.1038/nmat2461} {\bibfield  {journal} {\bibinfo  {journal} {Nat
  Mater}\ }\textbf {\bibinfo {volume} {8}},\ \bibinfo {pages} {568}
  (\bibinfo {year} {2009})}\BibitemShut {NoStop}%
\bibitem [{\citenamefont {Pendry}(2000)}]{Pendry2000}%
  \BibitemOpen
  \bibfield  {author} {\bibinfo {author} {\bibfnamefont {J.~B.}\ \bibnamefont
  {Pendry}},\ }\href {\doibase 10.1103/PhysRevLett.85.3966} {\bibfield
  {journal} {\bibinfo  {journal} {Phys Rev Lett}\ }\textbf {\bibinfo
  {volume} {85}},\ \bibinfo {pages} {3966} (\bibinfo {year}
  {2000})}\BibitemShut {NoStop}%
\bibitem [{\citenamefont {Fang}\ \emph {et~al.}(2005)\citenamefont {Fang},
  \citenamefont {Lee}, \citenamefont {Sun},\ and\ \citenamefont
  {Zhang}}]{Fang2005}%
  \BibitemOpen
  \bibfield  {author} {\bibinfo {author} {\bibfnamefont {N.}~\bibnamefont
  {Fang}}, \bibinfo {author} {\bibfnamefont {H.}~\bibnamefont {Lee}}, \bibinfo
  {author} {\bibfnamefont {C.}~\bibnamefont {Sun}}, \ and\ \bibinfo {author}
  {\bibfnamefont {X.}~\bibnamefont {Zhang}},\ }\href {\doibase
  10.1126/science.1108759} {\bibfield  {journal} {\bibinfo  {journal} {Science
  (New York, N.Y.)}\ }\textbf {\bibinfo {volume} {308}},\ \bibinfo {pages}
  {534} (\bibinfo {year} {2005})}\BibitemShut {NoStop}%
\bibitem [{\citenamefont {Smolyaninov}\ \emph {et~al.}(2007)\citenamefont
  {Smolyaninov}, \citenamefont {Hung},\ and\ \citenamefont
  {Davis}}]{Smolyaninov2007}%
  \BibitemOpen
  \bibfield  {author} {\bibinfo {author} {\bibfnamefont {I.~I.}\ \bibnamefont
  {Smolyaninov}}, \bibinfo {author} {\bibfnamefont {Y.-J.}\ \bibnamefont
  {Hung}}, \ and\ \bibinfo {author} {\bibfnamefont {C.~C.}\ \bibnamefont
  {Davis}},\ }\href {\doibase 10.1126/science.1138746} {\bibfield  {journal}
  {\bibinfo  {journal} {Science (New York, N.Y.)}\ }\textbf {\bibinfo {volume}
  {315}},\ \bibinfo {pages} {1699} (\bibinfo {year} {2007})}\BibitemShut
  {NoStop}%
\bibitem [{\citenamefont {Zhang}\ and\ \citenamefont {Liu}(2008)}]{Zhang2008}%
  \BibitemOpen
  \bibfield  {author} {\bibinfo {author} {\bibfnamefont {X.}~\bibnamefont
  {Zhang}}\ and\ \bibinfo {author} {\bibfnamefont {Z.}~\bibnamefont {Liu}},\
  }\href {\doibase 10.1038/nmat2141} {\bibfield  {journal} {\bibinfo  {journal}
  {Nat Mater}\ }\textbf {\bibinfo {volume} {7}},\ \bibinfo {pages} {435}
  (\bibinfo {year} {2008})}\BibitemShut {NoStop}%
\bibitem [{\citenamefont {Halevi}\ \emph {et~al.}(1999)\citenamefont {Halevi},
  \citenamefont {Krokhin},\ and\ \citenamefont {Arriaga}}]{Halevi1999}%
  \BibitemOpen
  \bibfield  {author} {\bibinfo {author} {\bibfnamefont {P.}~\bibnamefont
  {Halevi}}, \bibinfo {author} {\bibfnamefont {A.~A.}\ \bibnamefont {Krokhin}},
  \ and\ \bibinfo {author} {\bibfnamefont {J.}~\bibnamefont {Arriaga}},\ }\href
  {\doibase 10.1103/PhysRevLett.82.719} {\bibfield  {journal} {\bibinfo
  {journal} {Phys Rev Lett}\ }\textbf {\bibinfo {volume} {82}},\
  \bibinfo {pages} {719} (\bibinfo {year} {1999})}\BibitemShut {NoStop}%
\bibitem [{\citenamefont {Smith}\ and\ \citenamefont
  {Schurig}(2003)}]{Smith2003}%
  \BibitemOpen
  \bibfield  {author} {\bibinfo {author} {\bibfnamefont {D.~R.}\ \bibnamefont
  {Smith}}\ and\ \bibinfo {author} {\bibfnamefont {D.}~\bibnamefont
  {Schurig}},\ }\href {\doibase 10.1103/PhysRevLett.90.077405} {\bibfield
  {journal} {\bibinfo  {journal} {Phys Rev Lett}\ }\textbf {\bibinfo
  {volume} {90}},\ \bibinfo {pages} {077405} (\bibinfo {year}
  {2003})}\BibitemShut {NoStop}%
\bibitem [{\citenamefont {Francoeur}\ \emph {et~al.}(2011)\citenamefont
  {Francoeur}, \citenamefont {Basu},\ and\ \citenamefont
  {Petersen}}]{Francoeur2011}%
  \BibitemOpen
  \bibfield  {author} {\bibinfo {author} {\bibfnamefont {M.}~\bibnamefont
  {Francoeur}}, \bibinfo {author} {\bibfnamefont {S.}~\bibnamefont {Basu}}, \
  and\ \bibinfo {author} {\bibfnamefont {S.~J.}\ \bibnamefont {Petersen}},\
  }\href {\doibase 10.1364/OE.19.018774} {\bibfield  {journal} {\bibinfo
  {journal} {Opt Express}\ }\textbf {\bibinfo {volume} {19}},\ \bibinfo
  {pages} {18774} (\bibinfo {year} {2011})}\BibitemShut {NoStop}%
\bibitem [{\citenamefont {Liu}\ \emph {et~al.}(2011)\citenamefont {Liu},
  \citenamefont {Tyler}, \citenamefont {Starr}, \citenamefont {Starr},
  \citenamefont {Jokerst},\ and\ \citenamefont {Padilla}}]{Liu2011}%
  \BibitemOpen
  \bibfield  {author} {\bibinfo {author} {\bibfnamefont {X.}~\bibnamefont
  {Liu}}, \bibinfo {author} {\bibfnamefont {T.}~\bibnamefont {Tyler}}, \bibinfo
  {author} {\bibfnamefont {T.}~\bibnamefont {Starr}}, \bibinfo {author}
  {\bibfnamefont {A.~F.}\ \bibnamefont {Starr}}, \bibinfo {author}
  {\bibfnamefont {N.~M.}\ \bibnamefont {Jokerst}}, \ and\ \bibinfo {author}
  {\bibfnamefont {W.~J.}\ \bibnamefont {Padilla}},\ }\href {\doibase
  10.1103/PhysRevLett.107.045901} {\bibfield  {journal} {\bibinfo  {journal}
  {Phys Rev Lett}\ }\textbf {\bibinfo {volume} {107}},\ \bibinfo
  {pages} {045901} (\bibinfo {year} {2011})}\BibitemShut {NoStop}%
\bibitem [{\citenamefont {Mason}\ \emph {et~al.}(2011)\citenamefont {Mason},
  \citenamefont {Smith},\ and\ \citenamefont {Wasserman}}]{Mason2011}%
  \BibitemOpen
  \bibfield  {author} {\bibinfo {author} {\bibfnamefont {J.~A.}\ \bibnamefont
  {Mason}}, \bibinfo {author} {\bibfnamefont {S.}~\bibnamefont {Smith}}, \ and\
  \bibinfo {author} {\bibfnamefont {D.}~\bibnamefont {Wasserman}},\ }\href
  {\doibase 10.1063/1.3600779} {\bibfield  {journal} {\bibinfo  {journal}
  {Appl Phys Lett}\ }\textbf {\bibinfo {volume} {98}},\ \bibinfo
  {pages} {241105} (\bibinfo {year} {2011})}\BibitemShut {NoStop}%
\bibitem [{\citenamefont {Biehs}\ \emph {et~al.}(2012)\citenamefont {Biehs},
  \citenamefont {Tschikin},\ and\ \citenamefont {Ben-Abdallah}}]{Biehs2012}%
  \BibitemOpen
  \bibfield  {author} {\bibinfo {author} {\bibfnamefont {S.-A.}\ \bibnamefont
  {Biehs}}, \bibinfo {author} {\bibfnamefont {M.}~\bibnamefont {Tschikin}}, \
  and\ \bibinfo {author} {\bibfnamefont {P.}~\bibnamefont {Ben-Abdallah}},\
  }\href {\doibase 10.1103/PhysRevLett.109.104301} {\bibfield  {journal}
  {\bibinfo  {journal} {Phys Rev Lett}\ }\textbf {\bibinfo {volume}
  {109}},\ \bibinfo {pages} {104301} (\bibinfo {year} {2012})}\BibitemShut
  {NoStop}%
\bibitem [{\citenamefont {Guo}\ \emph {et~al.}(2012)\citenamefont {Guo},
  \citenamefont {Cortes}, \citenamefont {Molesky},\ and\ \citenamefont
  {Jacob}}]{Guo2012}%
  \BibitemOpen
  \bibfield  {author} {\bibinfo {author} {\bibfnamefont {Y.}~\bibnamefont
  {Guo}}, \bibinfo {author} {\bibfnamefont {C.~L.}\ \bibnamefont {Cortes}},
  \bibinfo {author} {\bibfnamefont {S.}~\bibnamefont {Molesky}}, \ and\
  \bibinfo {author} {\bibfnamefont {Z.}~\bibnamefont {Jacob}},\ }\href
  {\doibase 10.1063/1.4754616} {\bibfield  {journal} {\bibinfo  {journal}
  {Appl Phys Lett}\ }\textbf {\bibinfo {volume} {101}},\ \bibinfo
  {pages} {131106} (\bibinfo {year} {2012})}\BibitemShut {NoStop}%
\bibitem [{\citenamefont {Guo}\ and\ \citenamefont {Jacob}(2013)}]{Guo2013}%
  \BibitemOpen
  \bibfield  {author} {\bibinfo {author} {\bibfnamefont {Y.}~\bibnamefont
  {Guo}}\ and\ \bibinfo {author} {\bibfnamefont {Z.}~\bibnamefont {Jacob}},\
  }\href {\doibase 10.1364/OE.21.015014} {\bibfield  {journal} {\bibinfo
  {journal} {Opt Express}\ }\textbf {\bibinfo {volume} {21}},\ \bibinfo
  {pages} {15014} (\bibinfo {year} {2013})}\BibitemShut {NoStop}%
\bibitem [{\citenamefont {Liu}\ and\ \citenamefont {Shen}(2013)}]{Liu2013}%
  \BibitemOpen
  \bibfield  {author} {\bibinfo {author} {\bibfnamefont {B.}~\bibnamefont
  {Liu}}\ and\ \bibinfo {author} {\bibfnamefont {S.}~\bibnamefont {Shen}},\
  }\href {\doibase 10.1103/PhysRevB.87.115403} {\bibfield  {journal} {\bibinfo
  {journal} {Phys Rev B}\ }\textbf {\bibinfo {volume} {87}},\ \bibinfo
  {pages} {115403} (\bibinfo {year} {2013})}\BibitemShut {NoStop}%
\bibitem [{\citenamefont {Biehs}(2007)}]{Biehs2007}%
  \BibitemOpen
  \bibfield  {author} {\bibinfo {author} {\bibfnamefont {S.-A.}\ \bibnamefont
  {Biehs}},\ }\href {\doibase 10.1140/epjb/e2007-00254-8} {\bibfield  {journal}
  {\bibinfo  {journal} {EPJ B}\ }\textbf {\bibinfo
  {volume} {58}},\ \bibinfo {pages} {423} (\bibinfo {year} {2007})}\BibitemShut
  {NoStop}%
\bibitem [{\citenamefont {Fu}\ and\ \citenamefont {Tan}(2009)}]{Fu2009}%
  \BibitemOpen
  \bibfield  {author} {\bibinfo {author} {\bibfnamefont {C.}~\bibnamefont
  {Fu}}\ and\ \bibinfo {author} {\bibfnamefont {W.}~\bibnamefont {Tan}},\
  }\href {\doibase 10.1016/j.jqsrt.2009.02.007} {\bibfield  {journal} {\bibinfo
   {journal} {J Quant Spectrosc and Radiat Transfer}\
  }\textbf {\bibinfo {volume} {110}},\ \bibinfo {pages} {1027} (\bibinfo {year}
  {2009})}\BibitemShut {NoStop}%
\bibitem [{\citenamefont {Svetovoy}\ \emph {et~al.}(2012)\citenamefont
  {Svetovoy}, \citenamefont {van Zwol},\ and\ \citenamefont
  {Chevrier}}]{Svetovoy2012}%
  \BibitemOpen
  \bibfield  {author} {\bibinfo {author} {\bibfnamefont {V.~B.}\ \bibnamefont
  {Svetovoy}}, \bibinfo {author} {\bibfnamefont {P.~J.}\ \bibnamefont {van
  Zwol}}, \ and\ \bibinfo {author} {\bibfnamefont {J.}~\bibnamefont
  {Chevrier}},\ }\href {\doibase 10.1103/PhysRevB.85.155418} {\bibfield
  {journal} {\bibinfo  {journal} {Phys Rev B}\ }\textbf {\bibinfo
  {volume} {85}},\ \bibinfo {pages} {155418} (\bibinfo {year}
  {2012})}\BibitemShut {NoStop}%
\bibitem [{\citenamefont {Polder}\ and\ \citenamefont {{Van
  Hove}}(1971)}]{Polder1971}%
  \BibitemOpen
  \bibfield  {author} {\bibinfo {author} {\bibfnamefont {D.}~\bibnamefont
  {Polder}}\ and\ \bibinfo {author} {\bibfnamefont {M.}~\bibnamefont {{Van
  Hove}}},\ }\href {\doibase 10.1103/PhysRevB.4.3303} {\bibfield  {journal}
  {\bibinfo  {journal} {Phys Rev B}\ }\textbf {\bibinfo {volume} {4}},\
  \bibinfo {pages} {3303} (\bibinfo {year} {1971})}\BibitemShut {NoStop}%
\bibitem [{\citenamefont {Francoeur}\ and\ \citenamefont {{Pinar
  Meng{\"{u}}{\c{c}}}}(2008)}]{Francoeur2008}%
  \BibitemOpen
  \bibfield  {author} {\bibinfo {author} {\bibfnamefont {M.}~\bibnamefont
  {Francoeur}}\ and\ \bibinfo {author} {\bibfnamefont {M.}~\bibnamefont {{Pinar
  Meng{\"{u}}{\c{c}}}}},\ }\href {\doibase 10.1016/j.jqsrt.2007.08.017}
  {\bibfield  {journal} {\bibinfo  {journal} {J Quant
  Spectrosc and Radiat Transfer}\ }\textbf {\bibinfo {volume} {109}},\
  \bibinfo {pages} {280} (\bibinfo {year} {2008})}\BibitemShut {NoStop}%
\bibitem [{\citenamefont {Francoeur}\ \emph {et~al.}(2009)\citenamefont
  {Francoeur}, \citenamefont {{Pinar Meng{\"{u}}{\c{c}}}},\ and\ \citenamefont
  {Vaillon}}]{Francoeur2009}%
  \BibitemOpen
  \bibfield  {author} {\bibinfo {author} {\bibfnamefont {M.}~\bibnamefont
  {Francoeur}}, \bibinfo {author} {\bibfnamefont {M.}~\bibnamefont {{Pinar
  Meng{\"{u}}{\c{c}}}}}, \ and\ \bibinfo {author} {\bibfnamefont
  {R.}~\bibnamefont {Vaillon}},\ }\href {\doibase 10.1016/j.jqsrt.2009.05.010}
  {\bibfield  {journal} {\bibinfo  {journal} {J Quant
  Spectrosc and Radiat Transfer}\ }\textbf {\bibinfo {volume} {110}},\
  \bibinfo {pages} {2002} (\bibinfo {year} {2009})}\BibitemShut {NoStop}%
\bibitem [{\citenamefont {Song}\ \emph {et~al.}(2016)\citenamefont {Song},
  \citenamefont {Thompson}, \citenamefont {Fiorino}, \citenamefont {Ganjeh},
  \citenamefont {Reddy},\ and\ \citenamefont {Meyhofer}}]{Song2016}%
  \BibitemOpen
  \bibfield  {author} {\bibinfo {author} {\bibfnamefont {B.}~\bibnamefont
  {Song}}, \bibinfo {author} {\bibfnamefont {D.}~\bibnamefont {Thompson}},
  \bibinfo {author} {\bibfnamefont {A.}~\bibnamefont {Fiorino}}, \bibinfo
  {author} {\bibfnamefont {Y.}~\bibnamefont {Ganjeh}}, \bibinfo {author}
  {\bibfnamefont {P.}~\bibnamefont {Reddy}}, \ and\ \bibinfo {author}
  {\bibfnamefont {E.}~\bibnamefont {Meyhofer}},\ }\href {\doibase
  10.1038/nnano.2016.17} {\bibfield  {journal} {\bibinfo  {journal} {Nat
  Nanotechnol}\ }\textbf {\bibinfo {volume} {11}},\ \bibinfo {pages} {509}
  (\bibinfo {year} {2016})}\BibitemShut {NoStop}%
\bibitem [{\citenamefont {Narayanaswamy}\ and\ \citenamefont
  {Chen}(2008)}]{Narayanaswamy2008}%
  \BibitemOpen
  \bibfield  {author} {\bibinfo {author} {\bibfnamefont {A.}~\bibnamefont
  {Narayanaswamy}}\ and\ \bibinfo {author} {\bibfnamefont {G.}~\bibnamefont
  {Chen}},\ }\href {\doibase 10.1103/PhysRevB.77.075125} {\bibfield  {journal}
  {\bibinfo  {journal} {Phys Rev B}\ }\textbf {\bibinfo {volume} {77}},\
  \bibinfo {pages} {075125} (\bibinfo {year} {2008})} \BibitemShut {NoStop}%
\bibitem [{\citenamefont {Mackowski}\ and\ \citenamefont
  {Mishchenko}(2008)}]{Mackowski2008}%
  \BibitemOpen
  \bibfield  {author} {\bibinfo {author} {\bibfnamefont {D.~W.}\ \bibnamefont
  {Mackowski}}\ and\ \bibinfo {author} {\bibfnamefont {M.~I.}\ \bibnamefont
  {Mishchenko}},\ }\href {\doibase 10.1115/1.2957596} {\bibfield  {journal}
  {\bibinfo  {journal} {J Heat Transfer}\ }\textbf {\bibinfo {volume}
  {130}},\ \bibinfo {pages} {112702} (\bibinfo {year} {2008})}\BibitemShut
  {NoStop}%
\bibitem [{\citenamefont {Kr{\"{u}}ger}\ \emph {et~al.}(2012)\citenamefont
  {Kr{\"{u}}ger}, \citenamefont {Bimonte}, \citenamefont {Emig},\ and\
  \citenamefont {Kardar}}]{Kruger2012}%
  \BibitemOpen
  \bibfield  {author} {\bibinfo {author} {\bibfnamefont {M.}~\bibnamefont
  {Kr{\"{u}}ger}}, \bibinfo {author} {\bibfnamefont {G.}~\bibnamefont
  {Bimonte}}, \bibinfo {author} {\bibfnamefont {T.}~\bibnamefont {Emig}}, \
  and\ \bibinfo {author} {\bibfnamefont {M.}~\bibnamefont {Kardar}},\ }\href
  {\doibase 10.1103/PhysRevB.86.115423} {\bibfield  {journal} {\bibinfo
  {journal} {Phys Rev B}\ }\textbf {\bibinfo {volume} {86}},\ \bibinfo
  {pages} {115423} (\bibinfo {year} {2012})}\BibitemShut {NoStop}%
\bibitem [{\citenamefont {Otey}\ and\ \citenamefont {Fan}(2011)}]{Otey2011}%
  \BibitemOpen
  \bibfield  {author} {\bibinfo {author} {\bibfnamefont {C.}~\bibnamefont
  {Otey}}\ and\ \bibinfo {author} {\bibfnamefont {S.}~\bibnamefont {Fan}},\
  }\href {\doibase 10.1103/PhysRevB.84.245431} {\bibfield  {journal} {\bibinfo
  {journal} {Phys Rev B}\ }\textbf {\bibinfo {volume} {84}},\ \bibinfo
  {pages} {245431} (\bibinfo {year} {2011})} \BibitemShut {NoStop}%
\bibitem [{\citenamefont {Sasihithlu}\ and\ \citenamefont
  {Narayanaswamy}(2014)}]{Sasihithlu2014}%
  \BibitemOpen
  \bibfield  {author} {\bibinfo {author} {\bibfnamefont {K.}~\bibnamefont
  {Sasihithlu}}\ and\ \bibinfo {author} {\bibfnamefont {A.}~\bibnamefont
  {Narayanaswamy}},\ }\href {\doibase 10.1364/OE.22.014473} {\bibfield
  {journal} {\bibinfo  {journal} {Opt Express}\ }\textbf {\bibinfo {volume}
  {22}},\ \bibinfo {pages} {14473} (\bibinfo {year} {2014})}\BibitemShut
  {NoStop}%
\bibitem [{\citenamefont {Waterman}(1965)}]{Waterman1965}%
  \BibitemOpen
  \bibfield  {author} {\bibinfo {author} {\bibfnamefont {P.}~\bibnamefont
  {Waterman}},\ }\href {\doibase 10.1109/PROC.1965.4058} {\bibfield  {journal}
  {\bibinfo  {journal} {Proc IEEE}\ }\textbf {\bibinfo {volume}
  {53}},\ \bibinfo {pages} {805} (\bibinfo {year} {1965})}\BibitemShut
  {NoStop}%
\bibitem [{\citenamefont {Peterson}\ and\ \citenamefont
  {Str{\"{o}}m}(1974)}]{Peterson1974}%
  \BibitemOpen
  \bibfield  {author} {\bibinfo {author} {\bibfnamefont {B.}~\bibnamefont
  {Peterson}}\ and\ \bibinfo {author} {\bibfnamefont {S.}~\bibnamefont
  {Str{\"{o}}m}},\ }\href {\doibase 10.1103/PhysRevD.10.2670} {\bibfield
  {journal} {\bibinfo  {journal} {Phys Rev D}\ }\textbf {\bibinfo
  {volume} {10}},\ \bibinfo {pages} {2670} (\bibinfo {year}
  {1974})}\BibitemShut {NoStop}%
\bibitem [{\citenamefont {Rytov}(1967)}]{Rytov1967}%
  \BibitemOpen
  \bibfield  {author} {\bibinfo {author} {\bibfnamefont {S.~M.}\ \bibnamefont
  {Rytov}},\ }\href@noop {} {\emph {\bibinfo {title} {{Theory of Electric
  Fluctuations and Thermal Radiation}}}},\ \bibinfo {type} {Tech. Rep.} \ \bibinfo{number} {AD0226765} \ (\bibinfo{institution} {Air Force Research Center}, \bibinfo {year}
  {1959}), \ \bibinfo{url} {www.dtic.mil/docs/citations/AD0226765} \BibitemShut {NoStop}%
\bibitem [{\citenamefont {Eckhardt}(1982)}]{Eckhardt1982}%
  \BibitemOpen
  \bibfield  {author} {\bibinfo {author} {\bibfnamefont {W.}~\bibnamefont
  {Eckhardt}},\ }\href {\doibase 10.1016/0030-4018(82)90402-3} {\bibfield
  {journal} {\bibinfo  {journal} {Optics Commun}\ }\textbf {\bibinfo
  {volume} {41}},\ \bibinfo {pages} {305} (\bibinfo {year} {1982})}\BibitemShut
  {NoStop}%
\bibitem [{\citenamefont {Narayanaswamy}\ and\ \citenamefont
  {Zheng}(2013{\natexlab{a}})}]{Narayanaswamy2013a}%
  \BibitemOpen
  \bibfield  {author} {\bibinfo {author} {\bibfnamefont {A.}~\bibnamefont
  {Narayanaswamy}}\ and\ \bibinfo {author} {\bibfnamefont {Y.}~\bibnamefont
  {Zheng}},\ }\href {\doibase 10.1016/j.jqsrt.2013.01.002} {\bibfield
  {journal} {\bibinfo  {journal} {J Quant Spectrosc and Radiat Transfer}\ }
  \textbf {\bibinfo {volume} {132}},\ \bibinfo {pages}
  {12} (\bibinfo {year} {2013}{\natexlab{a}})}\BibitemShut {NoStop}%
\bibitem [{\citenamefont {Lai}\ \emph {et~al.}()\citenamefont {Lai},
  \citenamefont {Rubin},\ and\ \citenamefont {Krempl}}]{Lai2009}%
  \BibitemOpen
  \bibfield  {author} {\bibinfo {author} {\bibfnamefont {W.~M.}\ \bibnamefont
  {Lai}}, \bibinfo {author} {\bibfnamefont {D.}~\bibnamefont {Rubin}}, \ and\
  \bibinfo {author} {\bibfnamefont {E.}~\bibnamefont {Krempl}},\ }\href
  {\doibase https://doi.org/10.1016/B978-0-7506-8560-3.00010-4} {\emph
  {\bibinfo {title} {{Introduction to Continuum Mechanics}}}},\ \bibinfo
  {edition} {4th}\ ed.,\ edited by\ \bibinfo {editor} {\bibfnamefont
  {W.~M.}\ \bibnamefont {Lai}}, \bibinfo {editor} {\bibfnamefont
  {D.}~\bibnamefont {Rubin}}, \ and\ \bibinfo {editor} {\bibfnamefont
  {E.}~\bibnamefont {Krempl}}\ (\bibinfo  {publisher} {Butterworth-Heinemann},\
  \bibinfo {address} {Boston},\ \bibinfo {year} {2010})\BibitemShut {NoStop}%
\bibitem [{\citenamefont {Chew}(1992)}]{Chew1992}%
  \BibitemOpen
  \bibfield  {author} {\bibinfo {author} {\bibfnamefont {W.~C.}\ \bibnamefont
  {Chew}},\ }\href {\doibase 10.1163/156939392X01075} {\bibfield  {journal}
  {\bibinfo  {journal} {J of Electromagnet Wave}\
  }\textbf {\bibinfo {volume} {6}},\ \bibinfo {pages} {133} (\bibinfo {year}
  {1992})}\BibitemShut {NoStop}%
\bibitem [{\citenamefont {Chew}(1995)}]{Chew1995}%
  \BibitemOpen
  \bibfield  {author} {\bibinfo {author} {\bibfnamefont {W.~C.}\ \bibnamefont
  {Chew}},\ }\href@noop {} {\emph {\bibinfo {title} {{Waves and Fields in
  Inhomogeneous Media}}}},\ edited by\ \bibinfo {editor} {\bibfnamefont
  {D.~G.}\ \bibnamefont {Dudley}}\ (\bibinfo  {publisher} {IEEE Press},\
  \bibinfo {year} {1995})\BibitemShut {NoStop}%
\bibitem [{\citenamefont {Kim}(2004)}]{Kim2004a}%
  \BibitemOpen
  \bibfield  {author} {\bibinfo {author} {\bibfnamefont {K.~T.}\ \bibnamefont
  {Kim}},\ }\href@noop {} {\bibfield  {journal} {\bibinfo  {journal} {Prog
  Electromagn Res}\ }\textbf {\bibinfo {volume} {48}},\
  \bibinfo {pages} {45} (\bibinfo {year} {2004})}\BibitemShut {NoStop}%
\bibitem [{\citenamefont {Dufva}\ \emph {et~al.}(2008)\citenamefont {Dufva},
  \citenamefont {Sarvas},\ and\ \citenamefont {Sten}}]{Dufva2008}%
  \BibitemOpen
  \bibfield  {author} {\bibinfo {author} {\bibfnamefont {T.~J.}\ \bibnamefont
  {Dufva}}, \bibinfo {author} {\bibfnamefont {J.}~\bibnamefont {Sarvas}}, \
  and\ \bibinfo {author} {\bibfnamefont {J.~C.-E.}\ \bibnamefont {Sten}},\
  }\href {\doibase 10.2528/PIERB07121203} {\bibfield  {journal} {\bibinfo
  {journal} {Prog Electromagn Res}\ }\textbf {\bibinfo
  {volume} {4}},\ \bibinfo {pages} {79} (\bibinfo {year} {2008})}\BibitemShut
  {NoStop}%
\bibitem [{\citenamefont {Mackowski}(1991)}]{Mackowski1991}%
  \BibitemOpen
  \bibfield  {author} {\bibinfo {author} {\bibfnamefont {D.~W.}\ \bibnamefont
  {Mackowski}},\ }\href
  {http://rspa.royalsocietypublishing.org/content/433/1889/599} {\bibfield
  {journal} {\bibinfo  {journal} {Proc Roy Soc London Ser A}\ }\textbf {\bibinfo {volume}
  {433}},\ \bibinfo {pages} {599} (\bibinfo {year} {1991})}\BibitemShut {NoStop}%
\bibitem [{\citenamefont {Mackowski}(1994)}]{Mackowski1994}%
  \BibitemOpen
  \bibfield  {author} {\bibinfo {author} {\bibfnamefont {D.~W.}\ \bibnamefont
  {Mackowski}},\ }\href {\doibase 10.1364/JOSAA.11.002851} {\bibfield
  {journal} {\bibinfo  {journal} {J Opt Soc Am A}\
  }\textbf {\bibinfo {volume} {11}},\ \bibinfo {pages} {2851} (\bibinfo {year}
  {1994})}\BibitemShut {NoStop}%
\bibitem [{\citenamefont {Xu}\ and\ \citenamefont {Gustafson}(1996)}]{Xu1996}%
  \BibitemOpen
  \bibfield  {author} {\bibinfo {author} {\bibfnamefont {Y.-L.}\ \bibnamefont
  {Xu}}\ and\ \bibinfo {author} {\bibfnamefont {B.~A.~S.}\ \bibnamefont
  {Gustafson}},\ }in\ \href@noop {} {\emph {\bibinfo {booktitle} {IAU Colloq.
  150: Physics, Chemistry, and Dynamics of Interplanetary Dust}}} \ (\bibinfo  {publisher} {Astronomical Society of the Pacific Press},\
  \bibinfo {address} {San Francisco},\ \bibinfo {year} {1996}), \ Vol.\ \bibinfo {volume} {104}, \ p.\ \bibinfo {pages}
  {419}\BibitemShut {NoStop}%
\bibitem [{\citenamefont {Gumerov}\ and\ \citenamefont
  {Duraiswami}(2002)}]{Gumerov2002}%
  \BibitemOpen
  \bibfield  {author} {\bibinfo {author} {\bibfnamefont {N.~A.}\ \bibnamefont
  {Gumerov}}\ and\ \bibinfo {author} {\bibfnamefont {R.}~\bibnamefont
  {Duraiswami}},\ }\href {\doibase 10.1121/1.1517253} {\bibfield  {journal}
  {\bibinfo  {journal} {J Acoust Soc Am}\
  }\textbf {\bibinfo {volume} {112}},\ \bibinfo {pages} {2688} (\bibinfo {year}
  {2002})}\BibitemShut {NoStop}%
\bibitem [{\citenamefont {L{\'{e}}tourneau}\ \emph {et~al.}(2017)\citenamefont
  {L{\'{e}}tourneau}, \citenamefont {Wu}, \citenamefont {Papanicolaou},
  \citenamefont {Garnier},\ and\ \citenamefont {Darve}}]{Letourneau2017}%
  \BibitemOpen
  \bibfield  {author} {\bibinfo {author} {\bibfnamefont {P.-D.}\ \bibnamefont
  {L{\'{e}}tourneau}}, \bibinfo {author} {\bibfnamefont {Y.}~\bibnamefont
  {Wu}}, \bibinfo {author} {\bibfnamefont {G.}~\bibnamefont {Papanicolaou}},
  \bibinfo {author} {\bibfnamefont {J.}~\bibnamefont {Garnier}}, \ and\
  \bibinfo {author} {\bibfnamefont {E.}~\bibnamefont {Darve}},\ }\href
  {\doibase 10.1016/j.wavemoti.2016.08.012} {\bibfield  {journal} {\bibinfo
  {journal} {Wave Motion}\ }\textbf {\bibinfo {volume} {70}},\ \bibinfo {pages}
  {113} (\bibinfo {year} {2017})}\BibitemShut {NoStop}%
\bibitem [{\citenamefont {Mishchenko}\ \emph {et~al.}(1996)\citenamefont
  {Mishchenko}, \citenamefont {Travis},\ and\ \citenamefont
  {Mackowski}}]{Mishchenko1996}%
  \BibitemOpen
  \bibfield  {author} {\bibinfo {author} {\bibfnamefont {M.~I.}\ \bibnamefont
  {Mishchenko}}, \bibinfo {author} {\bibfnamefont {L.~D.}\ \bibnamefont
  {Travis}}, \ and\ \bibinfo {author} {\bibfnamefont {D.~W.}\ \bibnamefont
  {Mackowski}},\ }\href {\doibase 10.1016/0022-4073(96)00002-7} {\bibfield
  {journal} {\bibinfo  {journal} {J Quant
  Spectrosc and Radiat Transfer}\ }\textbf {\bibinfo {volume} {55}},\ \bibinfo {pages}
  {535} (\bibinfo {year} {1996})}\BibitemShut {NoStop}%
\bibitem [{\citenamefont {Mishchenko}\ and\ \citenamefont
  {Martin}(2013)}]{Mishchenko2013}%
  \BibitemOpen
  \bibfield  {author} {\bibinfo {author} {\bibfnamefont {M.~I.}\ \bibnamefont
  {Mishchenko}}\ and\ \bibinfo {author} {\bibfnamefont {P.}~\bibnamefont
  {Martin}},\ }\href {\doibase 10.1016/j.jqsrt.2012.10.025} {\bibfield
  {journal} {\bibinfo  {journal} {J Quant Spectrosc and
  Radiat Transfer}\ }\textbf {\bibinfo {volume} {123}},\ \bibinfo {pages}
  {2} (\bibinfo {year} {2013})}\BibitemShut {NoStop}%
\bibitem [{\citenamefont {Sasihithlu}\ and\ \citenamefont
  {Narayanaswamy}(2011{\natexlab{a}})}]{Sasihithlu2011a}%
  \BibitemOpen
  \bibfield  {author} {\bibinfo {author} {\bibfnamefont {K.}~\bibnamefont
  {Sasihithlu}}\ and\ \bibinfo {author} {\bibfnamefont {A.}~\bibnamefont
  {Narayanaswamy}},\ }\href {\doibase 10.1364/OE.19.00A772} {\bibfield
  {journal} {\bibinfo  {journal} {Opt Express}\ }\textbf {\bibinfo {volume}
  {19}},\ \bibinfo {pages} {A772} (\bibinfo {year}
  {2011}{\natexlab{a}})}\BibitemShut {NoStop}%
\bibitem [{\citenamefont {Kattawar}\ and\ \citenamefont
  {Eisner}(1970)}]{Kattawar1970}%
  \BibitemOpen
  \bibfield  {author} {\bibinfo {author} {\bibfnamefont {G.~W.}\ \bibnamefont
  {Kattawar}}\ and\ \bibinfo {author} {\bibfnamefont {M.}~\bibnamefont
  {Eisner}},\ }\href {\doibase 10.1364/AO.9.002685} {\bibfield  {journal}
  {\bibinfo  {journal} {Appl Opt}\ }\textbf {\bibinfo {volume} {9}},\
  \bibinfo {pages} {2685} (\bibinfo {year} {1970})}\BibitemShut {NoStop}%
\bibitem [{\citenamefont {Kim}\ \emph {et~al.}(2015)\citenamefont {Kim},
  \citenamefont {Song}, \citenamefont {Fern{\'{a}}ndez-Hurtado}, \citenamefont
  {Lee}, \citenamefont {Jeong}, \citenamefont {Cui}, \citenamefont {Thompson},
  \citenamefont {Feist}, \citenamefont {Reid}, \citenamefont
  {Garc{\'{i}}a-Vidal}, \citenamefont {Cuevas}, \citenamefont {Meyhofer},\ and\
  \citenamefont {Reddy}}]{Kim2015}%
  \BibitemOpen
  \bibfield  {author} {\bibinfo {author} {\bibfnamefont {K.}~\bibnamefont
  {Kim}}, \bibinfo {author} {\bibfnamefont {B.}~\bibnamefont {Song}}, \bibinfo
  {author} {\bibfnamefont {V.}~\bibnamefont {Fern{\'{a}}ndez-Hurtado}},
  \bibinfo {author} {\bibfnamefont {W.}~\bibnamefont {Lee}}, \bibinfo {author}
  {\bibfnamefont {W.}~\bibnamefont {Jeong}}, \bibinfo {author} {\bibfnamefont
  {L.}~\bibnamefont {Cui}}, \bibinfo {author} {\bibfnamefont {D.}~\bibnamefont
  {Thompson}}, \bibinfo {author} {\bibfnamefont {J.}~\bibnamefont {Feist}},
  \bibinfo {author} {\bibfnamefont {M.~T.~H.}\ \bibnamefont {Reid}}, \bibinfo
  {author} {\bibfnamefont {F.~J.}\ \bibnamefont {Garc{\'{i}}a-Vidal}}, \bibinfo
  {author} {\bibfnamefont {J.~C.}\ \bibnamefont {Cuevas}}, \bibinfo {author}
  {\bibfnamefont {E.}~\bibnamefont {Meyhofer}}, \ and\ \bibinfo {author}
  {\bibfnamefont {P.}~\bibnamefont {Reddy}},\ }\href {\doibase
  10.1038/nature16070} {\bibfield  {journal} {\bibinfo  {journal} {Nature}\
  }\textbf {\bibinfo {volume} {528}},\ \bibinfo {pages} {387} (\bibinfo {year}
  {2015})}\BibitemShut {NoStop}%
\bibitem [{\citenamefont {Yang}\ \emph {et~al.}(2015)\citenamefont {Yang},
  \citenamefont {D'Archangel}, \citenamefont {Sundheimer}, \citenamefont
  {Tucker}, \citenamefont {Boreman},\ and\ \citenamefont {Raschke}}]{Yang2015}%
  \BibitemOpen
  \bibfield  {author} {\bibinfo {author} {\bibfnamefont {H.~U.}\ \bibnamefont
  {Yang}}, \bibinfo {author} {\bibfnamefont {J.}~\bibnamefont {D'Archangel}},
  \bibinfo {author} {\bibfnamefont {M.~L.}\ \bibnamefont {Sundheimer}},
  \bibinfo {author} {\bibfnamefont {E.}~\bibnamefont {Tucker}}, \bibinfo
  {author} {\bibfnamefont {G.~D.}\ \bibnamefont {Boreman}}, \ and\ \bibinfo
  {author} {\bibfnamefont {M.~B.}\ \bibnamefont {Raschke}},\ }\href {\doibase
  10.1103/PhysRevB.91.235137} {\bibfield  {journal} {\bibinfo  {journal}
  {Phys Rev B}\ }\textbf {\bibinfo {volume} {91}},\ \bibinfo {pages}
  {235137} (\bibinfo {year} {2015})}\BibitemShut {NoStop}%
\bibitem [{\citenamefont {Palik}(1985)}]{Palik1985}%
  \BibitemOpen
  \bibfield  {author} {\bibinfo {author} {\bibfnamefont {E.~D.}\ \bibnamefont
  {Palik}}, \ }\href@noop {} {\emph {\bibinfo {title} {{Handbook of Optical
  Constants of Solids}}}}\ (\bibinfo  {publisher} {Academic Press}, \bibinfo {address} {New York},\ \bibinfo
  {year} {1985})\BibitemShut {NoStop}%
\bibitem [{\citenamefont {SupplementalMaterials}(1985)}]{SupplementalMaterials}%
  \BibitemOpen
  \bibfield  {author} {See Supplemental Material at link.aps.org/supplemental/ 10.1103/PhysRevB.96.125404 for select data points from Figs. 2-4 in tabular form}
  \BibitemShut {NoStop}%
\bibitem [{\citenamefont {Narayanaswamy}\ \emph {et~al.}(2014)\citenamefont
  {Narayanaswamy}, \citenamefont {Mayo},\ and\ \citenamefont
  {Canetta}}]{Narayanaswamy2014}%
  \BibitemOpen
  \bibfield  {author} {\bibinfo {author} {\bibfnamefont {A.}~\bibnamefont
  {Narayanaswamy}}, \bibinfo {author} {\bibfnamefont {J.}~\bibnamefont {Mayo}},
  \ and\ \bibinfo {author} {\bibfnamefont {C.}~\bibnamefont {Canetta}},\ }\href
  {\doibase 10.1063/1.4875699} {\bibfield  {journal} {\bibinfo  {journal}
  {Appl Phys Lett}\ }\textbf {\bibinfo {volume} {104}},\ \bibinfo
  {pages} {183107} (\bibinfo {year} {2014})}\BibitemShut {NoStop}%
\bibitem [{\citenamefont {Granqvist}\ and\ \citenamefont
  {Hjortsberg}(1980)}]{Granqvist1980}%
  \BibitemOpen
  \bibfield  {author} {\bibinfo {author} {\bibfnamefont {C.~G.}\ \bibnamefont
  {Granqvist}}\ and\ \bibinfo {author} {\bibfnamefont {A.}~\bibnamefont
  {Hjortsberg}},\ }\href {\doibase 10.1063/1.91406} {\bibfield  {journal}
  {\bibinfo  {journal} {Appl Phys Lett}\ }\textbf {\bibinfo {volume}
  {36}},\ \bibinfo {pages} {139} (\bibinfo {year} {1980})}\BibitemShut
  {NoStop}%
\bibitem [{\citenamefont {Narayanaswamy}\ \emph {et~al.}(2008)\citenamefont
  {Narayanaswamy}, \citenamefont {Shen},\ and\ \citenamefont
  {Chen}}]{Narayanaswamy2008a}%
  \BibitemOpen
  \bibfield  {author} {\bibinfo {author} {\bibfnamefont {A.}~\bibnamefont
  {Narayanaswamy}}, \bibinfo {author} {\bibfnamefont {S.}~\bibnamefont {Shen}},
  \ and\ \bibinfo {author} {\bibfnamefont {G.}~\bibnamefont {Chen}},\ }\href
  {\doibase 10.1103/PhysRevB.78.115303} {\bibfield  {journal} {\bibinfo
  {journal} {Phys Rev B}\ }\textbf {\bibinfo {volume} {78}},\ \bibinfo
  {pages} {115303} (\bibinfo {year} {2008})}\BibitemShut {NoStop}%
\bibitem [{\citenamefont {Shen}\ \emph {et~al.}(2009)\citenamefont {Shen},
  \citenamefont {Narayanaswamy},\ and\ \citenamefont {Chen}}]{Shen2009}%
  \BibitemOpen
  \bibfield  {author} {\bibinfo {author} {\bibfnamefont {S.}~\bibnamefont
  {Shen}}, \bibinfo {author} {\bibfnamefont {A.}~\bibnamefont {Narayanaswamy}},
  \ and\ \bibinfo {author} {\bibfnamefont {G.}~\bibnamefont {Chen}},\ }\href
  {\doibase 10.1021/nl901208v} {\bibfield  {journal} {\bibinfo  {journal} {Nano
  Lett}\ }\textbf {\bibinfo {volume} {9}},\ \bibinfo {pages} {2909}
  (\bibinfo {year} {2009})}\BibitemShut {NoStop}%
\bibitem [{\citenamefont {Guha}\ \emph {et~al.}(2012)\citenamefont {Guha},
  \citenamefont {Otey}, \citenamefont {Poitras}, \citenamefont {Fan},\ and\
  \citenamefont {Lipson}}]{Guha2012}%
  \BibitemOpen
  \bibfield  {author} {\bibinfo {author} {\bibfnamefont {B.}~\bibnamefont
  {Guha}}, \bibinfo {author} {\bibfnamefont {C.}~\bibnamefont {Otey}}, \bibinfo
  {author} {\bibfnamefont {C.~B.}\ \bibnamefont {Poitras}}, \bibinfo {author}
  {\bibfnamefont {S.}~\bibnamefont {Fan}}, \ and\ \bibinfo {author}
  {\bibfnamefont {M.}~\bibnamefont {Lipson}},\ }\href {\doibase
  10.1021/nl301708e} {\bibfield  {journal} {\bibinfo  {journal} {Nano Lett}\
  }\textbf {\bibinfo {volume} {12}},\ \bibinfo {pages} {4546} (\bibinfo {year}
  {2012})}\BibitemShut {NoStop}%
\bibitem [{\citenamefont {Rousseau}\ \emph {et~al.}(2009)\citenamefont
  {Rousseau}, \citenamefont {Siria}, \citenamefont {Jourdan}, \citenamefont
  {Volz}, \citenamefont {Comin}, \citenamefont {Chevrier},\ and\ \citenamefont
  {Greffet}}]{Rousseau2009}%
  \BibitemOpen
  \bibfield  {author} {\bibinfo {author} {\bibfnamefont {E.}~\bibnamefont
  {Rousseau}}, \bibinfo {author} {\bibfnamefont {A.}~\bibnamefont {Siria}},
  \bibinfo {author} {\bibfnamefont {G.}~\bibnamefont {Jourdan}}, \bibinfo
  {author} {\bibfnamefont {S.}~\bibnamefont {Volz}}, \bibinfo {author}
  {\bibfnamefont {F.}~\bibnamefont {Comin}}, \bibinfo {author} {\bibfnamefont
  {J.}~\bibnamefont {Chevrier}}, \ and\ \bibinfo {author} {\bibfnamefont
  {J.-J.}\ \bibnamefont {Greffet}},\ }\href {\doibase 10.1038/nphoton.2009.144}
  {\bibfield  {journal} {\bibinfo  {journal} {Nat Photon}\ }\textbf
  {\bibinfo {volume} {3}},\ \bibinfo {pages} {514} (\bibinfo {year}
  {2009})}\BibitemShut {NoStop}%
\bibitem [{\citenamefont {Sasihithlu}\ and\ \citenamefont
  {Narayanaswamy}(2011{\natexlab{b}})}]{Sasihithlu2011}%
  \BibitemOpen
  \bibfield  {author} {\bibinfo {author} {\bibfnamefont {K.}~\bibnamefont
  {Sasihithlu}}\ and\ \bibinfo {author} {\bibfnamefont {A.}~\bibnamefont
  {Narayanaswamy}},\ }\href {\doibase 10.1103/PhysRevB.83.161406} {\bibfield
  {journal} {\bibinfo  {journal} {Phys Rev B}\ }\textbf {\bibinfo
  {volume} {83}},\ \bibinfo {pages} {161406} (\bibinfo {year}
  {2011}{\natexlab{b}})}\BibitemShut {NoStop}%
\bibitem [{\citenamefont {Song}\ \emph {et~al.}(2015)\citenamefont {Song},
  \citenamefont {Ganjeh}, \citenamefont {Sadat}, \citenamefont {Thompson},
  \citenamefont {Fiorino}, \citenamefont {Fern{\'{a}}ndez-Hurtado},
  \citenamefont {Feist}, \citenamefont {Garc{\'{i}}a-Vidal}, \citenamefont
  {Cuevas}, \citenamefont {Reddy},\ and\ \citenamefont {Meyhofer}}]{Song2015a}%
  \BibitemOpen
  \bibfield  {author} {\bibinfo {author} {\bibfnamefont {B.}~\bibnamefont
  {Song}}, \bibinfo {author} {\bibfnamefont {Y.}~\bibnamefont {Ganjeh}},
  \bibinfo {author} {\bibfnamefont {S.}~\bibnamefont {Sadat}}, \bibinfo
  {author} {\bibfnamefont {D.}~\bibnamefont {Thompson}}, \bibinfo {author}
  {\bibfnamefont {A.}~\bibnamefont {Fiorino}}, \bibinfo {author} {\bibfnamefont
  {V.}~\bibnamefont {Fern{\'{a}}ndez-Hurtado}}, \bibinfo {author}
  {\bibfnamefont {J.}~\bibnamefont {Feist}}, \bibinfo {author} {\bibfnamefont
  {F.~J.}\ \bibnamefont {Garc{\'{i}}a-Vidal}}, \bibinfo {author} {\bibfnamefont
  {J.~C.}\ \bibnamefont {Cuevas}}, \bibinfo {author} {\bibfnamefont
  {P.}~\bibnamefont {Reddy}}, \ and\ \bibinfo {author} {\bibfnamefont
  {E.}~\bibnamefont {Meyhofer}},\ }\href {\doibase 10.1038/nnano.2015.6}
  {\bibfield  {journal} {\bibinfo  {journal} {Nat Nanotechnol}\ }\textbf
  {\bibinfo {volume} {10}},\ \bibinfo {pages} {253} (\bibinfo {year}
  {2015})}\BibitemShut {NoStop}%
\bibitem [{\citenamefont {Edalatpour}\ and\ \citenamefont
  {Francoeur}(2014)}]{Edalatpour2014}%
  \BibitemOpen
  \bibfield  {author} {\bibinfo {author} {\bibfnamefont {S.}~\bibnamefont
  {Edalatpour}}\ and\ \bibinfo {author} {\bibfnamefont {M.}~\bibnamefont
  {Francoeur}},\ }\href {\doibase 10.1016/j.jqsrt.2013.08.021} {\bibfield
  {journal} {\bibinfo  {journal} {J Quant
  Spectrosc and Radiat Transfer}\ }\textbf {\bibinfo {volume} {133}},\ \bibinfo {pages}
  {364} (\bibinfo {year} {2014})}\BibitemShut {NoStop}%
\bibitem [{\citenamefont {Edalatpour}\ and\ \citenamefont
  {Francoeur}(2014)}]{Edalatpour2016}%
  \BibitemOpen
  \bibfield  {author} {\bibinfo {author} {\bibfnamefont {S.}~\bibnamefont
  {Edalatpour}}, \bibinfo {author} {\bibfnamefont {M.}~\bibnamefont {Francoeur}},\ }\href
  {\doibase 10.1103/PhysRevB.94.045406} {\bibfield  {journal} {\bibinfo
  {journal} {Phys Rev B}\ }\textbf {\bibinfo {volume} {94}},\ \bibinfo
  {pages} {045406} (\bibinfo {year} {2016})}\BibitemShut {NoStop}%
\bibitem [{\citenamefont {Olver}\ \emph {et~al.}()\citenamefont {Olver},
  \citenamefont {{Olde Daalhuis}}, \citenamefont {Lozier}, \citenamefont
  {Schneider}, \citenamefont {Boisvert}, \citenamefont {Clark}, \citenamefont
  {Miller},\ and\ \citenamefont {Saunders}}]{Olver2017}%
  \BibitemOpen
  \bibfield  {author} {\bibinfo {author} {\bibfnamefont {F.~W.~J.}\
  \bibnamefont {Olver}}, \bibinfo {author} {\bibfnamefont {A.~B.}\ \bibnamefont
  {{Olde Daalhuis}}}, \bibinfo {author} {\bibfnamefont {D.~W.}\ \bibnamefont
  {Lozier}}, \bibinfo {author} {\bibfnamefont {B.~I.}\ \bibnamefont
  {Schneider}}, \bibinfo {author} {\bibfnamefont {R.~F.}\ \bibnamefont
  {Boisvert}}, \bibinfo {author} {\bibfnamefont {C.~W.}\ \bibnamefont {Clark}},
  \bibinfo {author} {\bibfnamefont {B.~R.}\ \bibnamefont {Miller}}, \ and\
  \bibinfo {author} {\bibfnamefont {B.~V.}\ \bibnamefont {Saunders}},\ }\href
  {http://dlmf.nist.gov/} {\enquote {\bibinfo {title} {{NIST Digital Library of
  Mathematical Functions}},}\ }\bibinfo {howpublished} {http://dlmf.nist.gov/,
  Release 1.0.15 of 2017-06-01}\BibitemShut {NoStop}%
\bibitem [{\citenamefont {Kaiser}\ and\ \citenamefont
  {Schweiger}(1993)}]{Kaiser1993}%
  \BibitemOpen
  \bibfield  {author} {\bibinfo {author} {\bibfnamefont {T.}~\bibnamefont
  {Kaiser}}\ and\ \bibinfo {author} {\bibfnamefont {G.}~\bibnamefont
  {Schweiger}},\ }\href {\doibase 10.1063/1.168475} {\bibfield  {journal}
  {\bibinfo  {journal} {Comput in Phys}\ }\textbf {\bibinfo {volume}
  {7}},\ \bibinfo {pages} {682} (\bibinfo {year} {1993})}\BibitemShut {NoStop}%
\bibitem [{\citenamefont {Bohren}\ and\ \citenamefont
  {Huffman}(2004)}]{Bohren2004}%
  \BibitemOpen
  \bibfield  {author} {\bibinfo {author} {\bibfnamefont {C.~F.}\ \bibnamefont
  {Bohren}}\ and\ \bibinfo {author} {\bibfnamefont {D.~R.}\ \bibnamefont
  {Huffman}},\ }\href@noop {} {\emph {\bibinfo {title} {{Absorption and
  Scattering of Light by Small Particles}}}}\ (\bibinfo  {publisher}
  {Wiley-VCH},\ \bibinfo {year} {2004})\BibitemShut {NoStop}%
\bibitem [{\citenamefont {Zhou}\ and\ \citenamefont {Hu}(2006)}]{Zhou2006}%
  \BibitemOpen
  \bibfield  {author} {\bibinfo {author} {\bibfnamefont {X.}~\bibnamefont
  {Zhou}}\ and\ \bibinfo {author} {\bibfnamefont {G.}~\bibnamefont {Hu}},\
  }\href {\doibase 10.1103/PhysRevE.74.026607} {\bibfield  {journal} {\bibinfo
  {journal} {Phys Rev E}\ }\textbf {\bibinfo {volume} {74}},\ \bibinfo
  {pages} {026607} (\bibinfo {year} {2006})}\BibitemShut {NoStop}%
\bibitem [{\citenamefont {Zheng}\ and\ \citenamefont
  {Ghanekar}(2015)}]{Zheng2015}%
  \BibitemOpen
  \bibfield  {author} {\bibinfo {author} {\bibfnamefont {Y.}~\bibnamefont
  {Zheng}}\ and\ \bibinfo {author} {\bibfnamefont {A.}~\bibnamefont
  {Ghanekar}},\ }\href {\doibase 10.1063/1.4907913} {\bibfield  {journal}
  {\bibinfo  {journal} {J Appl Phys}\ }\textbf {\bibinfo
  {volume} {117}},\ \bibinfo {pages} {064314} (\bibinfo {year}
  {2015})}\BibitemShut {NoStop}%
\bibitem [{\citenamefont {Narayanaswamy}\ and\ \citenamefont
  {Zheng}(2013{\natexlab{b}})}]{Narayanaswamy2013b}%
  \BibitemOpen
  \bibfield  {author} {\bibinfo {author} {\bibfnamefont {A.}~\bibnamefont
  {Narayanaswamy}}\ and\ \bibinfo {author} {\bibfnamefont {Y.}~\bibnamefont
  {Zheng}},\ }\href {\doibase 10.1103/PhysRevA.88.012502} {\bibfield  {journal}
  {\bibinfo  {journal} {Phys Rev A}\ }\textbf {\bibinfo {volume} {88}},\
  \bibinfo {pages} {012502} (\bibinfo {year} {2013}{\natexlab{b}})}\BibitemShut
  {NoStop}%
\bibitem [{\citenamefont {Tai}(1994)}]{tai1994dyadic}%
  \BibitemOpen
  \bibfield  {author} {\bibinfo {author} {\bibfnamefont {C.-T.}\ \bibnamefont
  {Tai}},\ }\href@noop {} {\emph {\bibinfo {title} {Dyadic Green Functions in
  Electromagnetic Theory}}}\ (\bibinfo  {publisher} {Institute of Electrical \&
  Electronics Engineers (IEEE)},\ \bibinfo {year} {1994})\BibitemShut {NoStop}%
\end{thebibliography}

%


\end{document}